\documentclass[preprint,aps,prd]{revtex4-1}
\usepackage[pdftex]{graphicx}
\usepackage{amsmath,amssymb}
\usepackage{bm}
\usepackage{tikz-feynman}

\newcommand{\Slash}[1]{{\ooalign{\hfil/\hfil\crcr$#1$}}}

\begin{document}

\preprint{OU-HET-1072}
\preprint{KEK-CP-0377}
\date{\today}

\title{Towards fully non-perturbative computation of inelastic
  $\ell N$ scattering cross sections from lattice QCD} 

\author{Hidenori Fukaya}
\affiliation{Department of Physics, Osaka University,
  Toyonaka 560-0043, Japan} 

\author{Shoji Hashimoto}
\affiliation{Theory Center, Institute of Particle and Nuclear Studies,
  High Energy Accelerator Research Organization (KEK), Tsukuba
  305-0801, Japan}
\affiliation{School of High Energy Accelerator Science,
  The Graduate University for Advanced Studies (SOKENDAI),
  Tsukuba 305-0801, Japan}

\author{Takashi Kaneko}
\affiliation{Theory Center, Institute of Particle and Nuclear Studies,
  High Energy Accelerator Research Organization (KEK), Tsukuba
  305-0801, Japan}
\affiliation{School of High Energy Accelerator Science,
  The Graduate University for Advanced Studies (SOKENDAI),
  Tsukuba 305-0801, Japan}

\author{Hiroshi Ohki}
\affiliation{Department of Physics, Nara Women’s University,
  Nara 630-8506, Japan}
\affiliation{RIKEN/BNL Research Center,
  Brookhaven National Laboratory, Upton,
  NY 11973, USA}

\begin{abstract}
  We propose a fully non-perturbative method to compute inelastic
  lepton-nucleon ($\ell N$) scattering cross sections using lattice
  QCD.
  The method is applicable even at low energies, such as
  the energy region relevant for the recent and future
  neutrino-nucleon ($\nu N$) scattering experiments,
  for which perturbative analysis is invalidated. 
  The basic building block is the forward Compton-scattering
  amplitude, or the hadronic tensor, computed on a Euclidean lattice.
  Total cross section is constructed from the hadronic tensor
  by multiplying a phase space factor and integrating over the energy
  and momentum of final hadronic states.
  The energy integral that induces a sum over all possible final states is 
  performed implicitly by promoting the phase space factor to an
  operator written in terms of the transfer matrix on the lattice. 
  The formalism is imported from that of the inclusive semileptonic
  $B$ meson decay \cite{Gambino:2020crt}.
  It can be generalized to compute the $\ell N$ scattering cross
  sections and their moments, as well as the virtual correction
  to the nuclear $\beta$-decay.
  Necessary quark-line contractions for two current insertions 
  corresponding to the Compton amplitude to be computed on the lattice
  are summarized.
\end{abstract}

\maketitle

\section{Introduction}
\label{sec:introduction}
Deep inelastic scattering (DIS) played an important role on the
emergence of the parton picture of nucleon and the discovery of the
asymptotic freedom, which lead to the fundamental theory of strong
interaction, Quantum Chromodynamics (QCD). 
The key finding was that the structure function $W^{\mu\nu}$
may well be described by a sum of parton distributions,
and the partons inside nucleon behave as if they are free particles
despite the strong force that binds them together.
And, one does not have to take account of the details of the
individual hadronic final states when calculating the total cross
section, which is called the quark-hadron duality.
Perturbative analysis of DIS is based on these observations.

Theoretically, the DIS process is factorized into perturbative scattering
amplitudes of elementary partons, {\it i.e.} quarks and gluons, and
non-perturbative functions called parton distribution functions (PDFs),
which have to be determined by fitting the experimental data.
(For a review of the early days of the development, see, for instance,
\cite{Altarelli:1981ax}.)
However, the separation of perturbative and non-perturbative
contributions, so-called the factorization \cite{Collins:1989gx}, is
not obvious, especially for higher twist contributions which become
relevant when one tries to calculate beyond the leading order of the 
operator product expansion (OPE).
There is even an intrinsic difficulty in the scale separation using
OPE due to the renormalon ambiguity (see, for instance,
\cite{Martinelli:1996pk,Beneke:1998ui}). 
Therefore, the theoretical analysis of the lepton-nucleon ($\ell N$)
scattering has been limited in the high-energy regime, where the
effects of higher twist operators are negligible.

The other limit, {\it i.e.} the low-energy limit, is given by the
elastic scattering of nucleon, whose form factors can be
calculated using lattice QCD, and the theoretical computation is based
on a solid ground. 
In the intermediate energy region, where extra pion(s) can be
generated but the process is still highly non-perturbative,
the details of the final state hadrons also become relevant, and one
finds resonances and other structures in the differential cross
sections, {\it i.e.} the duality is violated.
The theoretical analysis of such processes is far more complicated and
quantitative computation based on QCD has remained impractical. 

In this work we construct a formalism to compute total inelastic
cross-section of the $\ell N$ scattering fully non-perturbatively using
lattice QCD. 
The formalism does not involve the parton distribution functions.
Rather, we directly compute the (weighted) integral of the hadronic
structure function, that defines the total cross section.
Since the method does not rely on the separation of perturbative and
non-perturbative contributions, there is no fundamental limitation on
the range of $Q^2$ that can be treated.
Namely, the analysis is useful even when the momentum transfer $Q^2$ is
small, where the standard perturbative analysis is not applicable.
This is an advantage for the analysis of the neutrino-nucleon scattering
cross section at low energy, {\it i.e.} the initial neutrino energy in
the range from several hundred MeV to a few GeV,
which is the energy range relevant for the neutrino
experiments such as T2K, NO$\nu$A and DUNE \cite{Katori:2016yel}.
In fact, in this energy region, the cross section is neither dominated
by the quasi-elastic scattering, nor described by deep inelastic
scattering.
The contributions of inelastic processes including a few pions
is significant, and there has so far been no established theoretical
method that can reliably treat this region.

We propose to use the hadronic tensor, or equivalently the forward
Compton-scattering amplitude, computed using lattice QCD.
The hadronic tensor is defined as a matrix element of two
weak-currents inserted between nucleon states that represent the
initial target nucleon. 
On the Euclidean lattice, the two currents are placed away from each
other in space and \emph{imaginary} time directions, so that the
matrix element obtained after Fourier (or inverse-Laplace) transform
is limited in a 
unphysical kinematical region where the final states never become
on-shell. 
The physically relevant structure function corresponds to the
imaginary part of the hadronic tensor on the physical cut, which is
hard to obtain with ordinary lattice computation methods.
(One has to solve the inverse problem. See below.)
Our proposal is to use the method developed to calculate the total
semileptonic decay rate of the $B$ meson \cite{Gambino:2020crt}, that
enables us to compute the the sum over all possible states between the
two weak currents. 
With an appropriate weight of the energy, so-called the phase space, 
the sum corresponds to the total scattering cross section after
integrating over spatial momentum. 

An approximation is introduced for the phase space factor that appears
as a weight in the energy integral (or the sum over
all possible final states).
As we describe later, the approximation is very precise when the
weight is a smooth function of the energy. 
This is not the case in practice because of the sharp upper limit on
the energy transfer from the initial lepton to the hadronic system,
and the weight factor is actually a discontinuous function of energy.
Therefore, the associated error has to be carefully examined.
The experience in the study of the $B$ meson semileptonic decay
suggests that it is not substantial \cite{Gambino:2020crt}, and we
study the size of the potential systematic effect for the case of the
$\ell N$ scattering assuming a form of spectrum of the final states.

Recently, some attempts have been proposed to compute the inclusive
processes using lattice QCD
\cite{Hansen:2017mnd,Hansen:2019idp,Bulava:2019kbi,Briceno:2019opb,%
Liang:2019frk,Chambers:2017dov,Can:2020sxc}. 
In the context of the $\ell N$ scattering, they correspond to the
calculation of the forward Compton-scattering amplitude, which is the
same as what we treat.
The kinematical point accessible on the lattice is apart from the
physical cut, and the methods are devised to relate the lattice data
to the physical amplitude by solving the inverse problem, which is
extremely difficult and there is no satisfactory solution that
provides reliable quantitative results for the physical amplitudes.
The difference of our proposal is that we do \emph{not} try to solve the
inverse problem, but we advocate to compute only the energy integral of
the physical amplitude. 
In that way, the difficulty of the inverse problem is circumvented,
and the physical quantity of interest is accessible.

The method of energy integral with two current insertions
may potentially be applied also to the study of the Cottingham formula
that relates the electromagnetic contribution to the proton-neutron mass
difference to the forward Compton-scattering amplitude
\cite{Cottingham:1963zz}.
(See also \cite{WalkerLoud:2012bg,Gasser:2020mzy,Gasser:2020hzn} and
references therein.)
The formula has the form of an integral of the hadronic tensor in
terms of the inserted energy and momentum, which is the same structure
as the total cross section, but there is an additional complexity due to the
ultraviolet divergence and some dedicated analysis would be
necessary. 
It is also related to the two-photon exchange diagram in the $\ell N$
scattering, which is relevant to the precise computation of the 
electromagnetic radius of proton \cite{Borisyuk:2019gym}.

Another interesting application is the $\gamma W$ exchange
contribution to the nuclear $\beta$ decay.
At $O(\alpha)$, the nucleon $\gamma W$ box diagram gives rise to
a nuclear-structure independent correction to the super-allowed
nuclear $\beta$ decays. 
Its hadronic uncertainty limits the accuracy of the determination of
the Cabibbo-Kobayashi-Maskawa matrix element $|V_{ud}|$, for which 
recent phenomenological estimates suggest a tension with the CKM
unitarity \cite{Seng:2018yzq,Czarnecki:2019mwq}.
In the $\gamma W$ box diagram involving a lepton and a nucleon in the
initial/final states connected by a photon propagator and a $W$ boson
propagator,
the hadronic states between the weak and electromagnetic currents can 
be any excited states (with a corresponding quantum number),
and their contributions have to be taken into account.
The integral over the internal momentum resembles that of the total
$\ell N$ cross section, and the method developed in this work is
applicable. 

The lattice computation of the necessary four-point functions
including two current insertions is a major challenge especially when
flavor-changing currents are involved.
The calculation of the forward Compton-scattering amplitude has been
performed so far using the Feynman-Hellmann technique
\cite{Chambers:2017dov,Can:2020sxc}.
In this work we figure out all the necessary quark-line contractions
including the cases of the flavor-diagonal, flavor-changing, as well
as those for the $\beta$-decay for future computations

This paper is organized as follows.
In Sec.~\ref{sec:spectral_func} we outline the method to perform the
integral over all possible intermediate states with an appropriate
weight.
The method is essentially the same as those in \cite{Bailas:2020qmv}
for current two-point functions and in \cite{Gambino:2020crt} for $B$
meson inclusive semileptonic decays.
The kinematics of the $\ell N$ scattering is summarized in 
Sec.~\ref{sec:kinematics} and the master formula of the total cross
section is given.
An application to the $\beta$ decay is discussed in Sec.~\ref{sec:Wgamma}.
Explicit formula and some examples of the energy integral are then
described in Sec.~\ref{sec:energy_integral}.
Potential errors due to approximation are also investigated.
We provide some details of the quark-line contractions, which are
necessary to compute the hadronic tensor.
They are especially complicated when the charged current, which
involves the change of flavors, as described in
Sec.~\ref{sec:contraction}. 
Further discussions and future prospect are given in
Sec.~\ref{sec:discussions}, and our conclusions are in
Sec.~\ref{sec:conclusions}.

\section{Lattice correlators and spectral functions}
\label{sec:spectral_func}

We first outline the basic idea of the method we are proposing.
As in the standard analysis, the inelastic scattering cross section
can be written in terms of a product of the leptonic tensor and
hadronic tensor.
Using the spectral decomposition, the hadronic tensor can be viewed as
a spectral function; the total cross section is its integral over
final-state energy with a weight factor determined by the leptonic tensor.
Therefore, once the spectral function is extracted from the
lattice data, the cross section can be obtained as emphasized in
\cite{Hansen:2017mnd}.
In practice, the extraction of the spectral function needs a solution
of the inverse problem, which is a well-known example of ill-posed
problems; there have been no practical methods developed so far that
allow sufficiently accurate quantitative estimates  
(see, for instance, \cite{Hansen:2017mnd,Hansen:2019idp,Karpie:2019eiq}).
The problem can be overcome by combining the energy integral with the
computation of the forward Compton-scattering matrix element
\cite{Bailas:2020qmv,Gambino:2020crt},  
as outlined below.
The idea was developed from an analysis to relate the different
kinematical regions of the Compton amplitude using analytic
continuation \cite{Hashimoto:2017wqo}. 

Let us consider a matrix element of a nucleon with two current
insertions 
\begin{align}
  C(t;\bm{q}) =
  & \sum_{\bm{x}} e^{i\bm{q}\cdot\bm{x}}
    \langle N| J(\bm{x},t) J(\bm{0},0) |N\rangle
    \label{eq:Ct}
  \\
  =
  &\frac{1}{V}
    \langle N|\tilde{J}(-\bm{q})e^{-\hat{H}t}\tilde{J}(\bm{q})|N\rangle.
    \label{eq:Ct_transfer}
\end{align}
For the moment, we ignore the Lorentz and flavor indices of the
current $J$ as well as the spin of the nucleon state $|N\rangle$;
a more concrete definition will be given in the following sections.
The matrix element of the form (\ref{eq:Ct}) can be computed on the
lattice from four-point correlation functions including the source
operators to create and annihilate the external state $|N\rangle$.
On the second line, we introduce a Fourier transform of the current
$\tilde{J}(\bm{q})\equiv
\sum_{\bm{x}} e^{-i\bm{q}\cdot\bm{x}} J(\bm{x})$,
and $V$ is the spatial volume of the lattice.
We assume that the initial state $|N\rangle$ is at rest, so that
(\ref{eq:Ct}) describes the process where a momentum $\bm{q}$ is
injected at $t=0$ and taken out at $t$.
(An extension to the case of non-zero initial momentum of nucleon is
straightforward.) 
The time separation between the two currents is imaginary since the
calculation is performed on the Euclidean lattice. 
The time evolution is then described by a transfer matrix
$e^{-\hat{H}}$ with $\hat{H}$ the Hamiltonian of the system.
Here we use the lattice unit, {\it i.e.} the lattice spacing is $a=1$.
In the analysis of the lattice data, we do not need the explicit form
of the lattice Hamiltonian $\hat{H}$;
it is introduced to remind us that the evolution of individual
intermediate eigenstates with energy $\omega$ is given by $e^{-\omega t}$.
The eigenvalues of $z\equiv e^{-\hat{H}}$ are limited in the range
$[0,1]$.

The correlator (\ref{eq:Ct_transfer}) can be formally decomposed
into the contributions of individual energy eigenstates, 
\begin{equation}
  \label{eq:Ct_spectral}
  C(t;\bm{q})=\int_0^\infty d\omega\, \rho(\omega;\bm{q}) e^{-\omega t},
\end{equation}
with the spectral function
\begin{equation}
  \label{eq:rho}
  \rho(\omega;\bm{q}) \propto \sum_{X(\bm{q})} \delta(\omega-E_{X(\bm{q})})
  |\langle X(\bm{q})|J|N\rangle|^2,
\end{equation}
where the sum runs over all possible states $X(\bm{q})$ with a
specified momentum $\bm{q}$.
The $\delta$-function in (\ref{eq:rho}) picks the states of a certain
energy $\omega$ among all possible states with energy $E_{X(\bm{q})}$.

Since the spectral function $\rho(\omega;\bm{q})$ describes the
transition rate of the initial state $|N\rangle$ to the states with a
certain energy $\omega$ and momentum $\bm{q}$,
the total cross section can be written as an integral of the 
spectral function with an appropriate weight function
$K(\omega;\bm{q})$:
\begin{equation}
  \label{eq:Gamma}
  \Gamma = \int d^3\bm{q} \int_0^\infty d\omega\,
  K(\omega;\bm{q}) \rho(\omega;\bm{q}).
\end{equation}
The weight function is determined by the details of the process
of interest.
For the inelastic $\ell N$ scattering, the spectral function
corresponds to the hadronic tensor and the weight function 
$K(\omega;\bm{q})$ represents the phase space of the scattering
specified by the leptonic tensor.
Since the energy integral in (\ref{eq:Gamma}) extends up to infinity, 
the kinematical upper limit for the energy $\omega$ is also encoded in
the weight function by a step function.
The momentum $\bm{q}$ is also integrated over to obtain the total
rate. 

One approach to obtain the total rate $\Gamma$ from (\ref{eq:Gamma}) 
would be to first extract the spectral function $\rho(\omega;\bm{q})$
from the lattice correlator by solving the inverse problem 
(\ref{eq:Ct_spectral}) and then to use it in (\ref{eq:Gamma}).
The problem is, however, that the inverse problem is extremely
difficult in practice since the lattice data $C(t;\bm{q})$ are known
only at limited values of $t$ with non-negligible statistical noise.
In the context of the nucleon structure, several methods have been
proposed to solve the inverse problem, {\it e.g.}
a reconstruction through moments \cite{Chambers:2017dov,Can:2020sxc},
the Backus-Gilbert method \cite{Hansen:2017mnd}.
More extensive tests of various methods in the market,
including the Maximum Entropy Method, Bayesian reconstruction, and
neural network methods are found in
\cite{Karpie:2019eiq,Liang:2019frk}.
Unfortunately, none of them allows fully quantitative computation of
the spectral function $\rho(\omega;\bm{q})$ as a function of $\omega$.
This is only natural because of the complicated structure of the
spectrum including resonances and scattering states with interactions.
To circumvent the problem, smeared spectral function has been
considered.
It is defined as
$\bar\rho(\omega)=\int d\omega' \Delta(\omega,\omega')\rho(\omega')$
with a certain smearing kernel $\Delta(\omega,\omega')$
that typically has a peak at $\omega\approx\omega'$ and rapidly
decreases for larger $|\omega-\omega'|$.
The smeared spectrum $\bar{\rho}(\omega)$ is easier to reconstruct
with limited numerical data \cite{Hansen:2019idp,Bailas:2020qmv}
when the smearing width is sufficiently large.
The computation of the integral (\ref{eq:Gamma}) from
$\bar{\rho}(\omega)$ would then become another non-trivial problem
as it needs to take the limit of vanishing smearing width.

A practical method to actually compute the smeared spectral function
was proposed in \cite{Bailas:2020qmv}, and it has been applied for the
inclusive decay rate of $B$ meson \cite{Gambino:2020crt}.
The key idea was to identify the weight function $K(\omega;\bm{q})$
in (\ref{eq:Gamma}) as a smearing kernel,
and then to perform the $\omega$-integral using the correlator
computed on the Euclidean lattice. 
The $\omega$-integral can be carried out formally using the relation 
\begin{align}
  \int\! d\omega\, K(\omega;\bm{q}) \rho(\omega;\bm{q})
  \propto
  &
    \int\! d\omega\, K(\omega;\bm{q})
    \int\! \frac{d^3\bm{P}_X}{(2\pi)^3}
    \sum_{X(\bm{P}_X)} \delta(\omega-E_{X(\bm{P}_X)})
    (2\pi)^3\delta^{(3)}(\bm{P}_X-\bm{q})
    \nonumber\\ 
  & \;\;\;\;\;\;
    \times
    \langle N| \tilde{J}(-\bm{q})|X(\bm{P}_X)\rangle
    \langle X(\bm{P}_X)|\tilde{J}(\bm{q}) |N\rangle
    \\
  = &
      \label{eq:K(w)}
    \int\! d\omega\, K(\omega;\bm{q})
    \langle N|
    \tilde{J}(-\bm{q}) \delta(\hat{H}-\omega)
    \tilde{J}(\bm{q})|N\rangle
  \\
  = &
    \label{eq:K(H)}
    \langle N|
    \tilde{J}(-\bm{q}) K(\hat{H};\bm{q})
      \tilde{J}(\bm{q})|N(\bm{0})\rangle.
\end{align}
On the first line, 
the definition of the spectral function
(\ref{eq:rho}) and
$\int d^3\bm{P}_X \delta^{(3)}(\bm{P}_X-\bm{q})=1$
are inserted.
The sum over all possible states
$\sum_{X(\bm{q})}|X(\bm{q})\rangle\langle X(\bm{q})|$
is performed with the $\delta$-function
$\delta(\omega-E_{X(\bm{q})})$,
replacing the energy of individual states $E_{X(\bm{q})}$ by the
Hamiltonian $\hat{H}$.

The intermediate form (\ref{eq:K(w)}) can be viewed as a smeared
spectral function.
In fact, a spectral function
$\langle N|\tilde{J}(-\bm{q}) \delta(\hat{H}-\omega)
\tilde{J}(\bm{q})|N\rangle$
corresponding to the $\ell N$ scattering process is integrated over
the energy with a smearing factor $K(\omega,\bm{q})$.
This smearing kernel does not have a peak structure around $\omega$,
but the mathematical form is equivalent.
On the last line, the $\omega$-integral is carried out by introducing
an \emph{operator} $K(\hat{H};\bm{q})$.
The remaining task is then to find an expression of
$K(\hat{H};\bm{q})$ that can be practically implemented in the lattice
calculation.

The use of the smeared spectral function has been extended towards a
different direction, {\it i.e.} to compute the scattering amplitude 
\cite{Bulava:2019kbi,Briceno:2019opb}, where the smearing kernel is
identified as a factor that appears in the LSZ reduction formula, so
that the necessary scattering amplitude is directly obtained.
The expression (\ref{eq:K(H)}) should also be applicable in such
a case. 

Comparing (\ref{eq:K(H)}) with (\ref{eq:Ct_transfer}), we notice that
the integral can be evaluated using the lattice correlators
if the operator $K(\hat{H};\bm{q})$ is approximated by a
polynomial of the form
\begin{equation}
  K(\hat{H};\bm{q}) \simeq
  k_0(\bm{q}) + k_1(\bm{q}) e^{-\hat{H}}+ k_2(\bm{q}) e^{-2\hat{H}} + \cdots
  + k_N(\bm{q}) e^{-N\hat{H}},
  \label{eq:Kapprox}
\end{equation}
because the matrix element of the right-hand side is nothing but
$C(t;\bm{q})$'s up to a normalization factor.
There may be various ways to construct this approximation, and the
method introduced in \cite{Bailas:2020qmv} uses the Chebyshev
approximation, which is described in the following.

The Chebyshev approximation is \emph{not} an expansion in terms of
some small parameters. 
Rather, it attempts to approximate the whole function
$K(\omega,\bm{q})$ in $\omega\in[0,\infty]$
by a set of orthogonal functions constructed as polynomials of
$z\equiv e^{-\omega}$. 
In other words, one may consider the kernel as a function of $z$,
which can take values between 0 and 1, and expand $K(-\ln z,\bm{q})$
using an orthonormal set of functions $\{T_j^*(z)\}$ with $j=0$, 1, 2,
$\cdots, N$.
Here, $T_j^*(x)$ represents the shifted Chebyshev polynomials, which
are related to the standard Chebyshev polynomials of the first kind
$T_j(x)$ as $T_j^*(x)\equiv T_j(2x-1)$.
The shifted Chebyshev polynomials $T_j^*(x)$'s are defined in
$0\le x\le 1$.
The polynomial order $N$ controls the precision of the approximation.

\begin{figure}[tbp]
  \centering
  \includegraphics[width=10cm]{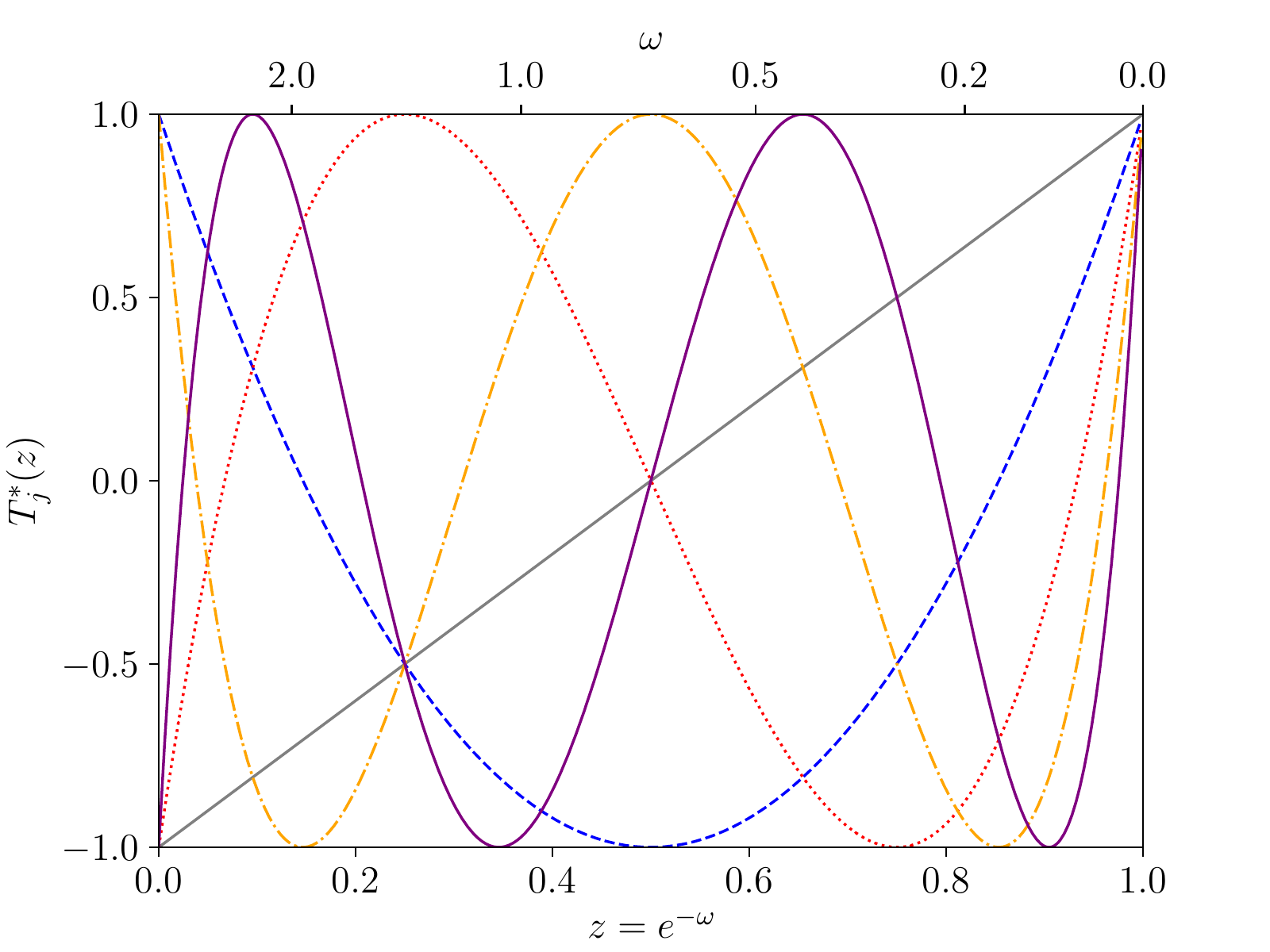}
  \caption{
    Shifted Chebyshev polynomials $T_j^*(z)$ for
    $j$ = 1 (gray line), 2 (red dotted), 3 (blue dashed), 4 (orange
    dot-dashed), 5 (purple solid). 
    The corresponding values of $\omega$ for $z=e^{-\omega}$ are shown
    on the horizontal axis (top).
  }
  \label{fig:cheb}
\end{figure}

The Chebyshev polynomials are obtained by a recurrence relation:
\begin{align}
  T_0(x)&=1, & T_1(x)&=x, & T_{j+1}(x)&=2xT_j(x)-T_{j-1}(x),
\end{align}
so that the shifted ones are derived as
\begin{align}
  T_0^*(x)&=1, & T_1^*(x)&=2x-1,
  & T^*_{j+1}(x)&=2(2x-1)T_j^*(x)-T_{j-1}^*(x).
\end{align}
The first few of the shifted Chebyshev polynomials are then
$T_2^*(x)=8x^2-8x+1$,
$T_3^*(x)=32x^3-48x^2+18x-1$, 
$T_4^*(x)=128x^4-256x^3+160x^2-32x+1$, and so on.
They are shown in Fig.~\ref{fig:cheb} as a function of $x$ (or $z$) as
well as $\omega$ through $z=e^{-\omega}$.

The Chebyshev polynomials are designed such that they evenly oscillate
between $-1$ and $+1$ with $j$ the number of nodes.
The Chebyshev approximation of a function $f(x)$ defined in $[0,1]$
has the form $f(x)\simeq c_0^*/2+\sum_{j=1}^N c_j^* T_j^*(x)$, and
each order represents (a sort of) frequency component of the function.
As the higher order terms are added, the approximation can reproduce
finer details of the original function.

The Chebyshev approximation of the matrix element of the operator
$K(\hat{H};\bm{q})$ 
and thus of the matrix element of $\hat{z}\equiv e^{-\hat{H}}$
can be obtained as follows. 
We define a state $|\psi(\bm{q})\rangle$ as
$|\psi(\bm{q})\rangle\equiv e^{-\hat{H}t_0}\tilde{J}(\bm{q})|N\rangle$
with $t_0$ a small time separation introduced to avoid any divergence
when evaluating $\langle\psi(\bm{q})|\psi(\bm{q})\rangle$.
The approximation may then be written as
\begin{equation}
  \frac{\langle\psi(\bm{q})|K(\hat{H};\bm{q})|\psi(\bm{q})\rangle}{
    \langle\psi(\bm{q})|\psi(\bm{q})\rangle}
  \simeq
  \frac{c_0^*(\bm{q})}{2} +
  \sum_{j=1}^N c_j^*(\bm{q})
  \frac{\langle\psi(\bm{q})|T_j^*(\hat{z})|\psi(\bm{q})\rangle}{
    \langle\psi(\bm{q})|\psi(\bm{q})\rangle},
  \label{eq:cheb_approx}
\end{equation}
which is based on the corresponding approximation formula
\begin{equation}
  K(\omega;\bm{q}) \simeq \frac{c_0^*(\bm{q})}{2} +
  \sum_{j=1}^N c_j^*(\bm{q}) T_j^*(z)
  \label{eq:cheb_approx_c}
\end{equation}
with $z=e^{-\omega}$.

The approximation evaluated on the state $|\psi(\bm{q})\rangle$, as
in (\ref{eq:cheb_approx}), may be obtained from that for a $c$-number
(\ref{eq:cheb_approx_c}) by considering a decomposition into energy
eigenstates $|i\rangle$ of energy $\omega_i$, {\it i.e.}
$|\psi\rangle=\sum_i a_i|i\rangle$.
Then, the matrix element of $T_j^*(\hat{z})$ may be written as a sum
$\sum_i |a_i|^2 T_j^*(e^{-\omega_i})$, while the matrix element of
the kernel operator is also given as
$\sum_i |a_i|^2 K(\omega_i)$,
each term of which may be approximated using (\ref{eq:cheb_approx_c}).

Each term of the right-hand side of (\ref{eq:cheb_approx}), the matrix
element of the Chebyshev polynomials $T_j^*(\hat{z})$ may be
constructed from those of $\hat{z}=e^{-\hat{H}}$ using
\begin{equation}
  \frac{C(t+2t_0)}{C(2t_0)} =
  \frac{
    \langle\psi(\bm{q})|e^{-\hat{H}t}|\psi(\bm{q})\rangle
  }{
    \langle\psi(\bm{q})|\psi(\bm{q})\rangle
  },
  \label{eq:C(t+2t0)/C(2t0)}
\end{equation}
which is immediately obtained from the lattice data
(\ref{eq:Ct_transfer}).
Then, the right-hand side of (\ref{eq:cheb_approx}) is nothing but a
linear combination of (\ref{eq:C(t+2t0)/C(2t0)}) with different $t$'s.

The coefficients $c_j^*(\bm{q})$ in (\ref{eq:cheb_approx}) are
obtained by an integral
\begin{equation}
  c_j^*(\bm{q}) = \frac{2}{\pi} \int_0^\pi d\theta\,
  K\left(-\ln\frac{1+\cos\theta}{2};\bm{q}\right) \cos(j\theta)
\end{equation}
according to the general formula of the Chebyshev approximation.
Since $K(\omega;\bm{q})=K(-\ln z;\bm{q})$ is a known function, the
coefficients can be obtained easily using numerical integration.

The Chebyshev approximation provides the \emph{best} possible
approximation at a given order of the polynomials of the form
(\ref{eq:cheb_approx}) and thus of any polynomials of that order.
It is the best in the sense that the maximum deviation from the
true function is minimal in the range
$0\le z\,(=e^{-\omega})\le 1$,
which covers all positive energy eigenvalues of the final
states. 

For smooth kernel functions $K(\omega;\bm{q})$, the coefficients
$c_j^*(\bm{q})$ rapidly decrease (often exponentially) for larger $j$,
so that the contributions from higher order terms are suppressed,
since the Chebyshev polynomials are bounded:
$|T_j^*(x)|\leq 1$.
When the kernel function $K(\omega;\bm{q})$ is non-smooth or even
discontinuous, the approximation becomes highly oscillatory near the
discontinuity and higher order terms become necessary to suppress them.
Examples relevant for the computation of the inelastic scattering
cross sections are discussed in Sec.~\ref{sec:energy_integral}. 

One may also consider to use the Chebyshev approximation to obtain
more information of the spectral function $\rho(\omega;\bm{q})$.
In fact, it is possible to introduce a kernel function that has a
finite value only in a small bin of $\omega$ and vanishes otherwise.
With such a filtering function, one can stochastically count the
number of energy eigenvalues in that bin, if the Chebyshev
approximation works. 
Unfortunately, the filtering function is highly discontinuous and thus
needs higher order Chebyshev polynomial terms to achieve good
approximation, which is not practical for this particular
application.
The Chebyshev eigenvalue filtering technique has been used in the
context of lattice QCD computation for the calculation of the Dirac
operator eigenvalue spectram, through which one can extract the chiral
condensate of QCD \cite{Cossu:2016eqs}
as well as the spectral function in the full energy
range \cite{Nakayama:2018ubk}.

Our proposal is to combine the operator representation
(\ref{eq:K(H)}) of the $\omega$-integral with the Chebyshev
polynomials in order to write it using the correlators computed on
the lattice.
The remaining integral over $\bm{q}$, see (\ref{eq:Gamma}), has to be
carried out with the lattice data obtained at several values of $\bm{q}$.
This program has been demonstrated for the $B$ meson inclusive
semileptonic decays in \cite{Gambino:2020crt}.
The formulation for the $\ell N$ scattering is described in the
following sections.

In (\ref{eq:Gamma}) the integral over the energy
$\omega$ of the hadronic final state corresponds to the sum over all
possible final states with a given spatial momentum $\bm{q}^2$.
Many of them are multi-particle states such as $N\pi$, $N\pi\pi$,
{\it etc.}, which have continuous spectra in the infinite volume
limit.
On the lattice of finite spatial extent, they are discretized to
satisfy the periodic boundary condition, so that the $\omega$-integral
is actually a sum over various allowed states.
The energy of each state receives power corrections of the form
$1/V$ \cite{Luscher:1986pf} and the limit of $V\to\infty$ has to be
taken.
We expect that the power-like finite volume effect is marginalized by
the integrals over $\omega$ and then $\bm{q}^2$, because the
finite-volume correction should be most significant for the low-energy
and low-momentum states while the total cross section receives more
contributions from higher energy regions due to the phase-space
enhancement. 
The problem remains severe when the upper limit of the energy integral
is relatively low so that only a limited number of states are
kinematically allowed due to the finite volume effect.
Studies of individual states, {\it e.g.} $N\pi$ states, in the finite
volume would be more useful in such cases.

\section{$\nu N$ scattering: kinematics}
\label{sec:kinematics}
In this section, we summarize the kinematics of the inelastic $\ell N$
scattering partly to establish our notations and to identify the phase
space factor that plays the role of the weight function of the energy
integral. 
We are particularly interested in the $\nu N$ scattering, which is 
relevant to the recent and future neutrino experiments, but the
formulation can also be applied for electromagnetic scattering of
electron (or muon) with a slight modification. 

\begin{figure}
  \centering
  \begin{tikzpicture}\begin{feynman}
      \vertex(nu){\(\nu\)};
      \vertex[below right=of nu] (vl);
      \vertex[above right=of vl] (ell){\(\ell\)};
      \vertex[below =of vl] (vh);
      \vertex[left=of vh] (N){\(N\)};
      \vertex[below right=of vh] (X){\(X\)};
      \diagram* {(nu)-- [fermion,edge label'=\(p\)] (vl)--
        [fermion,edge label'=\(p'\)] (ell),
        (vl)-- [boson,edge label'=\(W^{-}\)] (vh),
        (N)-- [fermion,thick,edge label'=\(P\)] (vh) -- [fermion,very thick,edge label'=\(P_X\)] (X),};
    \end{feynman}\end{tikzpicture}
  \caption{$\ell N$ scattering}
  \label{fig:lN_diagram}
\end{figure}
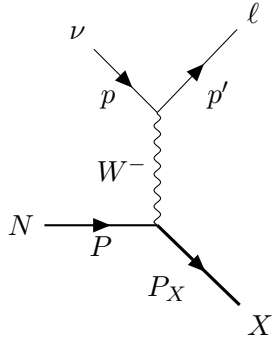

The diagram for the $\nu N$ scattering is shown in Fig.~\ref{fig:lN_diagram}.
We assign the energy-momentum as
$p^\mu=(E,\bm{p})$, $p'^\mu=(E',\bm{p}')$
for the incoming ($\nu$) and outgoing ($\ell$) leptons, and
$P^\mu=(M_N,\bm{0})$, $P_X^\mu=(\omega,\bm{P}_X)$
for the target nucleon ($N$) and outgoing hadronic system ($X$),
respectively.
The momentum transfer is then
$q^\mu=(p-p')^\mu=(E-E',\bm{p}-\bm{p}')$.
The rest frame of the target nucleon is assumed for simplicity, and
$M_N$ denotes the nucleon mass.
The weak current is denoted as
$J_\mu=\bar{\nu}\gamma_\mu\frac{1+\gamma_5}{2}e+
\bar{u}\gamma_\mu\frac{1+\gamma_5}{2}d + \cdots$,
while the electromagnetic current is
$J_\mu^{\mathrm{(em)}}=-\bar{e}\gamma_\mu e
  +\frac{2}{3}\bar{u}\gamma_\mu u
  -\frac{1}{3}\bar{d}\gamma_\mu d$.

For a lepton scattering off unpolarized nucleon through a $W$-boson
exchange, the differential cross section is given by
\begin{equation}
  \label{eq:cross_section}
  \frac{d^2\sigma}{dE' d\Omega} = \frac{1}{4\pi^2} \frac{E'}{E}
  \left(\frac{1}{q^2-M_W^2}\right)^2
  \left(\frac{g^2}{2} L_{\mu\nu}\right)
  \left(\frac{g^2}{2} W^{\mu\nu}\right),
\end{equation}
where $d\Omega$ represents a solid angle of the final lepton measured
with respect to the direction of the incoming lepton.
The leptonic tensor $L_{\mu\nu}$ is
\begin{equation}
  \label{eq:Lmunu}
  L_{\mu\nu} = 2(p_\mu'p_\nu+p_\mu p_\nu'-p'\cdot p \,g_{\mu\nu}
  -i\epsilon_{\mu\nu\alpha\beta}p^\alpha p'^\beta)
\end{equation}
for the weak current.
The last term with the totally anti-symmetric tensor
$\epsilon_{\mu\nu\alpha\beta}$ arises from the cross term between the
vector and axial-vector currents.
For the electromagnetic interaction the $W$-boson propagator
$1/(q^2-M_W^2)$ in (\ref{eq:cross_section}) has to be replaced by the
photon propagator $1/q^2$.
Also, the weak coupling $g^2/2$ needs to be replaced by the electric
charge $e^2$, and the last term in the leptonic tensor (\ref{eq:Lmunu})
has to be omitted. 

The hadronic tensor contains the contribution from various hadronic
states:
\begin{align}
  \label{eq:Wmunu}
  W_{\mu\nu}(P\cdot P_X,q^2) =
  \frac{1}{2} \sum_{\mathrm{pol.}} \int\frac{d^3\bm{P}_X}{(2\pi)^3}
  \sum_{X(\bm{P}_X)}
  &
  (2\pi)^3 \delta^{(4)}(P_X-P-q)
  \nonumber\\
  & \times
    \langle N(\bm{P})|J_\mu(0)|X(\bm{P}_X)\rangle
    \langle X(\bm{P}_X)|J_\nu^\dagger(0)|N(\bm{P})\rangle.
\end{align}
Here, $|X(\bm{P}_X)\rangle$ represents arbitrary hadronic final states with
total spatial momentum $\bm{P}_X$.
The states are normalized such that
$\int d^3\bm{P}_X\sum_{X(\bm{P}_X)}|X(\bm{P}_X)\rangle\langle X(\bm{P}_X)|=1$.
Since the nucleon is unpolarized, its polarizations (pol.) are
averaged.
The mass dimension of $W_{\mu\nu}$ is $-1$.

The total cross section may be obtained by integrating out 
the differential cross section (\ref{eq:cross_section}).
We choose $\omega$ and $\bm{q}^2=(\bm{p}-\bm{p}')^2$ as the
kinematical variables as they are convenient for the lattice
calculation.
Using $dE'd\Omega=(\pi/EE')d\omega d\bm{q}^2$,
which assumes an integral over a cylindrical angle,
we obtain the total cross section 
\begin{equation}
  \label{eq:total_cross_section}
  \sigma(E) =
  \frac{1}{4\pi E^2}
  \int_0^{E^2}\!d\bm{q}^2
  \int_{\sqrt{M_N^2+\bm{q}^2}}^{M_N+|\bm{q}|}d\omega\,
  \left(\frac{1}{q^2-M_W^2}\right)^2
  \left(\frac{g^2}{2} L_{\mu\nu}\right)
  \left(\frac{g^2}{2} W^{\mu\nu}\right).
\end{equation}
In terms of these variables, $\omega$ and $\bm{q}^2$,
the standard kinematical variables are written as
\begin{align}
  Q^2 & \equiv -q^2 = \bm{q}^2-(\omega-M_N)^2,
       \label{eq:Q2}
  \\
  \nu & \equiv M_N(E-E') = M_N(\omega-M_N),
        \label{eq:nu}
\end{align}
and the Bjorken scaling variables are
\begin{align}
  x & \equiv \frac{Q^2}{2\nu} = \frac{Q^2}{2P\cdot q}
      = \frac{\bm{q}^2-(\omega-M_N)^2}{2M_N(\omega-M_N)},
      \label{eq:x}
  \\
  y & \equiv \frac{\nu}{M_NE} = \frac{P\cdot q}{P\cdot p}
      = \frac{\omega-M_N}{E}.
      \label{eq:y}
\end{align}
The lower and upper limits of the $\omega$-integral in
(\ref{eq:total_cross_section}) correspond to the kinematical limits of
$x=1$ and $x=0$, respectively.

\begin{figure}[tbp]
  \centering
  \includegraphics[width=10cm]{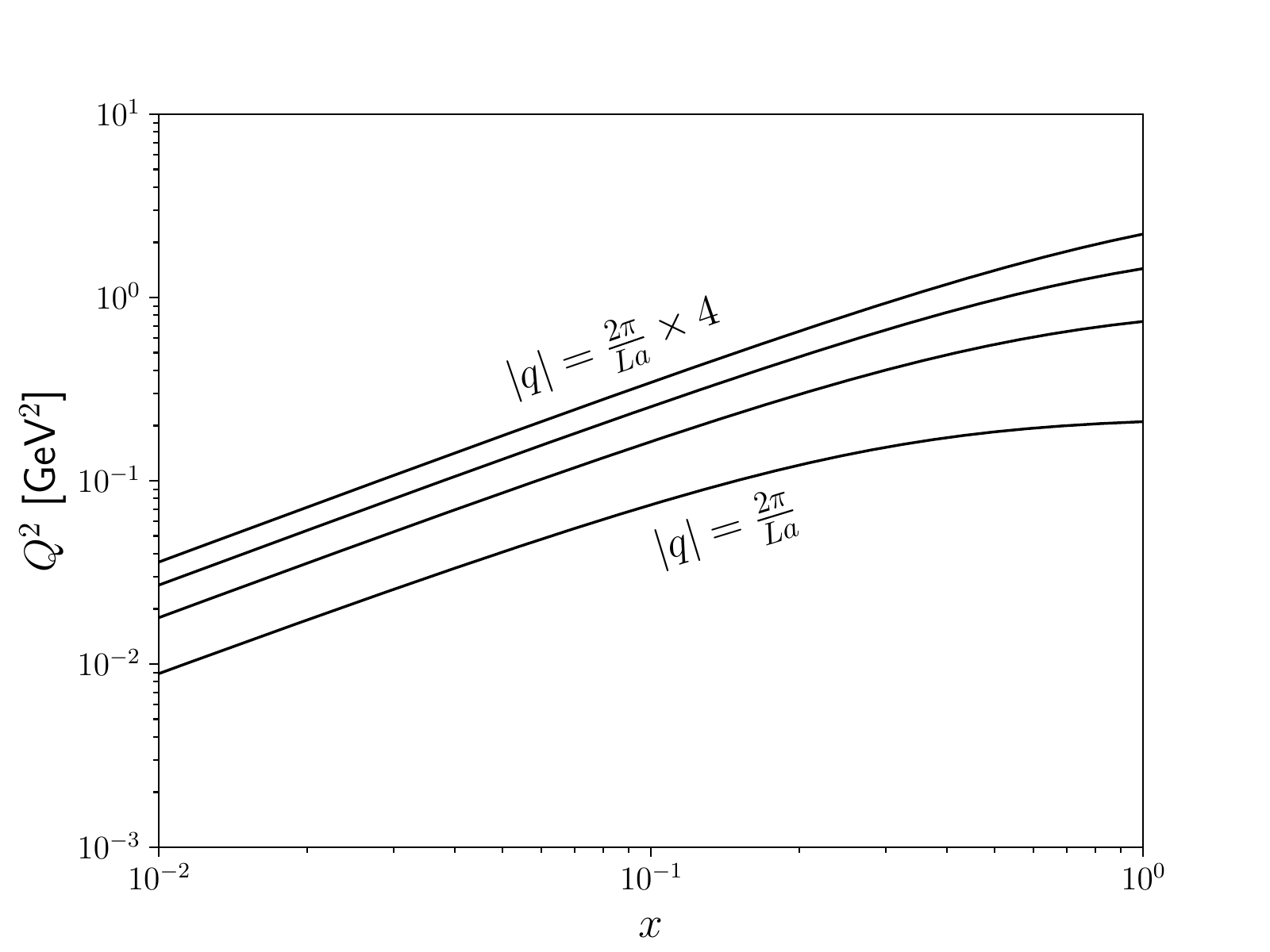}
  \caption{
    Integral paths in terms of $\bm{q}^2$ and $\omega$ on the plane of
    $x$ and $Q^2$.
    The spatial momentum is chosen as $|\bm{q}|=2\pi/La\times k$ with
    $k$ = 1, 2, 3, 4 (from bottom to top).
    With a lattice cutoff $1/a$ = 2.4~GeV and the lattice extent
    $L/a=32$, the physical lattice extent is $L$ = 2.62~fm.
    The momentum $|\bm{q}|$ drawn in this plot corresponds to the range
    between 0.47~GeV and 1.88~GeV from bottom to top.
  }
  \label{fig:kin}
\end{figure}

Figure~\ref{fig:kin} shows the integral paths of $\omega$ and
$\bm{q}^2$ in the plane of $x$ and $Q^2$.
For a fixed $|\bm{q}|$, the $\omega$ integral forms a trajectory shown
in the plot from $x=1$ down to $x=0$.
Since the upper limit of $|\bm{q}|$ is fixed by the initial lepton
energy $E$, the integral region is the range below a line of
$|\bm{q}|=E$.

Here we write down the explicit forms of the leptonic tensor
$L_{\mu\nu}/2$.
The metric is chosen as $g_{\mu\nu}=\mathrm{diag}(1,-1,-1,-1)$.
We set the direction of the initial neutrino on the $z$-axis,
{\it i.e.} $p^\mu=(E,0,0,E)$.
In the following, we label the $z$-direction by $\mu=k$.
Other two spatial directions are denoted as $i$ and $j$, which are
either $x$- or $y$-axis .
Each component of $L_{\mu\nu}$ is written in terms of $\omega$ and
$\bm{q}^2$ as 
\begin{align}
  L_{00}/2 = & 2E(E-q_0)-\frac{1}{2}(\bm{q}^2-q_0^2),
               \nonumber\\
  L_{0k}/2= L_{k0}/2 = & E(2E-q_0-q_k),
                        \nonumber\\
  L_{0i}/2= L_{i0}/2 = & -Eq_i,\nonumber\\
  L_{kk}/2 = & 2E(E-q_k)+\frac{1}{2}(\bm{q}^2-q_0^2),
               \nonumber\\
  L_{ii}/2= L_{jj}/2 = & \frac{1}{2}(\bm{q}^2-q_0^2),
                        \nonumber\\
  L_{ij}/2= -L_{ji}/2 = & -\frac{i}{2}(\bm{q}^2-q_0^2),
                         \nonumber
\end{align}
where $q_0=\omega-M_N$ and
$q_k=q_0+(\bm{q}^2-q_0^2)/2E$.
Here, $L_{0i}$ will be combined with $W_{0i}$, which supplies another
factor of $q_i$, so that the sum $L_{0i}W^{0i}+L_{0j}W^{0j}$ is
proportional to $q_i^2+q_j^2=\bm{q}^2-q_k^2$, which is written as a
function of $\omega$ and $\bm{q}^2$.

The $\nu N$ scattering cross section is thus obtained 
from the master formula (\ref{eq:total_cross_section}).
The dependence on the initial neutrino energy appears only through
the upper limit of the $\bm{q}^2$ integral.
The neutrino energy in the recent neutrino experiments are
typically in the range of several hundred MeV to a few GeV.
(For a review, see \cite{Katori:2016yel}.)
When the initial neutrino energy cannot be controlled event by event,
an weighted average of $\sigma(E)$ with respect to the incoming
neutrino energy distribution is actually observed.
Such an average can be easily obtained once the total cross section is
calculated as a function of the initial neutrino energy $E$.

\section{$W\gamma$ exchange contribution to $\beta$-decay}
\label{sec:Wgamma}
Another application of our formulation is about the higher order
corrections to the neutron $\beta$ decay.
It could also be considered as a nuclear-structure independent
correction to the nuclear $\beta$ decay, which is relevant for the
precise determination of $|V_{ud}|$.

At the leading order the $\beta$ decay occurs through a virtual $W$
exchange.
At the next order in $\alpha$, there may be another exchange of a
virtual photon between the final state proton and electron, that makes
a box diagram.
An estimate of such diagram can be obtained easily using the
electromagnetic form factor of proton if the intermediate
state can be assumed to be a ground-state proton,
but actually the contribution from excited states has to be taken into
account, and it is a source of significant uncertainty.
In our formulation, the integral over the inner-loop momentum
can be carried out with the effects of all possible intermediate states
included. 

The correction to the tree-level amplitude of the nucleon
$\beta$-decay may be written as \cite{Seng:2018yzq}
\begin{equation}
  \label{eq:Wgamma}
  \square_{\gamma W}^{VA}=\frac{3\alpha}{2\pi}
  \int_0^\infty \frac{dQ^2}{Q^2}\frac{M_W^2}{M_W^2+Q^2}
  M_3^{(0)}(1,Q^2),
\end{equation}
where $M_3^{(0)}(1,Q^2)$ is the first Nachtmann moment of the structure
function $F_3^{(0)}$ \cite{Nachtmann:1973mr,Nachtmann:1974aj}:
\begin{equation}
  \label{eq:Nachtmann}
  M_3^{(0)}(1,Q^2) = \frac{4}{3}
  \int_0^1 dx \frac{1+2r}{(1+r)^2} F_3^{(0)}(x,Q^2)
\end{equation}
with $r=\sqrt{1+4M_N^2x^2/Q^2}$.
Again, we may rewrite the integrals over $Q^2$ and $x$ by those of
$\omega$ and $\bm{q}^2$.
The Jacobian is given by
$dQ^2dx=2M_Nx/\nu d\omega d\bm{q}^2$.
Then, the formula (\ref{eq:Wgamma}) can be written in the form
\begin{equation}
  \label{eq:Wgamma_formula}
  \square_{\gamma W}^{VA}=\frac{2\alpha}{\pi}
  \int_0^\infty d\bm{q}^2
  \int_{\sqrt{M_N^2+\bm{q}^2}}^{M_N+|\bm{q}|}
  \frac{d\omega}{M_N(\omega-M_N)^2}
  \frac{M_W^2}{M_W^2+Q^2}
  \frac{1+2r}{(1+r)^2}
  F_3^{(0)}(x,Q^2)
\end{equation}
with $r$ and $Q^2$ also rewritten using $\omega$ and $\bm{q}^2$.
(See (\ref{eq:Q2}) and (\ref{eq:x}).)

The structure function $F_3^{(0)}$ is defined as the part including
the $\epsilon$-tensor $i\epsilon_{\mu\nu\alpha\beta}$ in $W^{\mu\nu}$
as
\begin{equation}
  W^{\mu\nu} \ni -\frac{i}{2M_N^2} \epsilon_{\mu\nu\alpha\beta}
  P^\alpha q^\beta W_3
\end{equation}
and $F_3\equiv\nu/M_N W_3$.
(The $F_3^{(0)}$ is an isospin singlet component of $F_3$.)
By looking at the component of $(\mu,\nu)=(i,j)=(1,2)$ we obtain
\begin{equation}
  iW_{ij}=\frac{|\bm{q}|}{2M_N(\omega-M_N)} F_3,
\end{equation}
which is to be combined with (\ref{eq:Wgamma_formula}).

There is a recent lattice computation of the same quantity but for
pion \cite{Feng:2020zdc}, which has been used to estimate the
contribution for nucleon \cite{Seng:2020wjq}.
They used the coordinate-space integral instead of the momentum space
integral given above.
Another method to use the Feynman-Hellmann theorem has also been
proposed \cite{Seng:2019plg}.

\section{Energy integral}
\label{sec:energy_integral}
We apply the method outlined in Sec.~\ref{sec:spectral_func} to the
computation of the total $\ell N$ cross section
(\ref{eq:total_cross_section}) 
or the loop correction (\ref{eq:Wgamma_formula}).

On the lattice, one can calculate the forward Compton-scattering
amplitude
\begin{align}
  C_{\mu\nu}^{JJ}(t;\bm{q})
  &
    = \sum_{\bm{x}} e^{i\bm{q}\cdot\bm{x}}
    \frac{1}{2} \sum_{\mathrm{pol.}}
    \langle N(\bm{0})|J_\mu^\dagger(\bm{x},t) J_\nu(\bm{0},0)
    |N(\bm{0})\rangle
    \nonumber\\
  &
    = \frac{1}{V}
    \frac{1}{2} \sum_{\mathrm{pol.}}
    \langle N(\bm{0})|\tilde{J}_\mu^\dagger(-\bm{q})
    e^{-\hat{H}t}
    \tilde{J}_\nu(\bm{q})
    |N(\bm{0})\rangle
    \label{eq:Cmunu}
\end{align}
from a four-point function that includes the operators to create
and annihilate the initial-state nucleon $|N(\bm{0})\rangle$.
We omit the spin index of the nucleon state, but an avarage over the
nucleon spin is assumed as indicated by
$\frac{1}{2}\sum_{\mathrm{pol.}}$.
On the second line, the correlation function is rewritten using the
transfer matrix $e^{-\hat{H}t}$,
and a Fourier transform of the current is introduced:
$\tilde{J}_\nu(\bm{q})=\sum_{\bm{x}}e^{-i\bm{q}\cdot\bm{x}}J_\nu(\bm{x})$.

Let us consider the $\omega$-integral in
(\ref{eq:total_cross_section}).
The factor $L_{\mu\nu}/(q^2-M_W^2)^2$ in front of $W_{\mu\nu}$ has the
form $\omega^l$ with $l$ either 0, 1, or 2, depending on the
components $\mu$ and $\nu$, up to an $\omega$-independent factor.
The $W$-boson propagator can be approximated by a constant, $1/M_W^4$,
for low-energy scatterings.
For the electromagnetic scattering, on the other hand,
$1/(q^2)^2=1/((\omega-M_N)^2-\bm{q}^2)^2$
has to be multiplied.
We write the factor in front of $W_{\mu\nu}$ collectively as
$K(\omega,\bm{q})$, so that the integral to be performed is written as
$\int d\omega\, K(\omega) W(\omega)$,
where the indices other than $\omega$ are omitted for simplicity.
(The definition of the integral kernel $K(\omega)$ will be slightly
modified. See below.)

Along the line of (\ref{eq:K(H)}), the $\omega$-integral at a fixed
$\bm{q}$ can be rewritten in the form 
\begin{equation}
  \label{eq:KW}
  \int\! d\omega\, K(\omega) W(\omega)
  = 
    \frac{1}{2}\sum_{\mathrm{pol.}}
    \langle N(\bm{0})|
    \tilde{J}_\mu^\dagger(-\bm{q}) K(\hat{H})
    \tilde{J}_\nu(\bm{q})|N(\bm{0})\rangle.
\end{equation}
The indices $\mu$, $\nu$ as well as $\bm{q}$ are omitted on the
left-hand side.
The upper and lower limits of the $\omega$-integral has to be included
in the definition of the kernel operator $K(\hat{H})$.
For the total cross section (\ref{eq:total_cross_section}), the kernel
$K(\omega)$ is proportional to $\omega^l$ ($l=$ 0, 1, or 2) until it 
hits the upper limit $M_N+|\bm{q}|$ where it vanishes
discontinuously.
The power $l$ depends on the form of the leptonic tensor $L_{\mu\nu}$,
so that (\ref{eq:KW}) corresponds to each term appearing in
(\ref{eq:total_cross_section}) and we have to add them together in the
end. 

As we have already mentioned, the approximation of $K(\omega)$ using the
Chebyshev polynomials becomes more difficult when the target function
is discontinuous.
In order to avoid this problem, we propose to smooth out the
discontinuity. 
For example, we can replace the Heaviside step function
$\theta(M_N+|\bm{q}|-\omega)$ to realize the upper limit
by $\theta_\sigma(M_N+|\bm{q}|-\omega)$ with $\theta_\sigma(x)$ a
smoothed step function, such as
$\theta_\sigma(x)=1/(1+\exp(-x/\sigma))$.
A parameter $\sigma$ is introduced to specify the range of the
smoothing (or smearing).
To be explicit, we may take the kernel function of the form
\begin{equation}
  \bar{K}(\omega) = e^{2\omega t_0} \omega^l \times
  \theta_\sigma(M_N+|\bm{q}|-\omega),
  \label{eq:kernel}
\end{equation}
where the factor $e^{2\omega t_0}$ is introduced to compensate the
small time evolution that appears when we define
$|\psi_\mu(\bm{q})\rangle\equiv
e^{-\hat{H}t_0}\tilde{J}_\mu(\bm{q})|N(\bm{0})\rangle$.
The kernel function (\ref{eq:kernel}) does not reflect the lower limit
of the $\omega$-integral at $\sqrt{M_N^2+\bm{q}^2}$.
That is because there is no state contributing to the integral below
the lower limit, which corresponds to the elastic scattering.
In fact, the forward Compton-scattering amplitude
$C_{\mu\nu}^{JJ}(t;\bm{q})$ in (\ref{eq:Cmunu}) behaves as
$e^{-\sqrt{M_N^2+\bm{q}^2}\,t}$ at large time separations.
Therefore, we can safely extend the lower limit of the
$\omega$-integral to zero.

The kernel function should be adjusted to treat the electromagnetic
scattering with a photon exchange or the $\gamma W$-box correction
to the $\beta$-decay (\ref{eq:Wgamma_formula}), as they have
complicated prefactors.
But, the basic strategy is unchanged.

\begin{figure}[tbp]
  \centering
  \includegraphics[width=8cm]{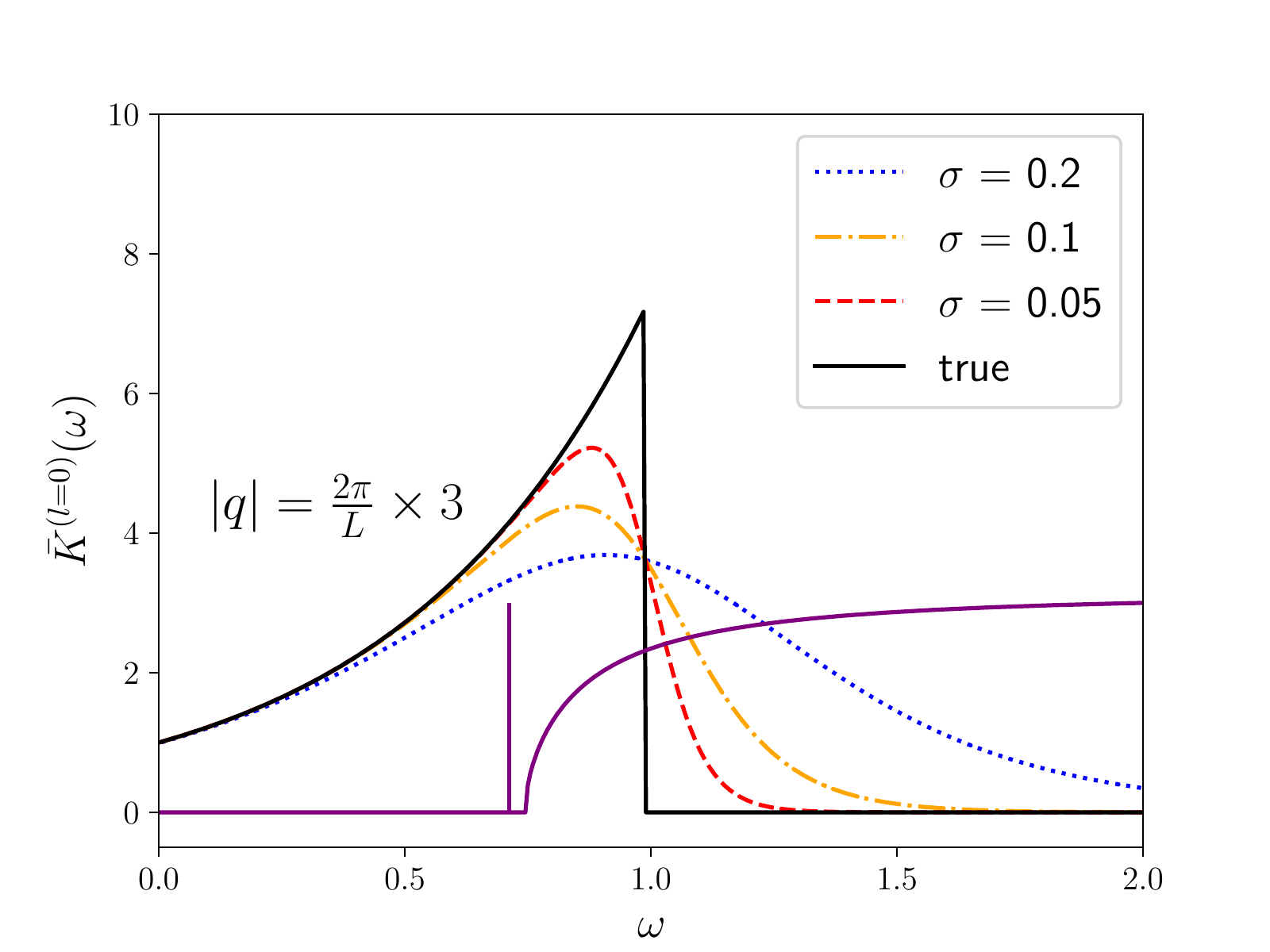}\\
  \includegraphics[width=8cm]{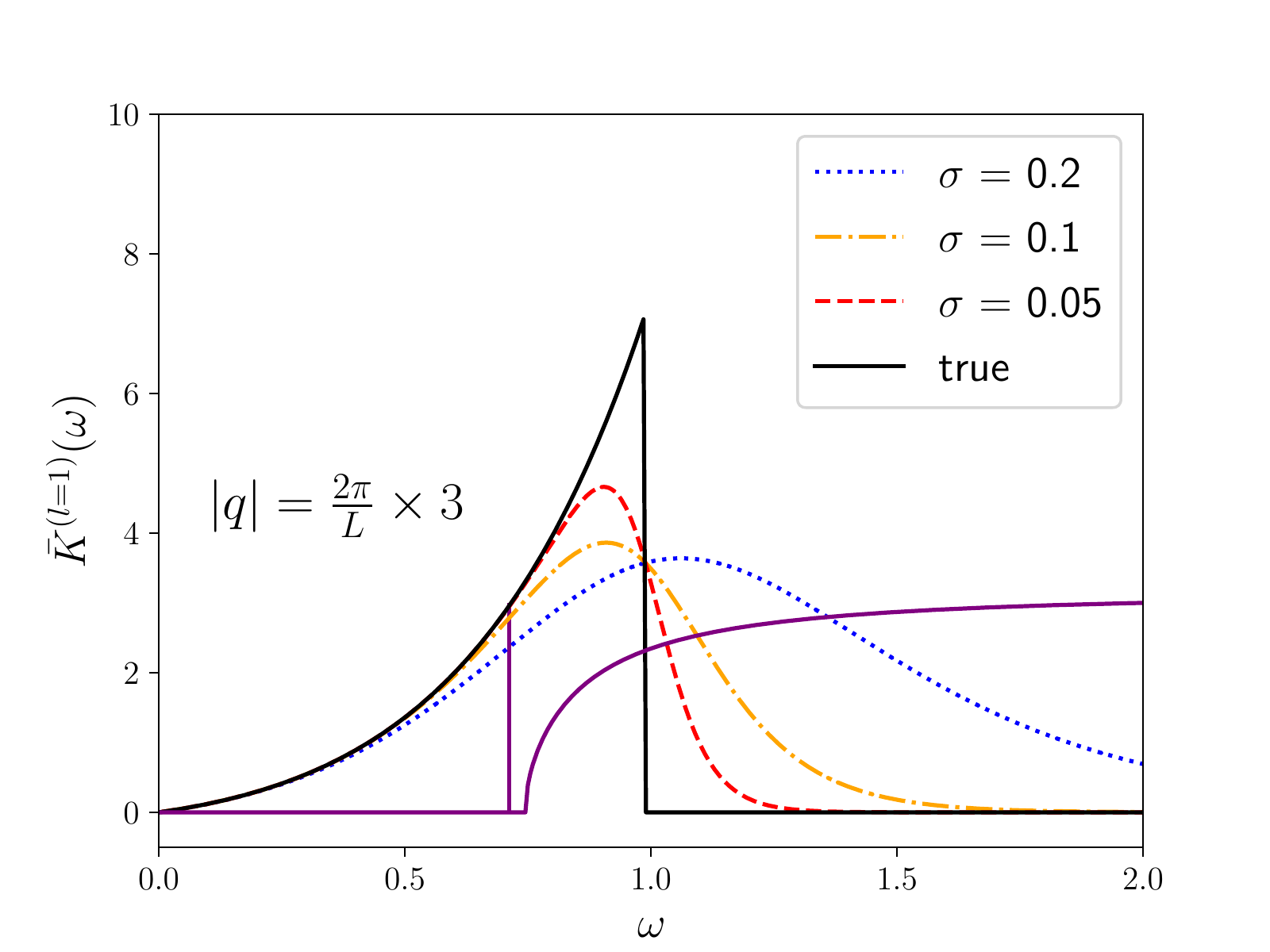}\\
  \includegraphics[width=8cm]{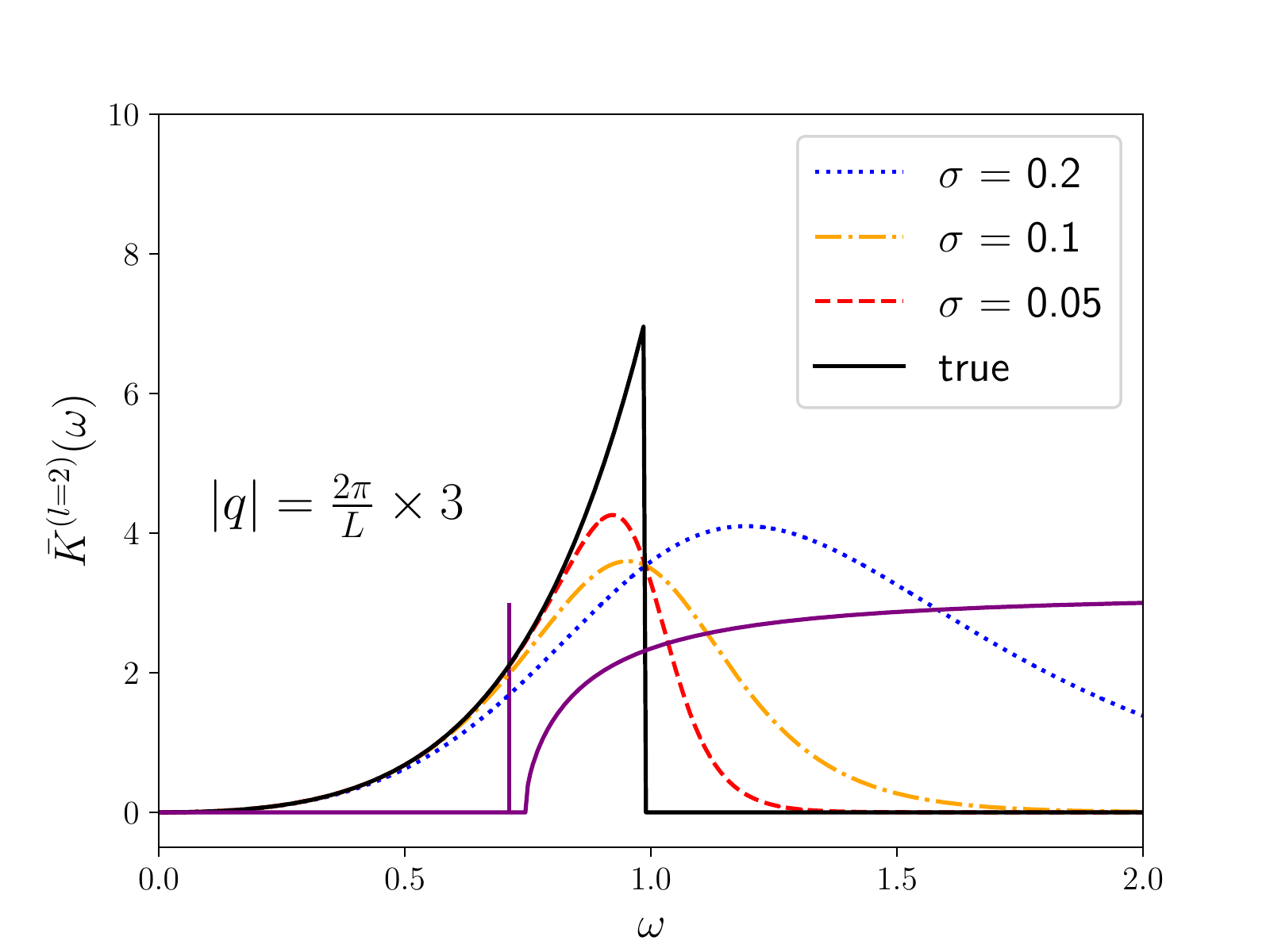}
  \caption{
    Kernel functions $\bar{K}(\omega)$ with $l=0$ (top), 1 (middle), 2
    (bottom) and their modification due to the smearing.
    The modified functions are shown with $\sigma$ = 0.2, 0.1 and 0.05
    in the lattice unit.
    Other parameters are described in the text.
    The solid purple curve represents a mock data for the spectrum used
    to test the method (in an arbitrary unit).
  }
  \label{fig:mock}
\end{figure}

In order to demonstrate how the Chebyshev approximation of
$\bar{K}(\omega)$ 
works, we show some examples of $\bar{K}(\omega)$ and its
approximations in the following. 
We set the lattice cutoff $1/a$ = 2.4~GeV and the
lattice size $L$ = 32, which are typical in today's lattice QCD
simulations.
The nucleon mass is taken at the physical value $M_N$ = 0.96~GeV.
The small time duration $t_0$ is taken to be minimum, $t_0=1$, in the
lattice unit.
The plots in the following are all in the lattice unit.

We choose the kernel function $\bar{K}(\omega)$ in (\ref{eq:kernel})
with the smoothing parameter
$\sigma$ = 0.2, 0.1 and 0.05 (lattice unit).
Fig.~\ref{fig:mock} shows how the smearing modifies the true function,
which has a discontinuity at $\omega=M_N+|\bm{q}|$ (black curve).
Here we choose the momentum insertion
$|\bm{q}|=2\pi/L\times 3$, which is roughly 1.4~GeV$/c$.
With these choices of the smearing width $\sigma$, the kernel function
is smoothed around the point of the discontinuity.
As $\sigma$ decreases, the curve becomes closer to the true function.
Also shown is a mock data for the spectral function (purple), which
models the elastic $N$ as well as $N\pi$ continuum contributions in
the infinite volume.
It will be used to estimate potential errors due to the smoothing (see
below).

\begin{figure}[tbp]
  \centering
  \includegraphics[width=5.2cm]{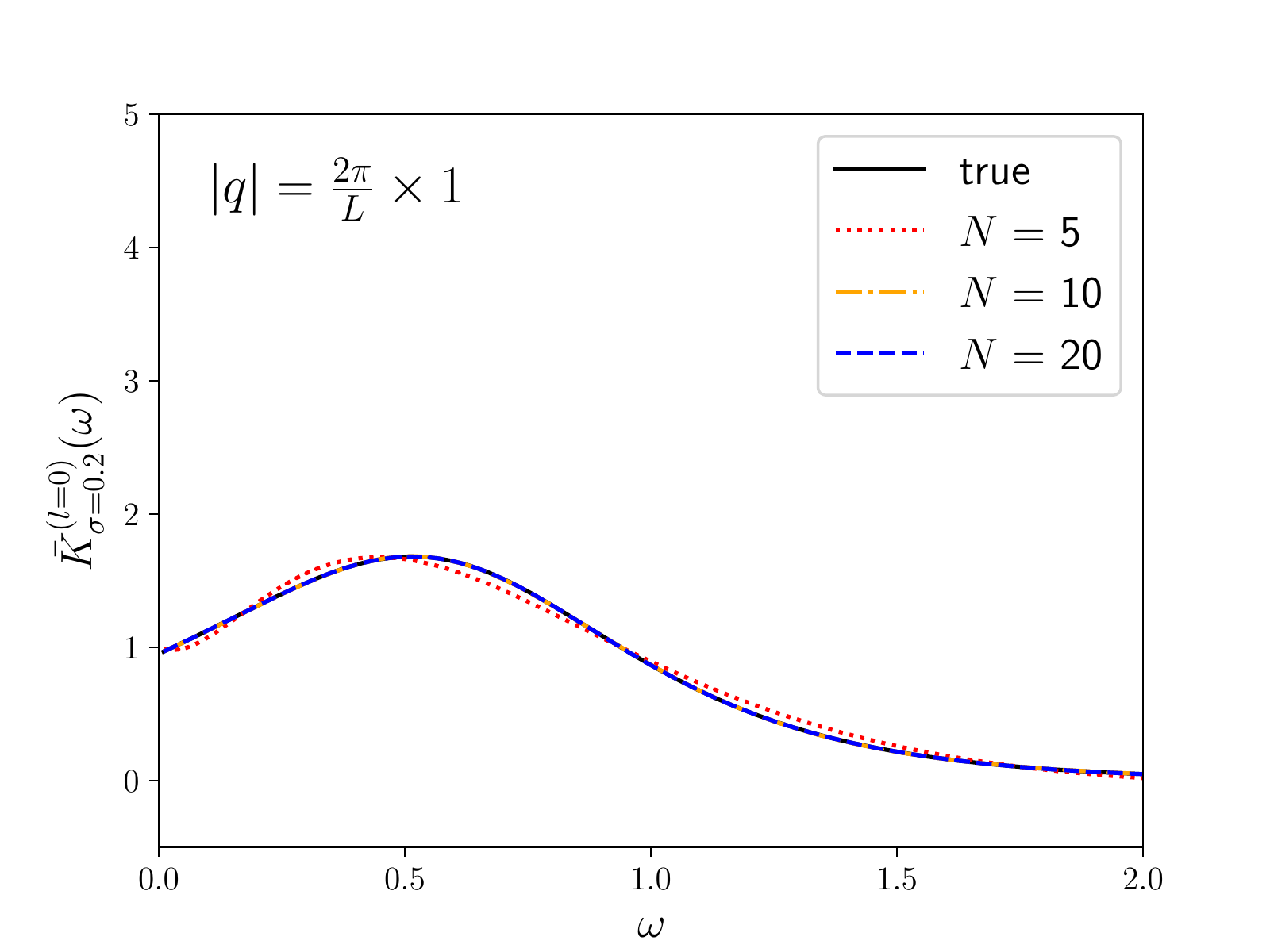}
  \includegraphics[width=5.2cm]{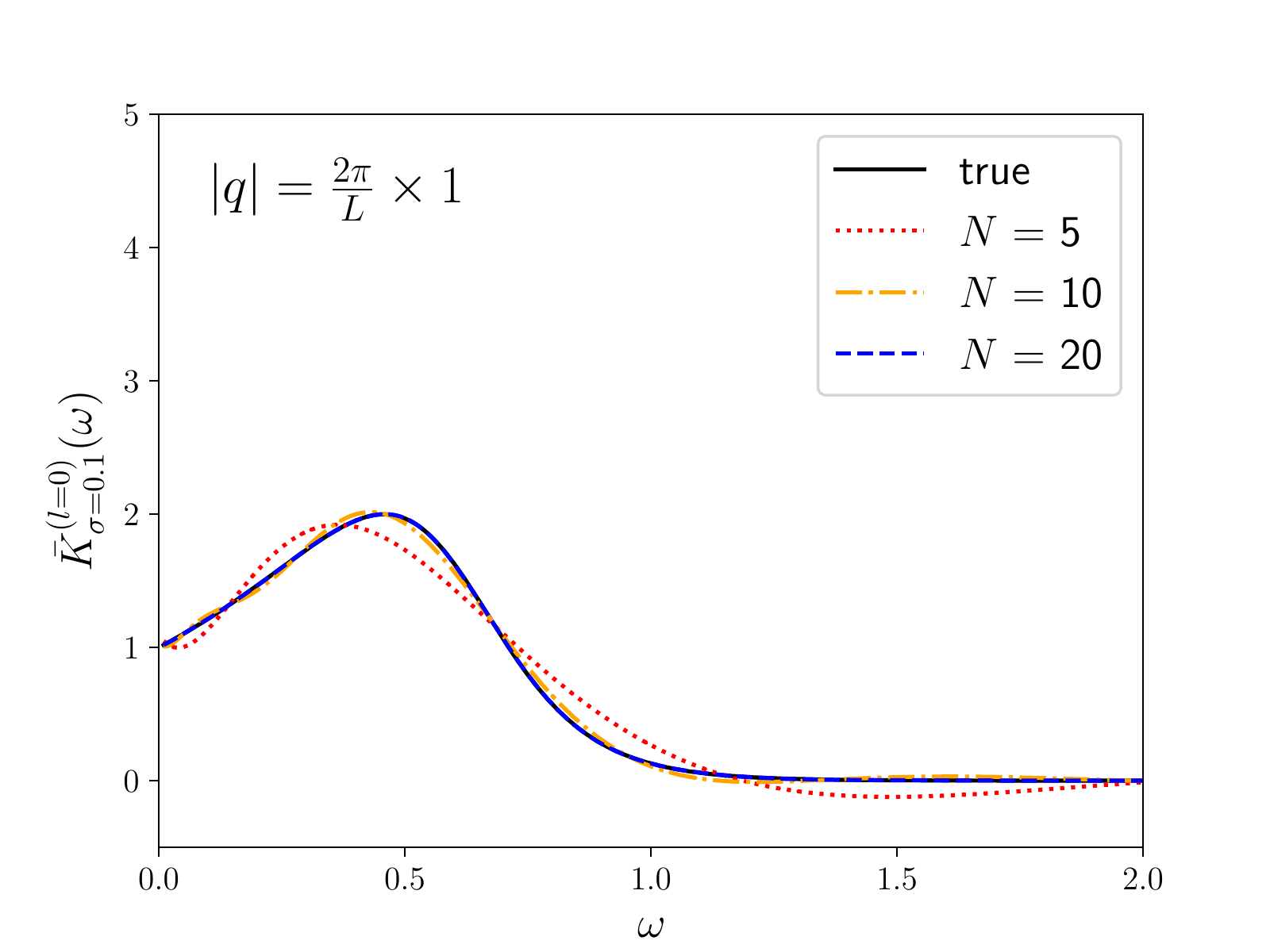}
  \includegraphics[width=5.2cm]{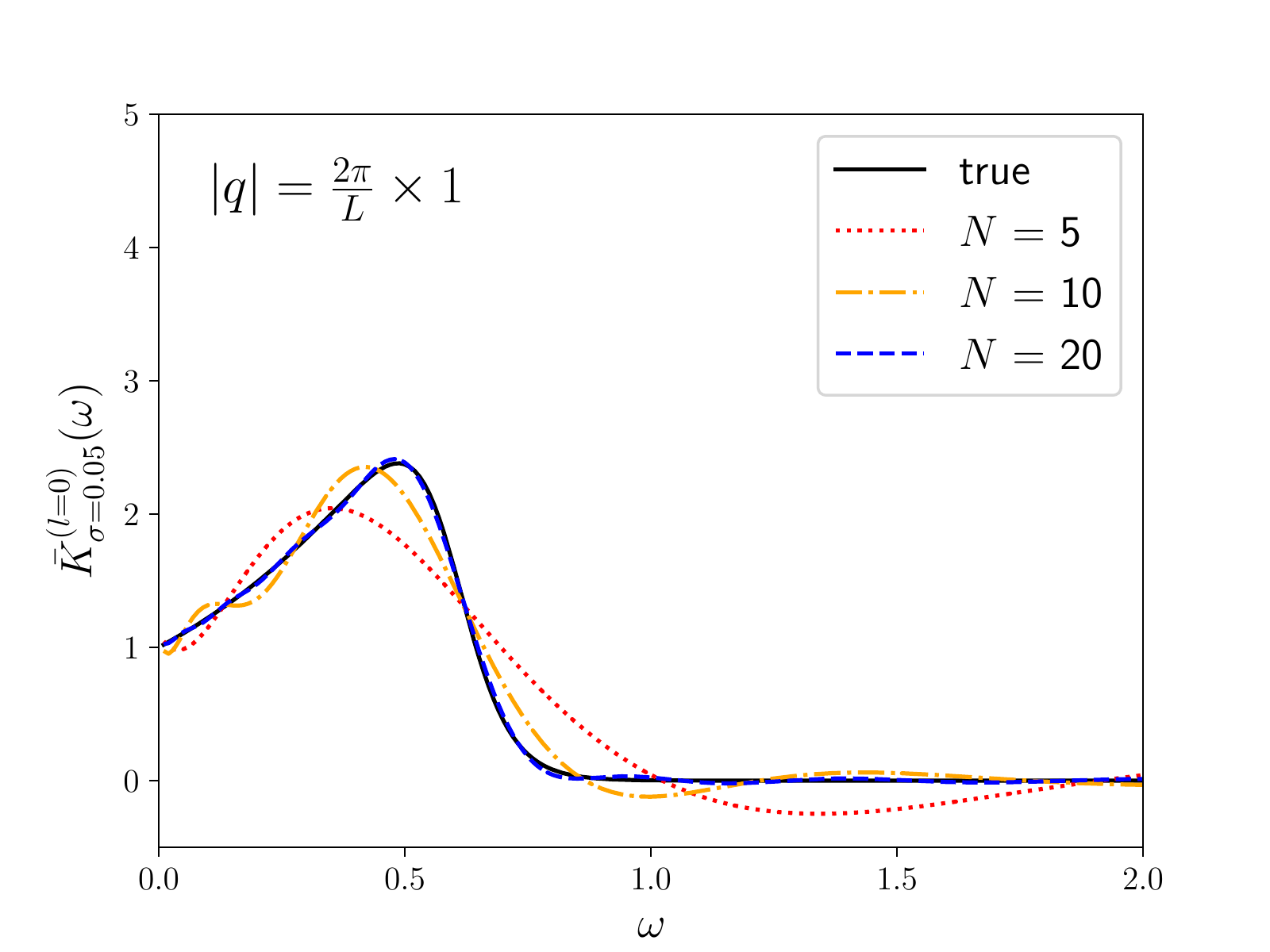}
  \includegraphics[width=5.2cm]{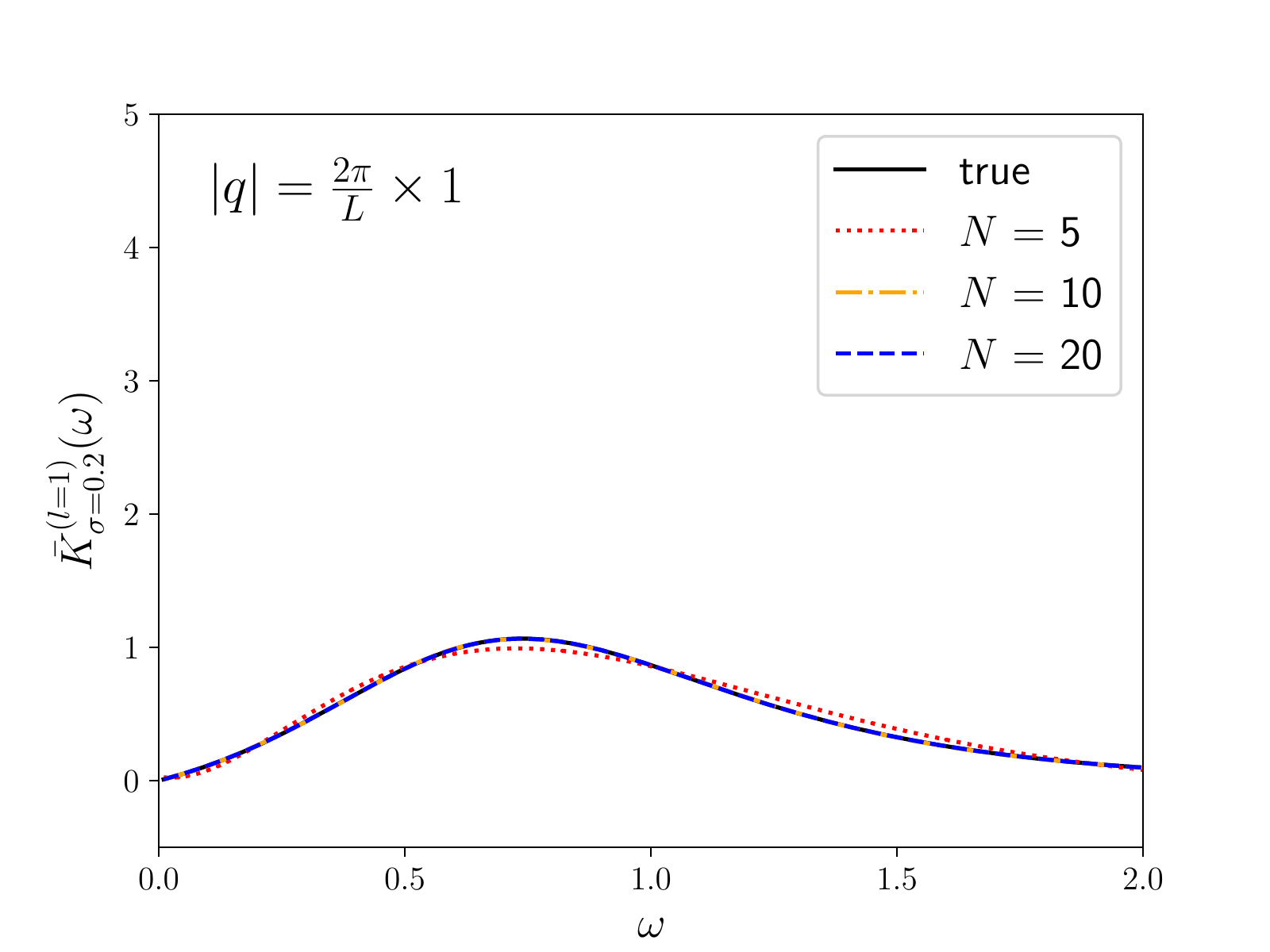}
  \includegraphics[width=5.2cm]{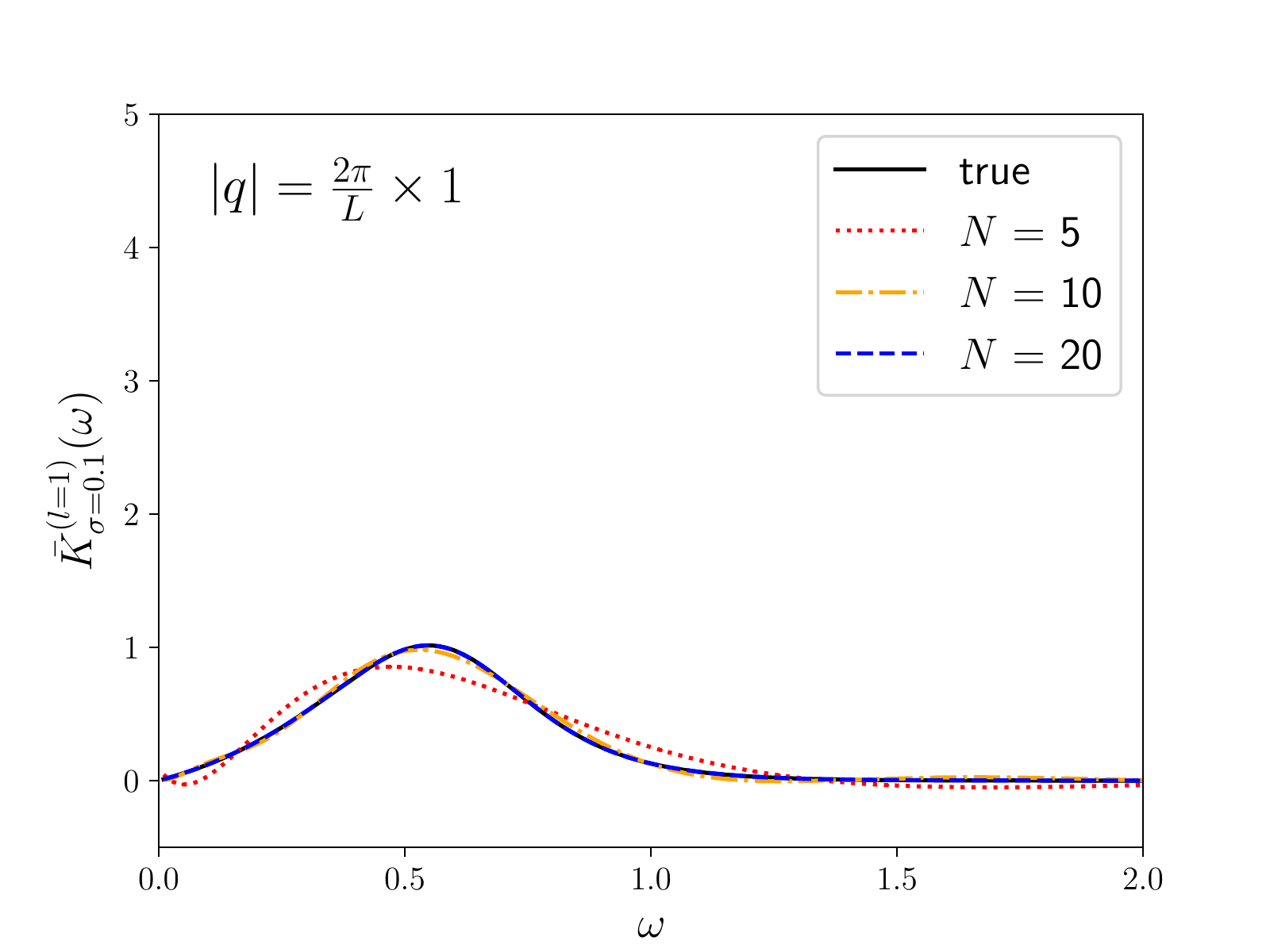}
  \includegraphics[width=5.2cm]{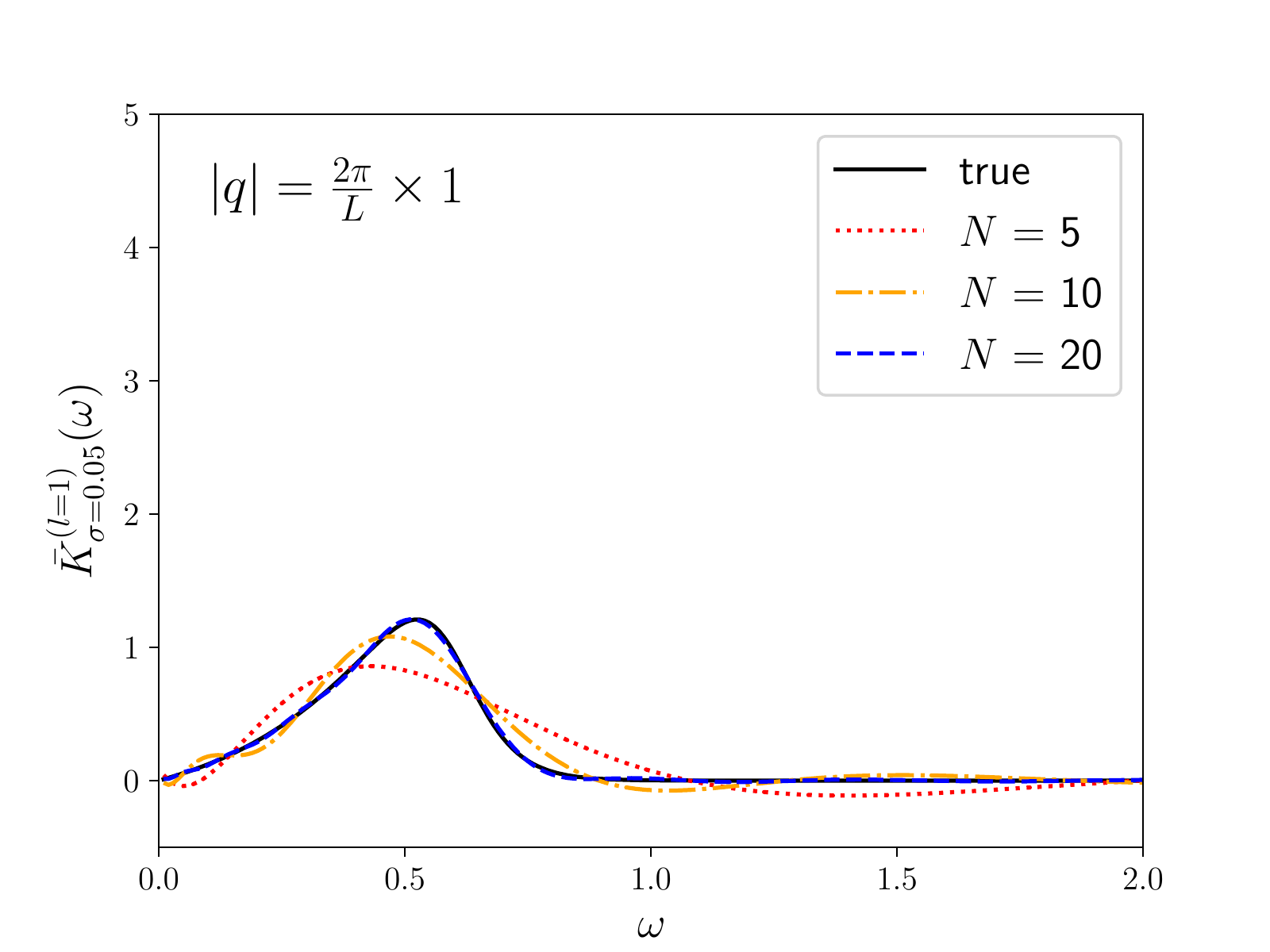}
  \includegraphics[width=5.2cm]{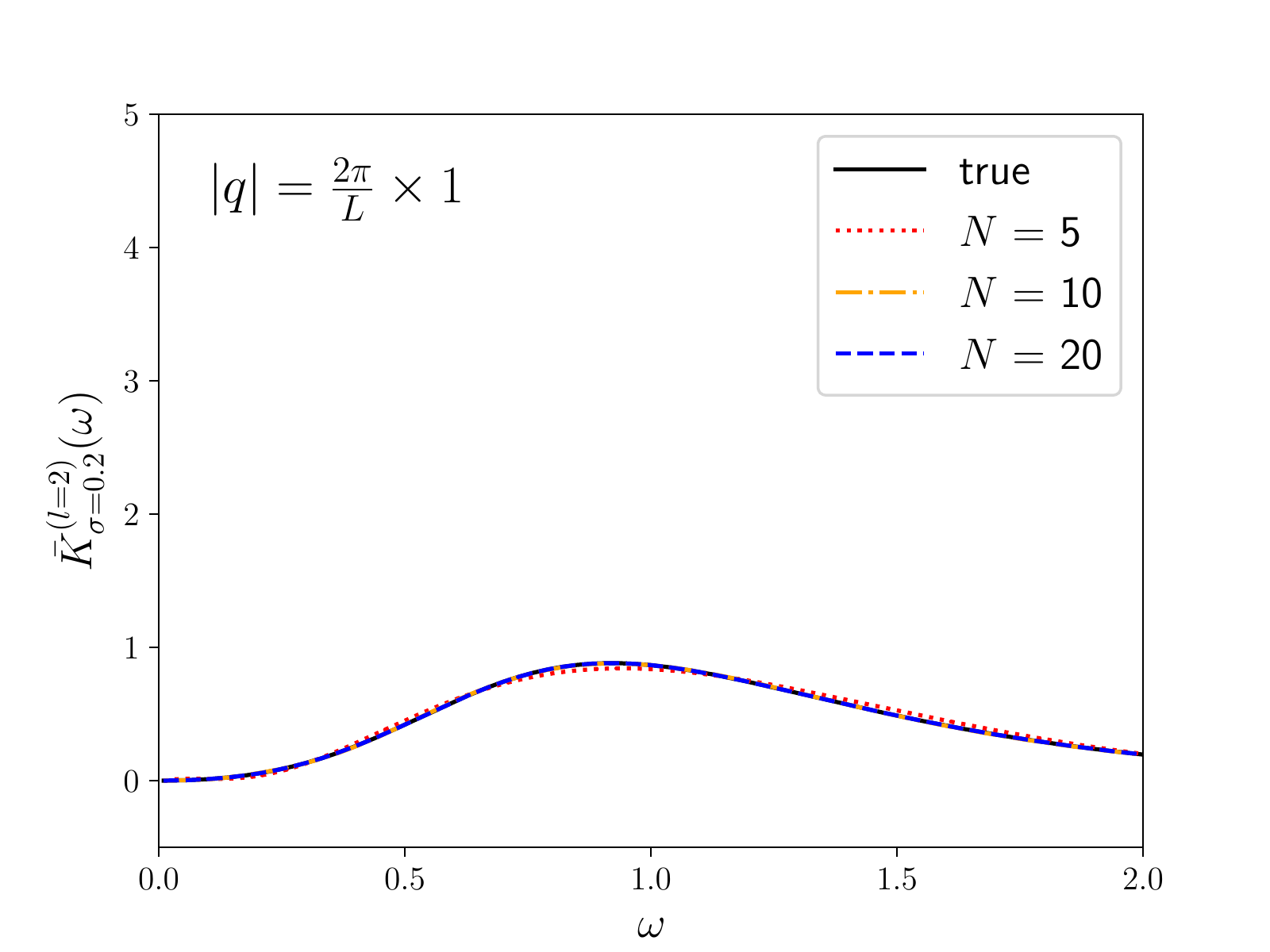}
  \includegraphics[width=5.2cm]{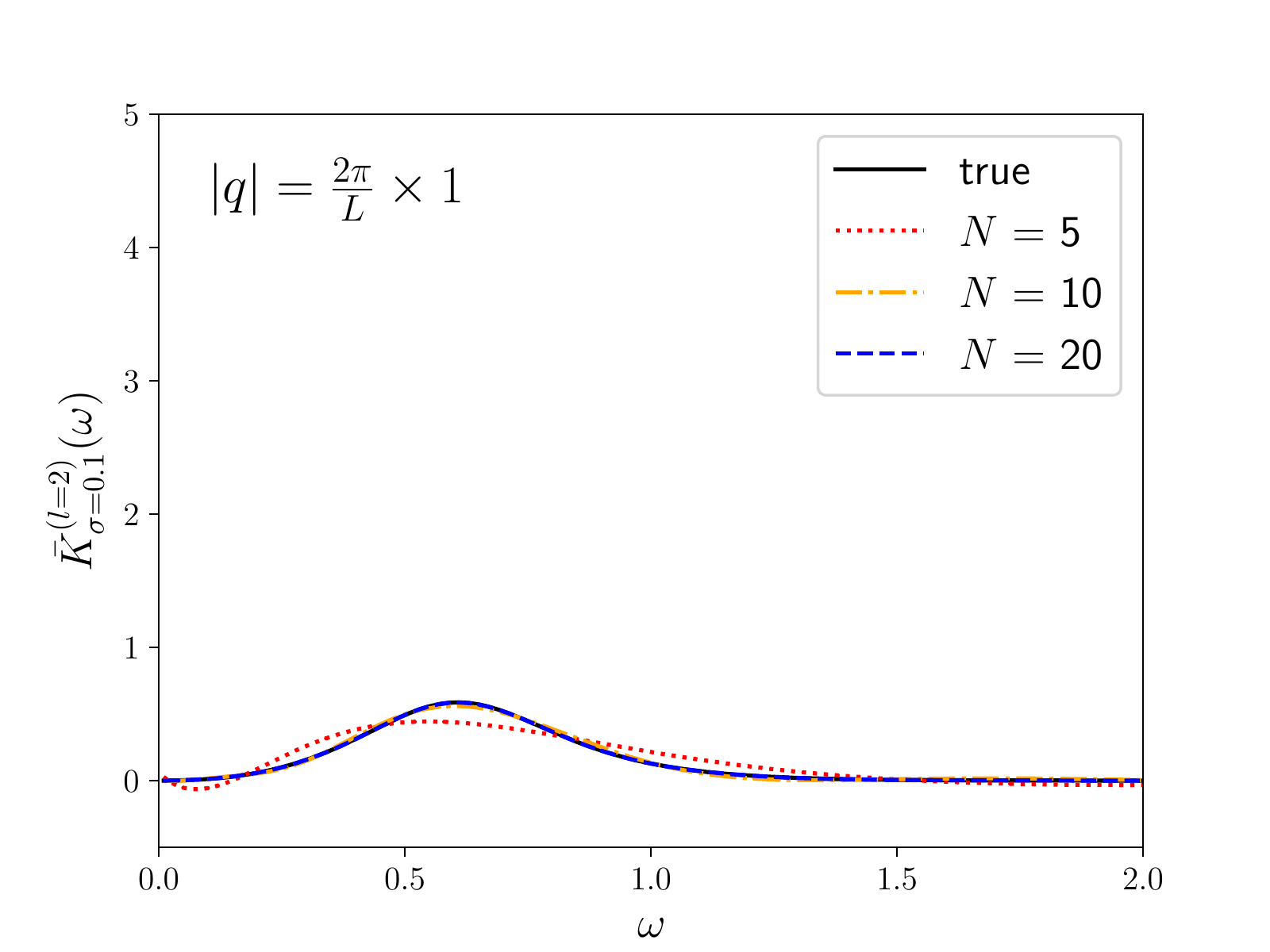}
  \includegraphics[width=5.2cm]{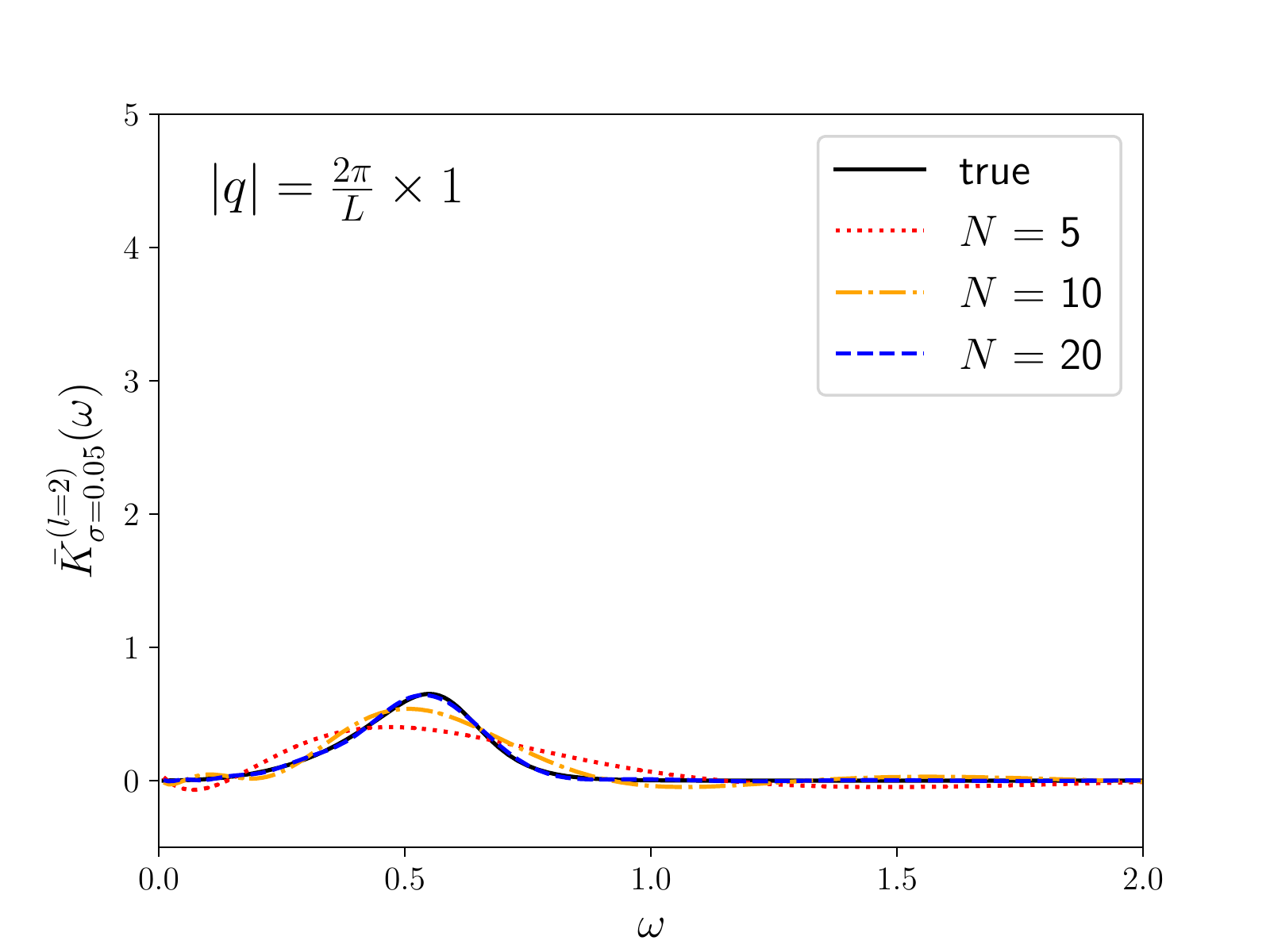}
  \caption{
    Kernel function $\bar{K}(\omega)$ and its Chebyshev approximations 
    plotted in the lattice unit.
    Top panels are for $l=0$ while the middle and bottom panels show
    them for $l=1$ and 2.
    From left to right, the smearing width gets narrower 
    ($\sigma$ = 0.2, 0.1 and 0.05).
    The approximations are written with the polynomial order $N$ = 5,
    10, and 20.
    The energy integral is truncated at $\omega=M_N+|\bm{q}|$;
    $|\bm{q}|$ is taken as the smallest non-zero momentum on the
    lattice: $|\bm{q}|=2\pi/L$,
    so that the upper limit is at $\omega$ = 0.60.
  }
  \label{fig:K(omega)q1}
\end{figure}

\begin{figure}[tbp]
  \centering
  \includegraphics[width=5.2cm]{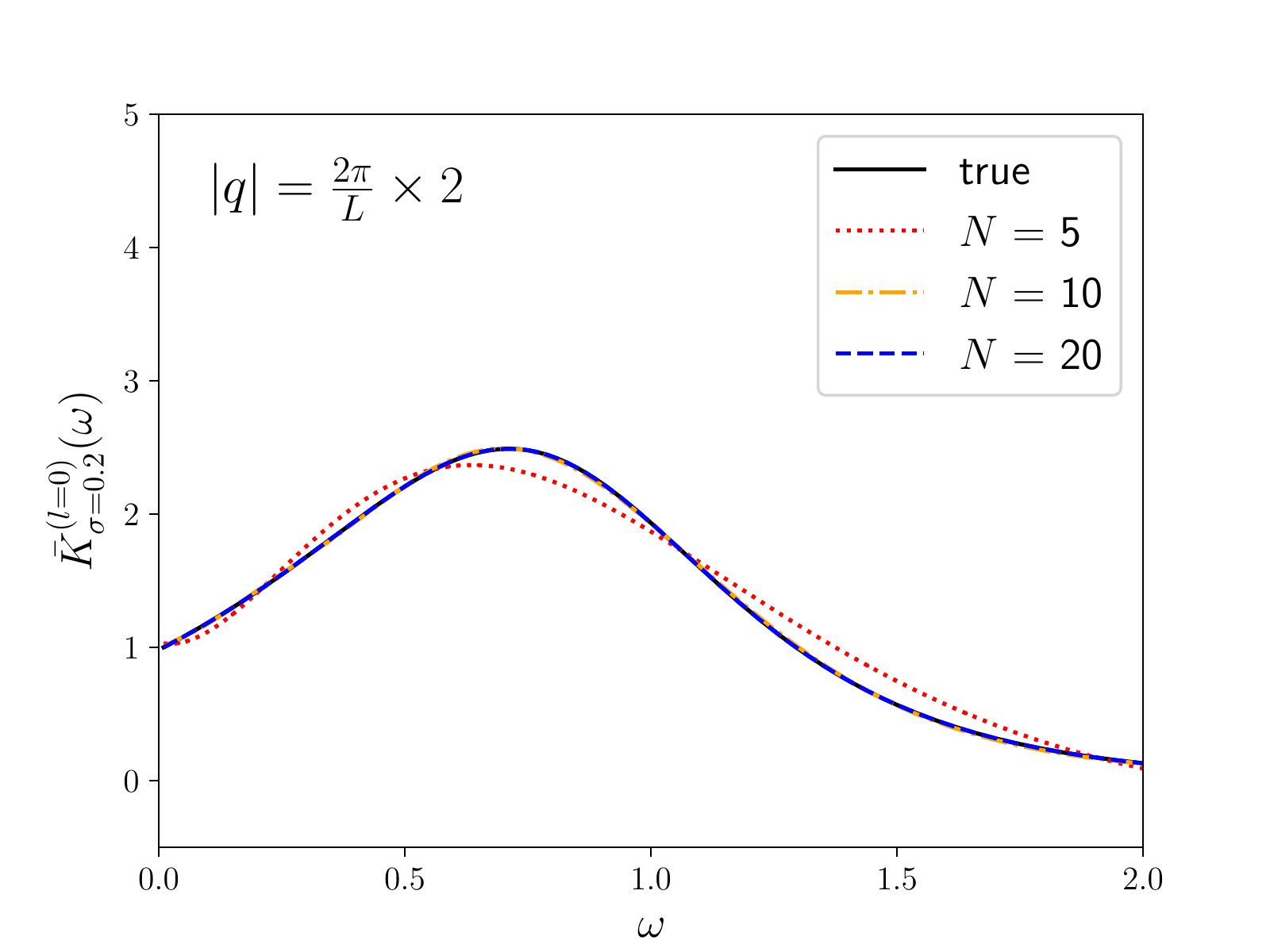}
  \includegraphics[width=5.2cm]{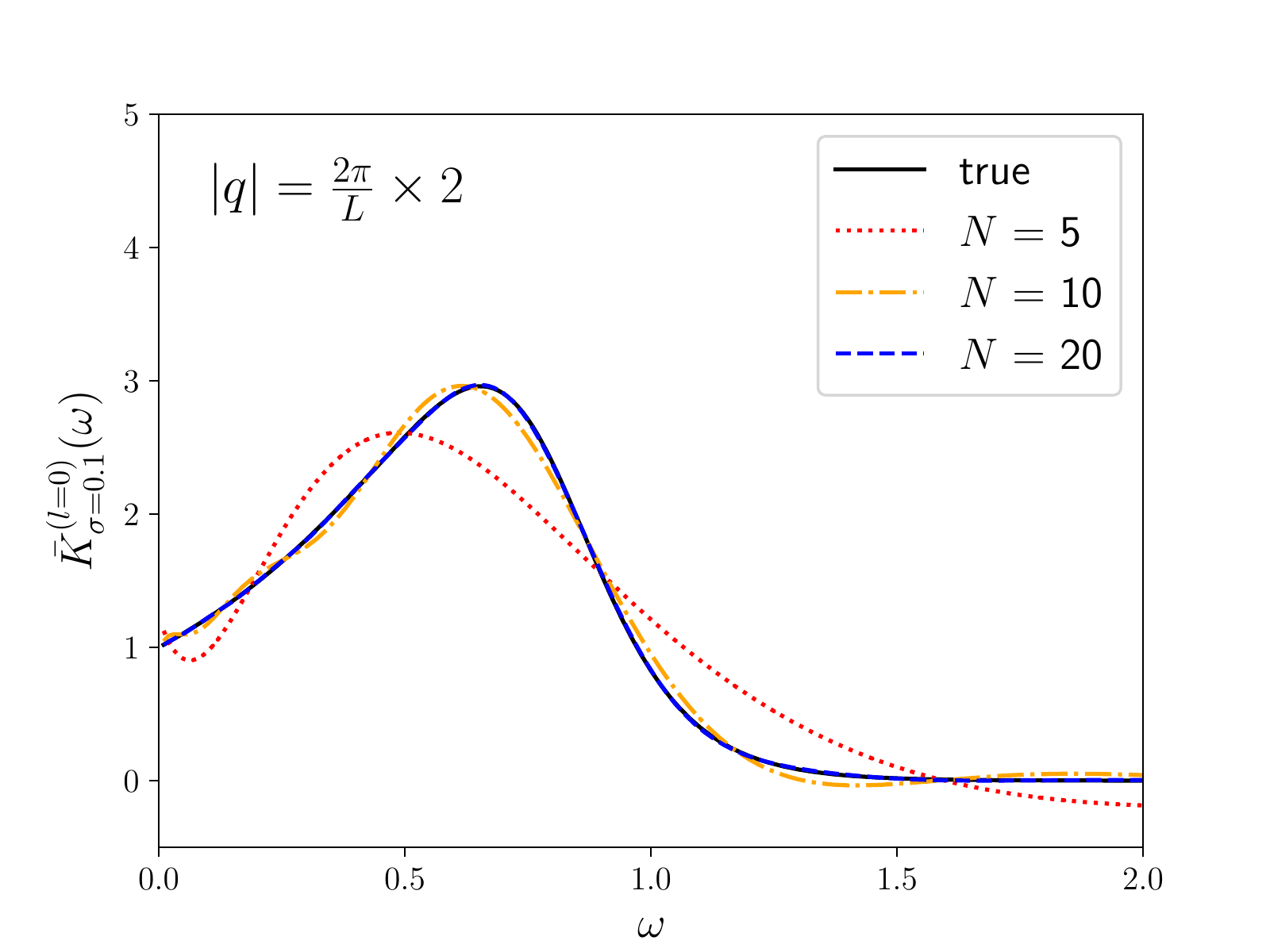}
  \includegraphics[width=5.2cm]{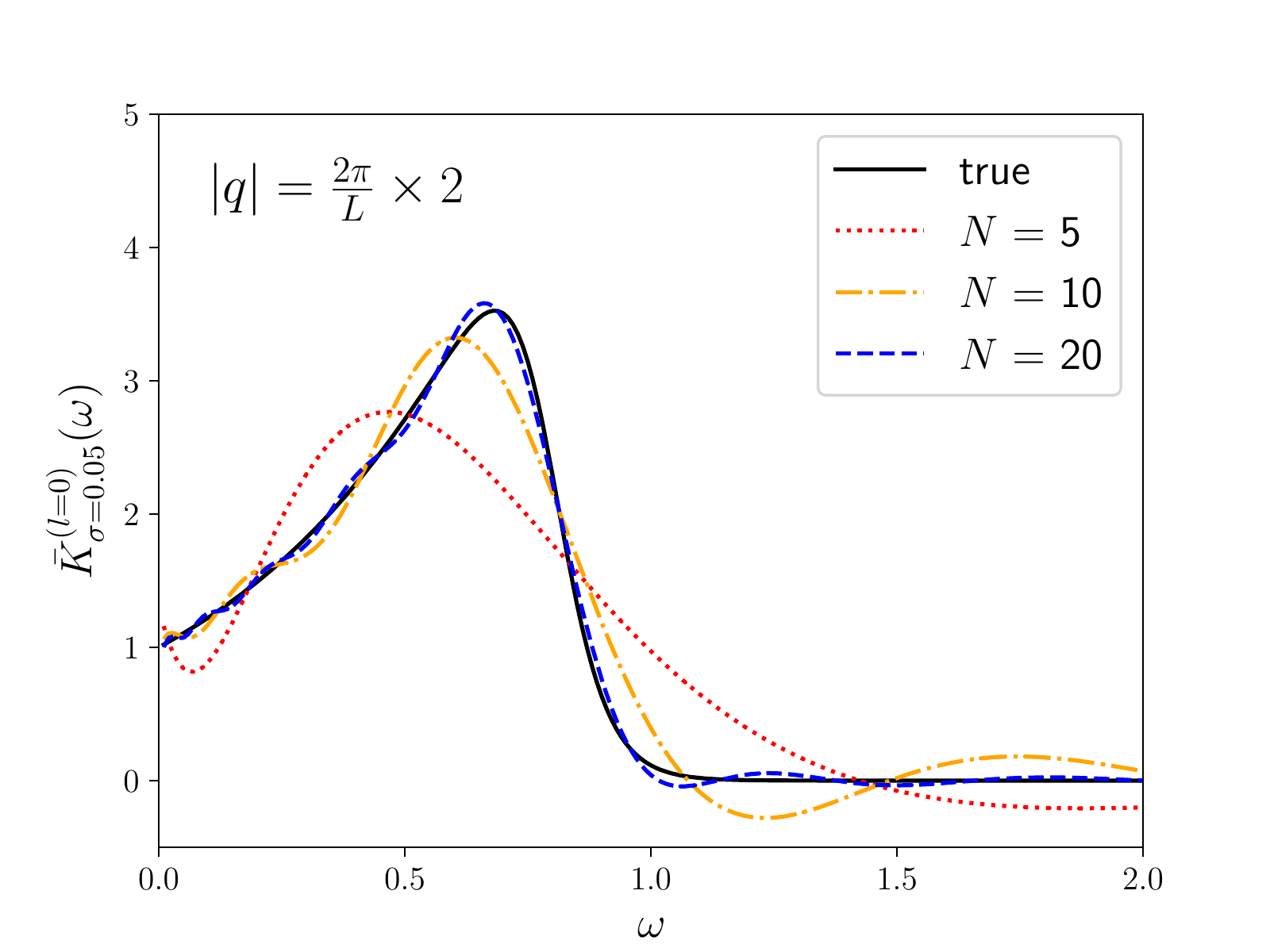}
  \includegraphics[width=5.2cm]{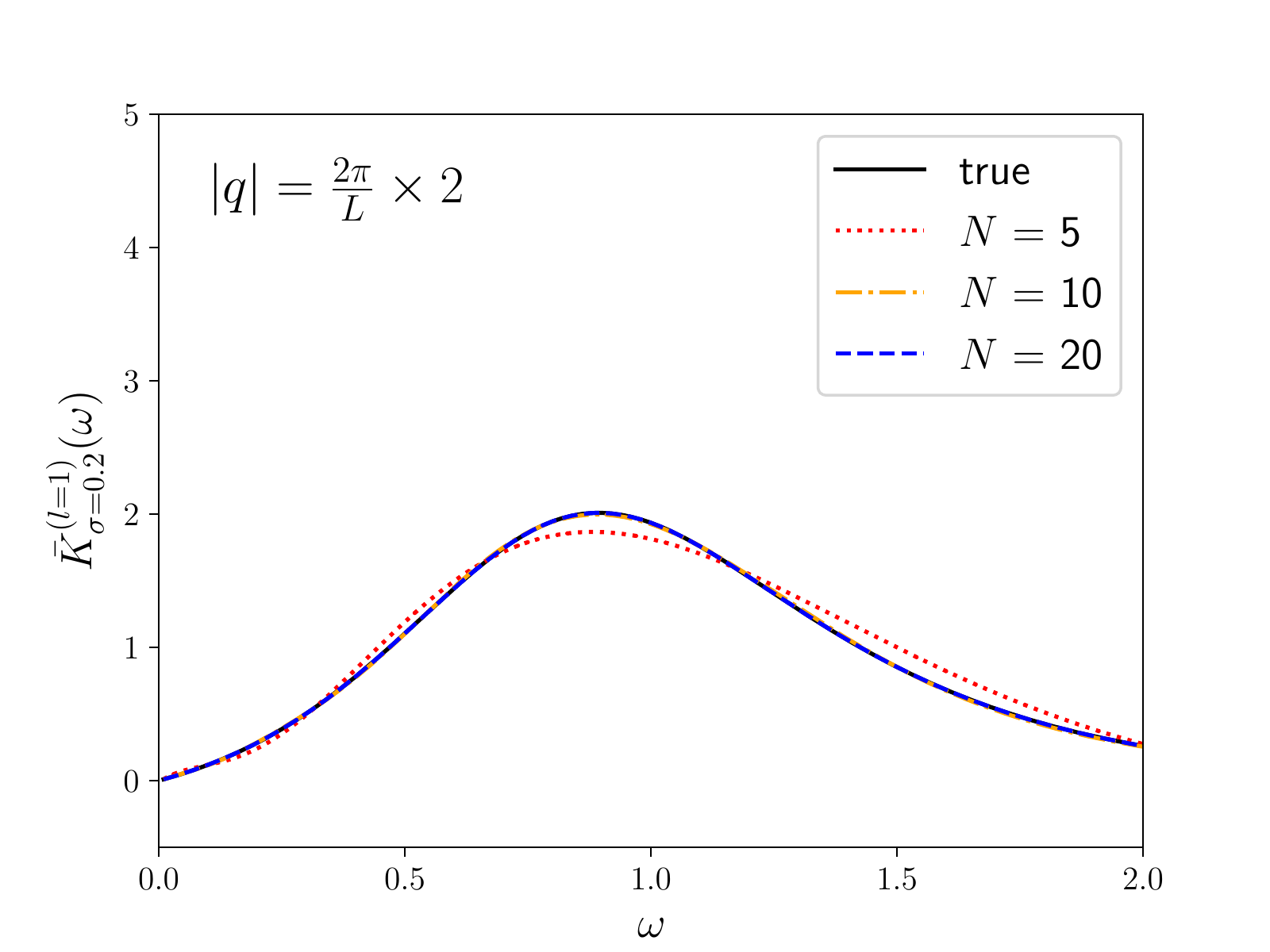}
  \includegraphics[width=5.2cm]{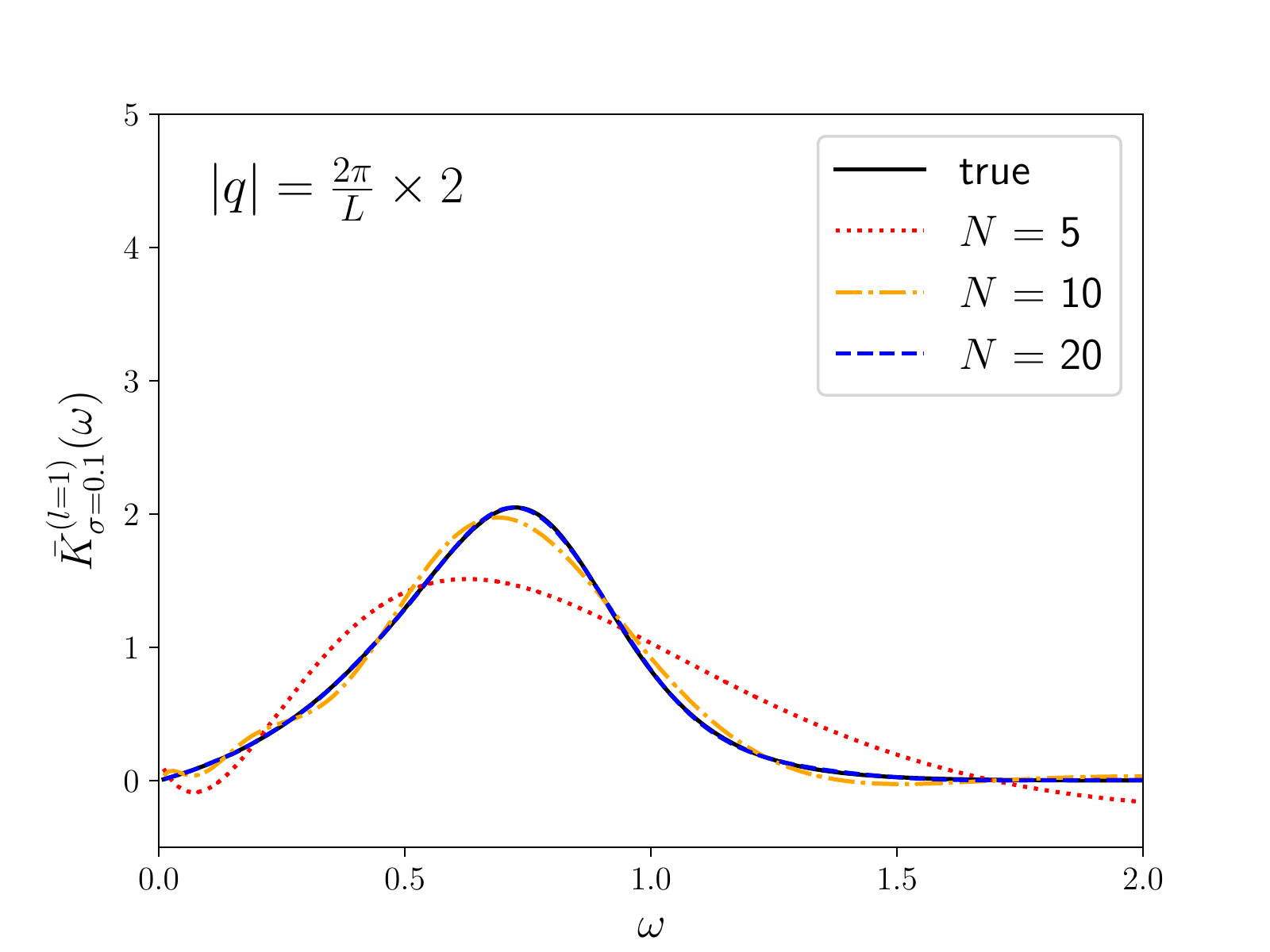}
  \includegraphics[width=5.2cm]{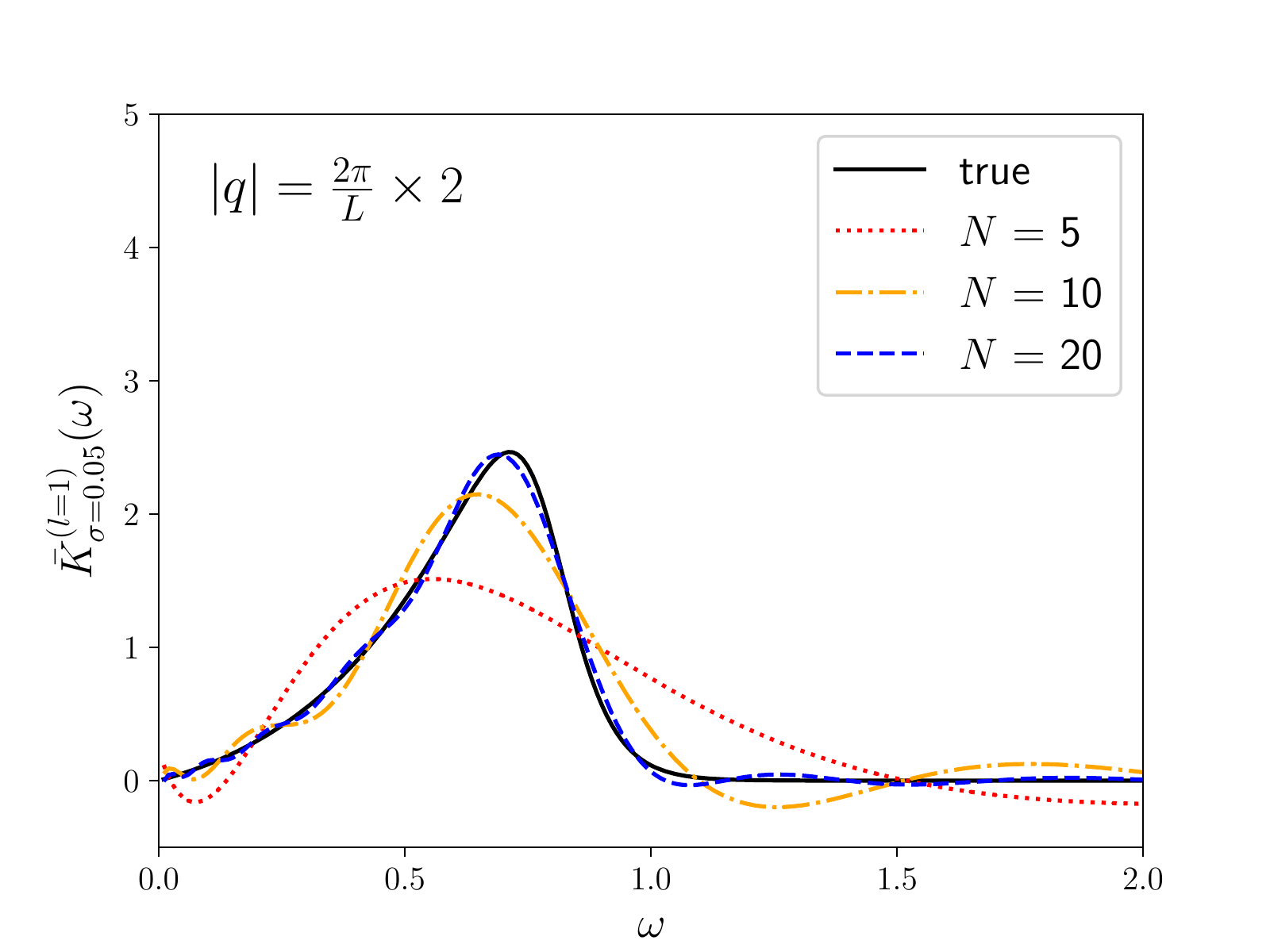}
  \includegraphics[width=5.2cm]{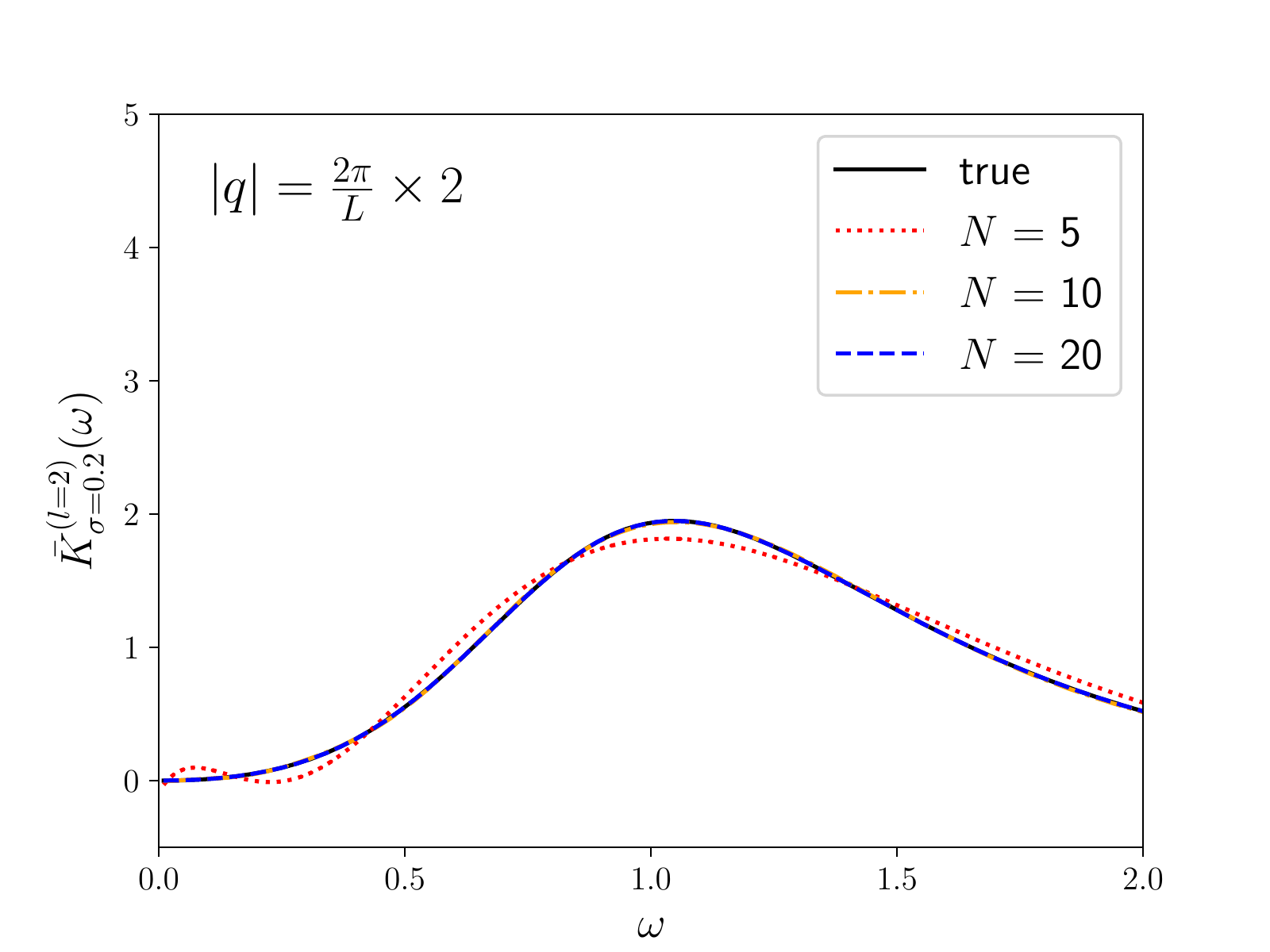}
  \includegraphics[width=5.2cm]{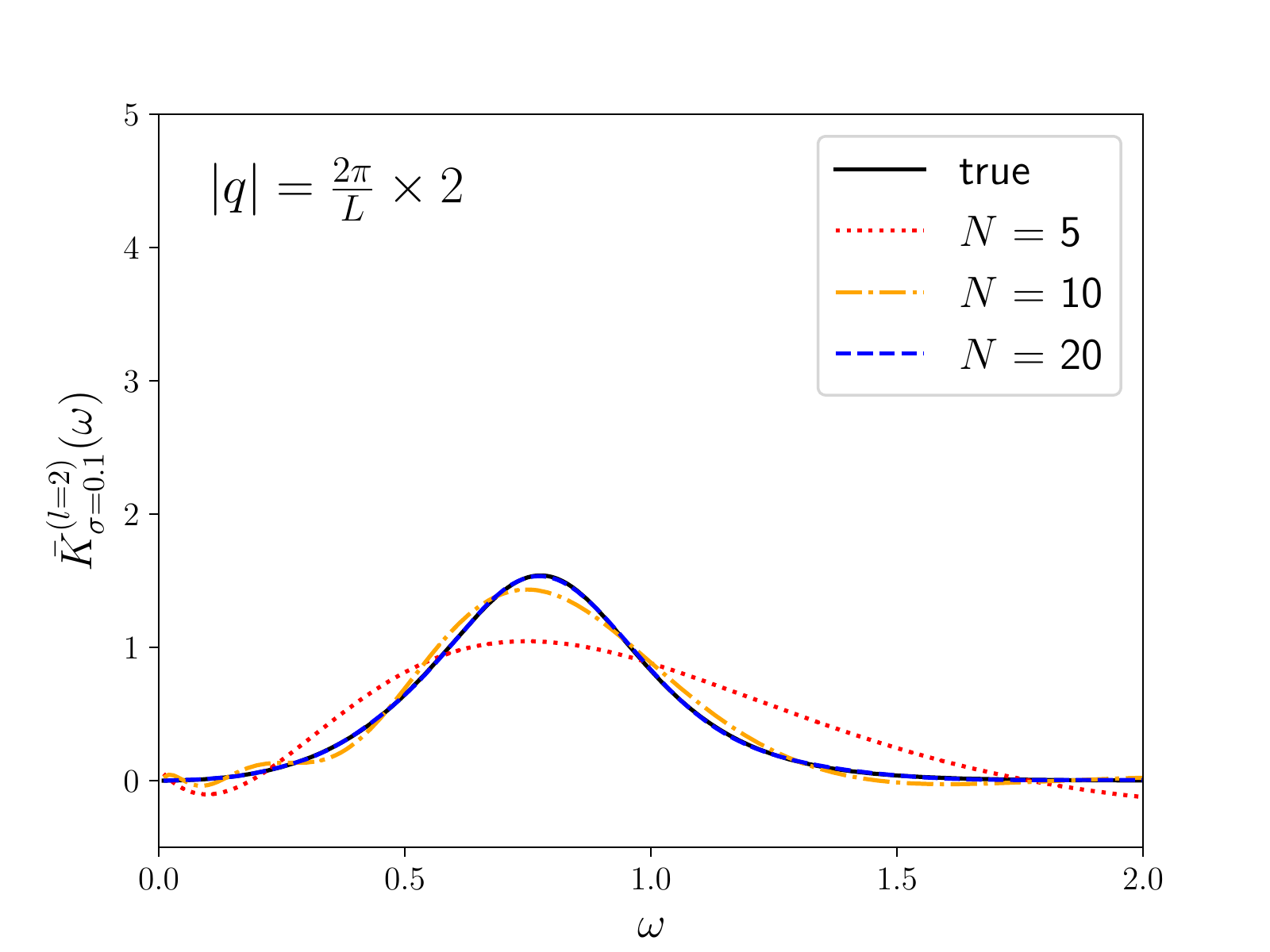}
  \includegraphics[width=5.2cm]{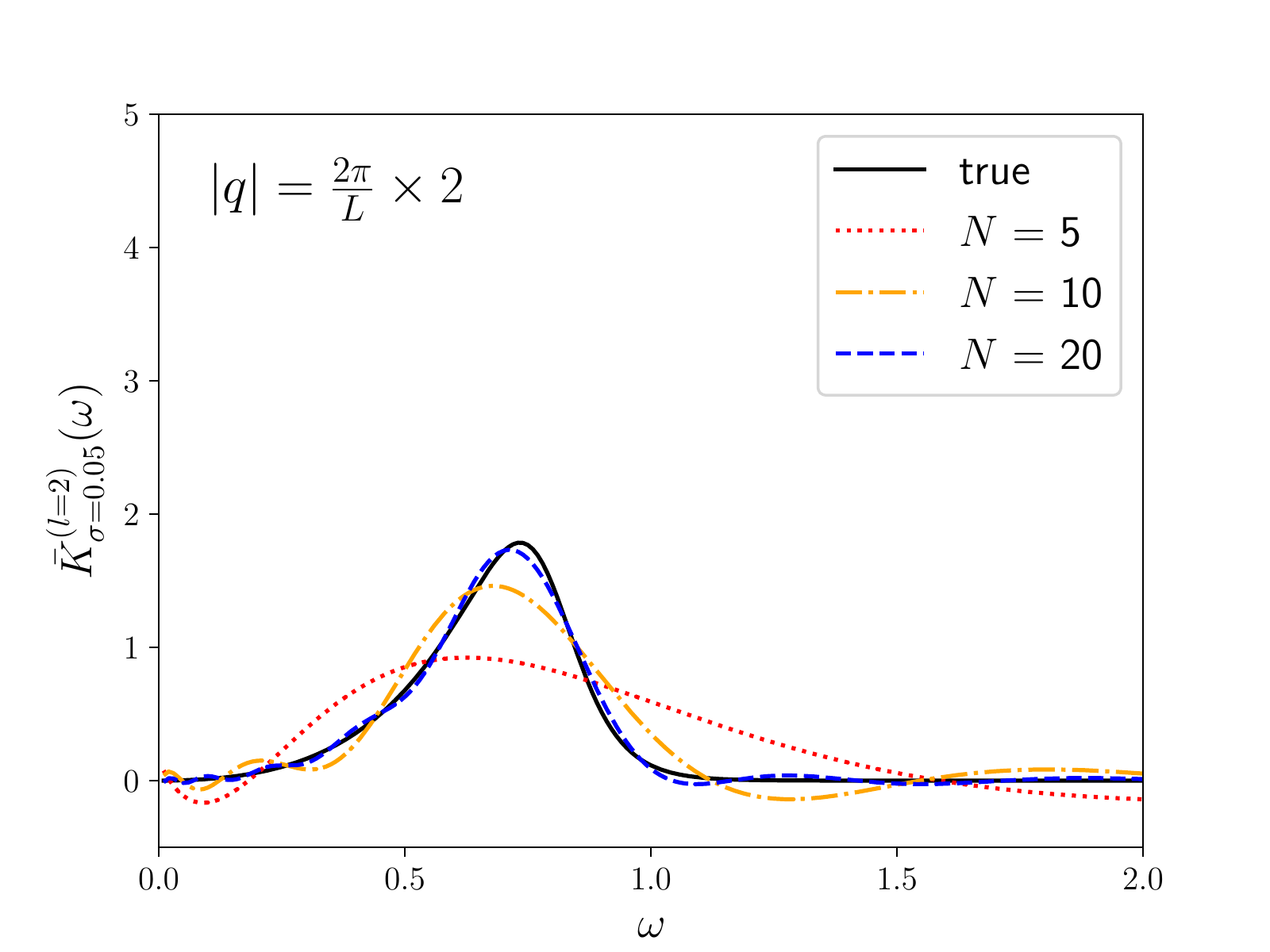}
  \caption{
    Same as Fig.~\ref{fig:K(omega)q1}, but with
    $|\bm{q}|=2\pi/L \times 2$.
    The upper limit of the kernel function is
    $\omega$ = 0.79.
  }
  \label{fig:K(omega)q2}
\end{figure}

\begin{figure}[tbp]
  \centering
  \includegraphics[width=5.2cm]{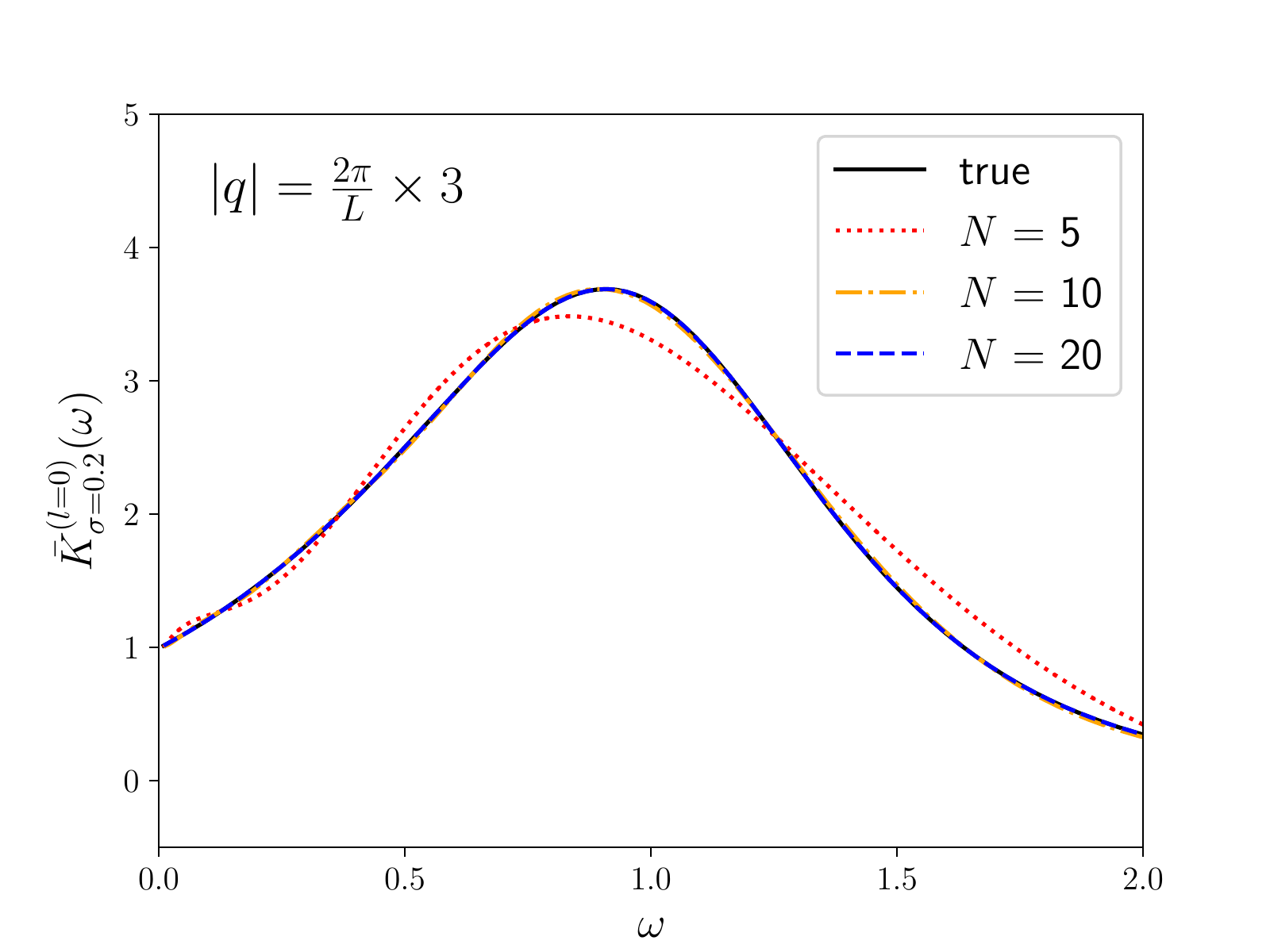}
  \includegraphics[width=5.2cm]{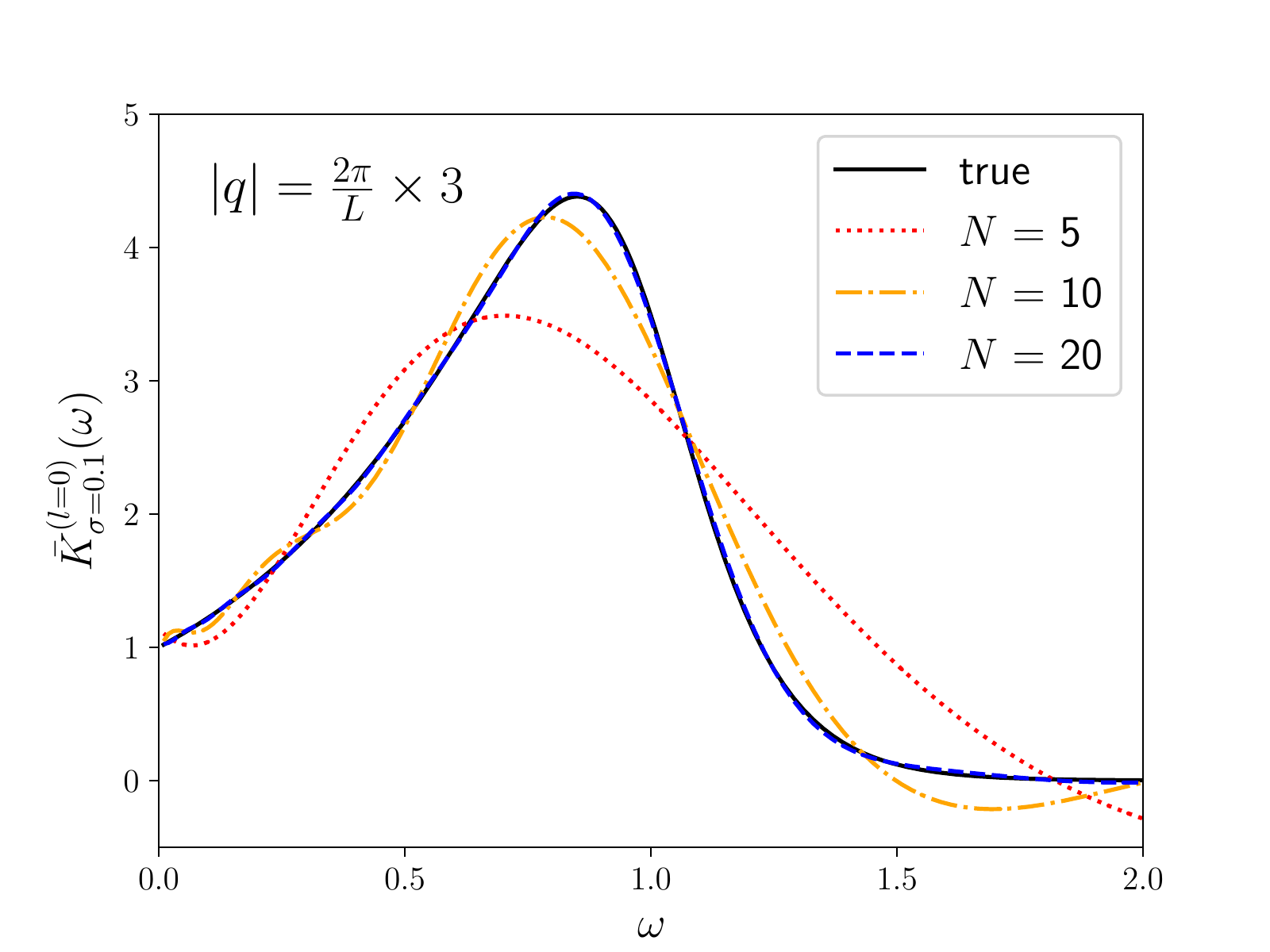}
  \includegraphics[width=5.2cm]{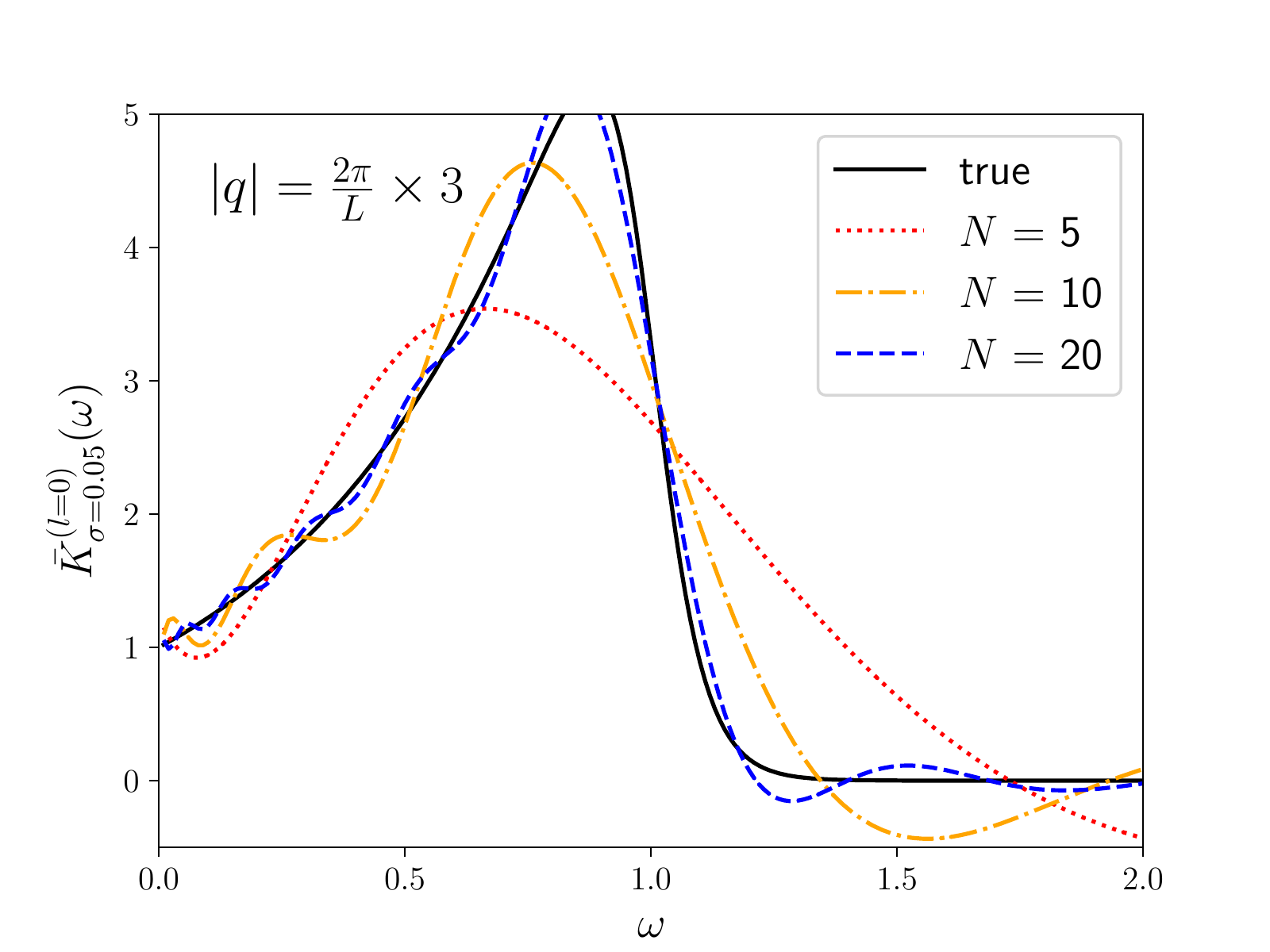}
  \includegraphics[width=5.2cm]{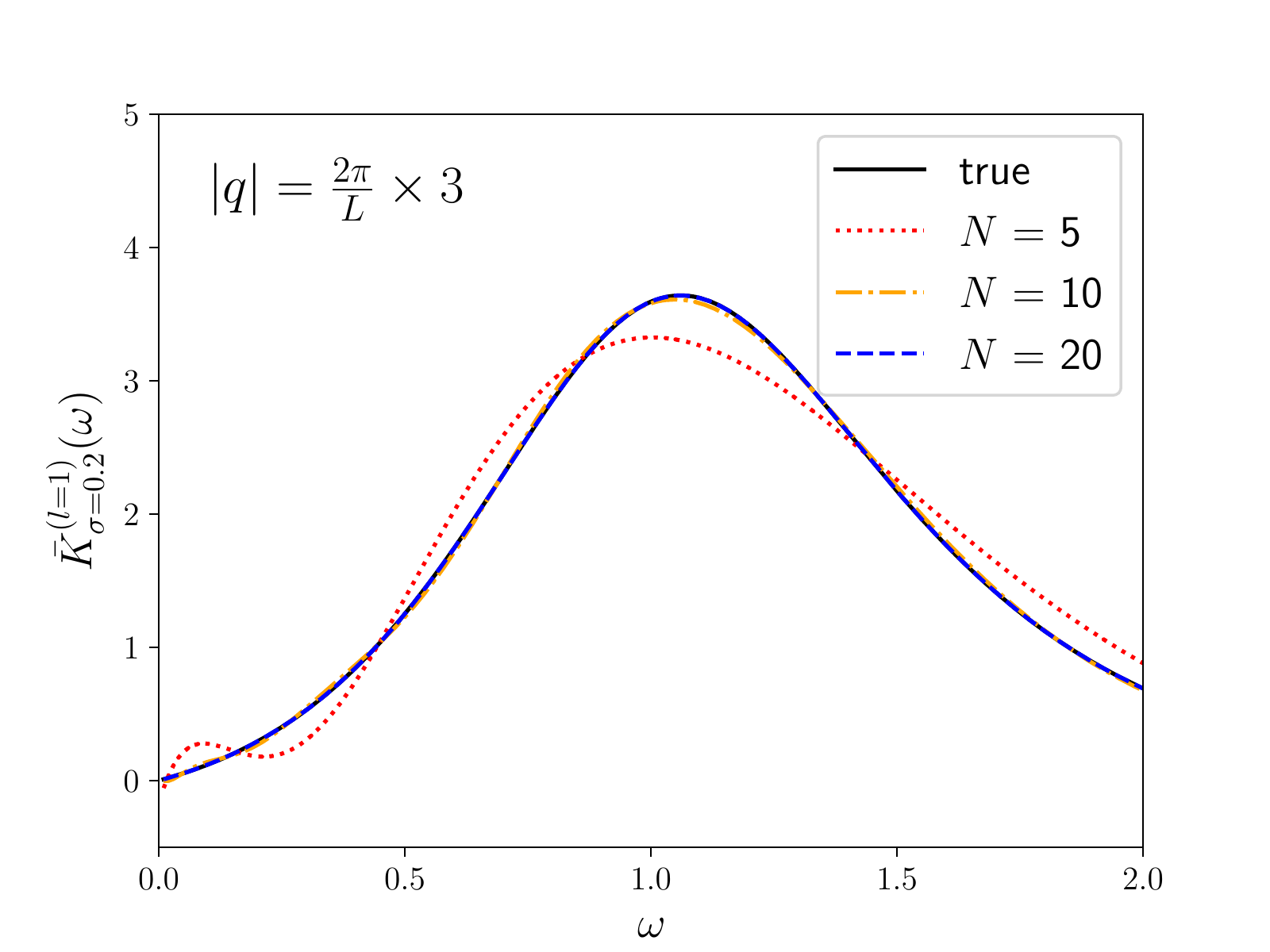}
  \includegraphics[width=5.2cm]{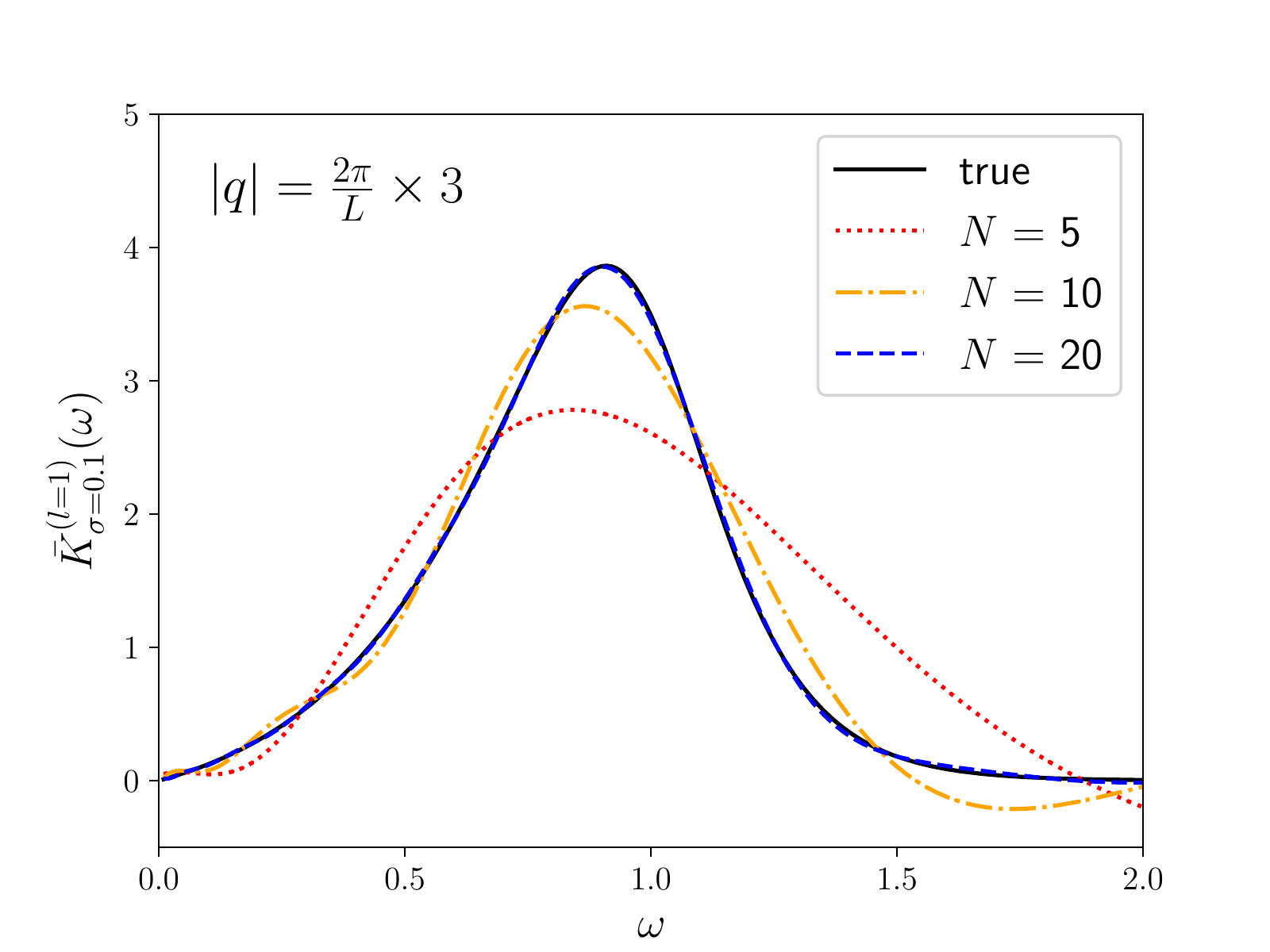}
  \includegraphics[width=5.2cm]{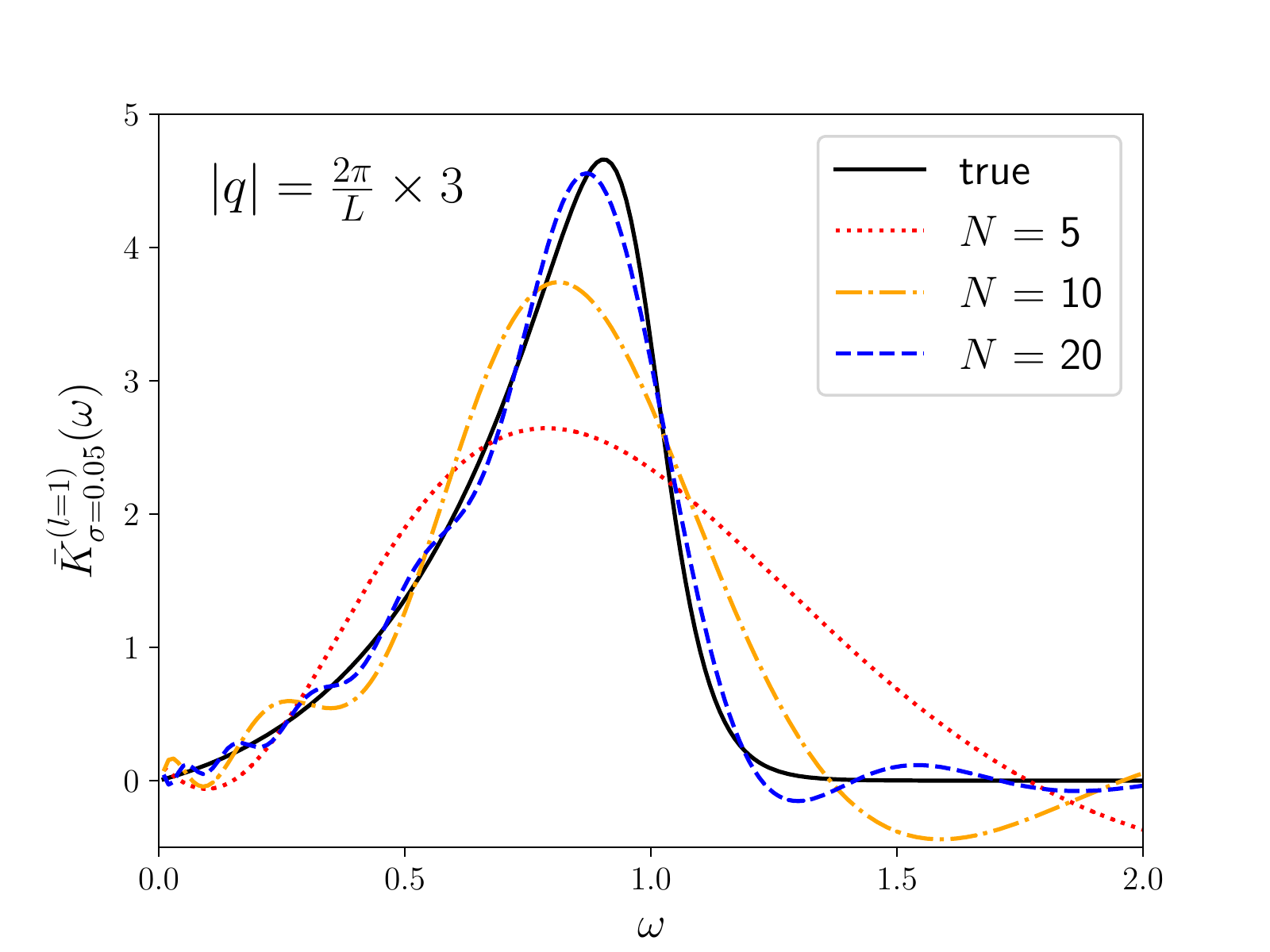}
  \includegraphics[width=5.2cm]{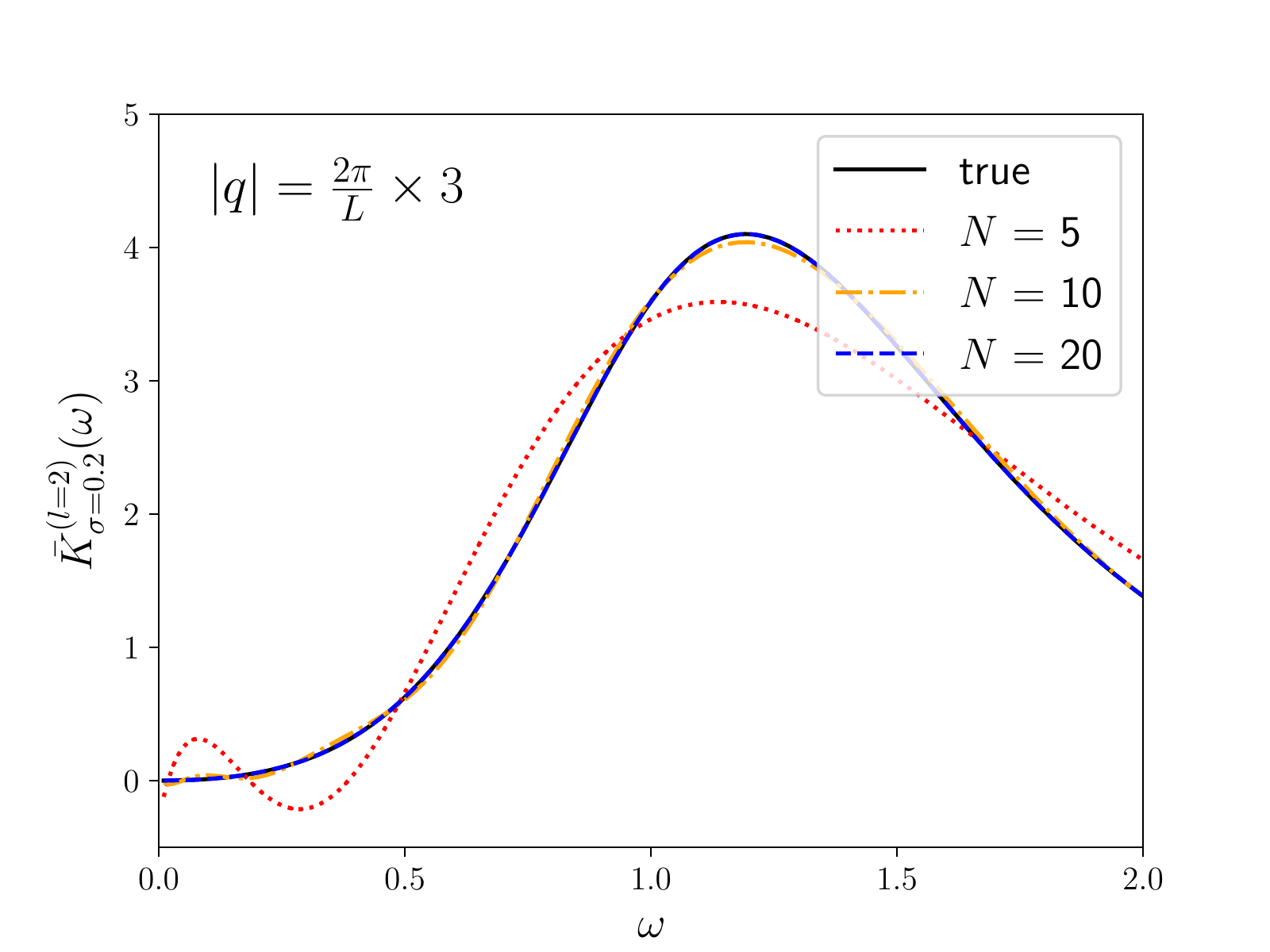}
  \includegraphics[width=5.2cm]{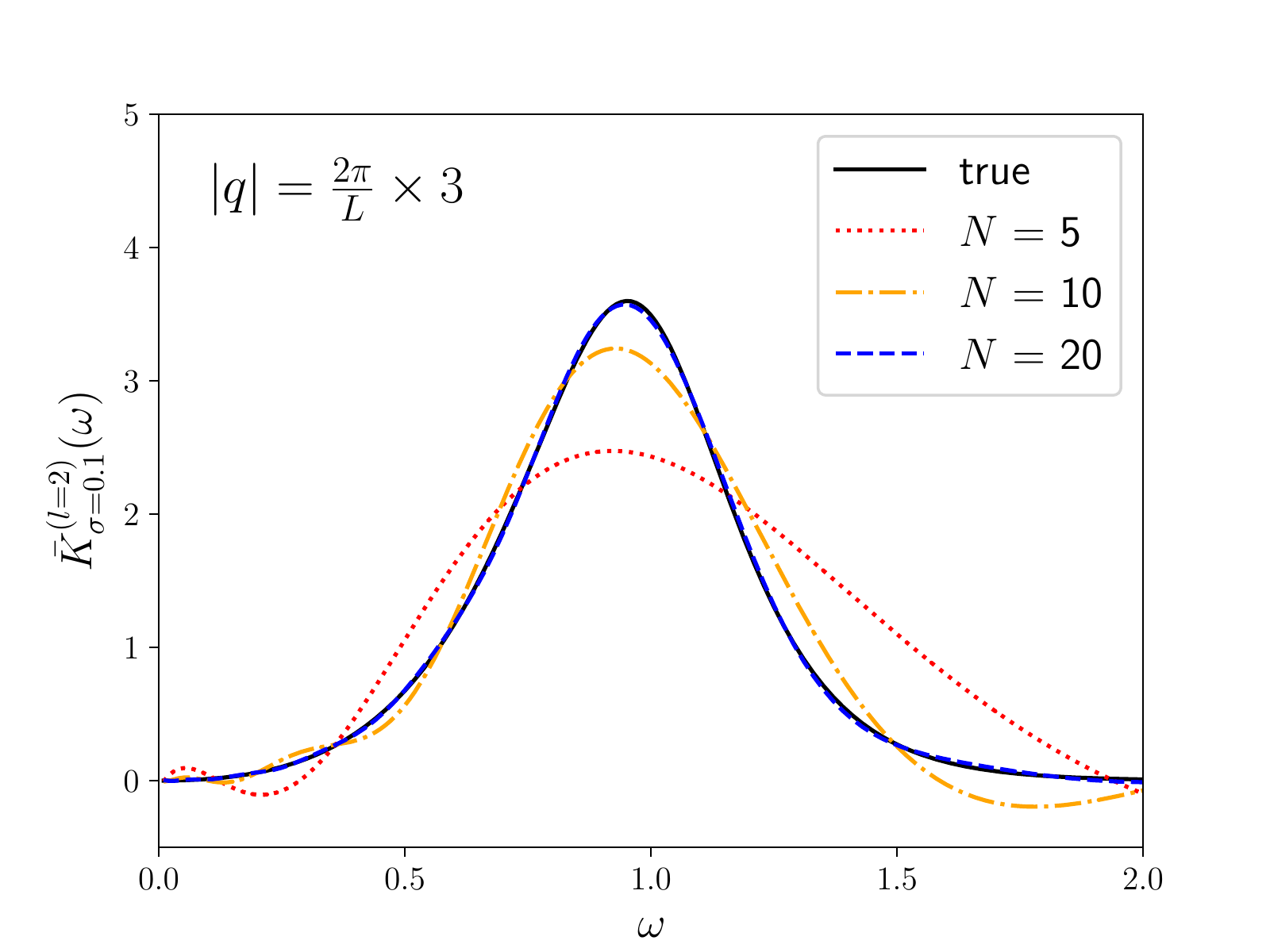}
  \includegraphics[width=5.2cm]{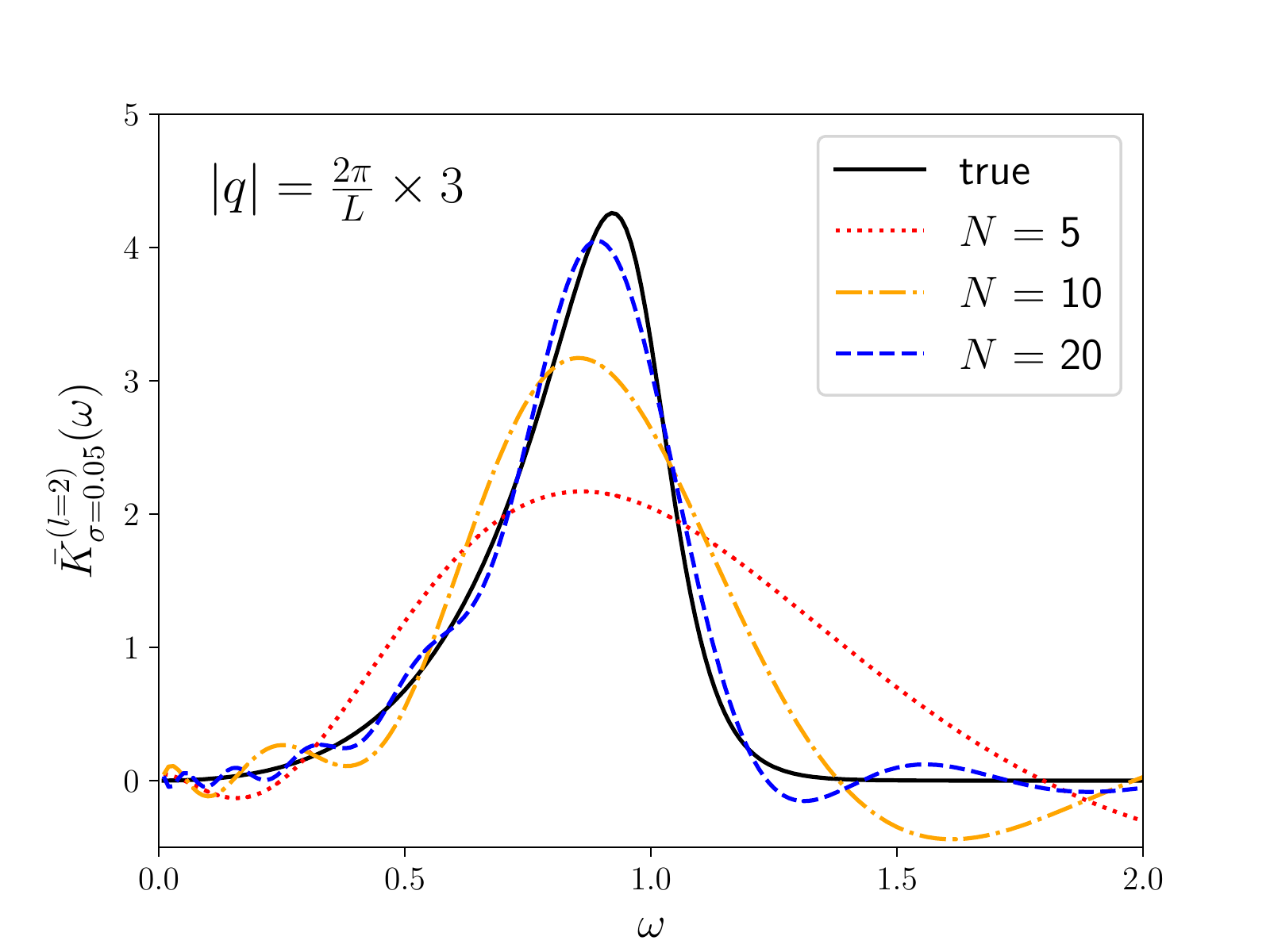}
  \caption{
    Same as Fig.~\ref{fig:K(omega)q1}, but with
    $|\bm{q}|=2\pi/L \times 3$.
    The upper limit of the kernel function is
    $\omega$ = 0.99.
  }
  \label{fig:K(omega)q3}
\end{figure}

The Chebyshev approximations of the kernel with polynomial orders
$N$ = 5, 10, and 20 
are shown in Fig.~\ref{fig:K(omega)q1}--\ref{fig:K(omega)q3}.
A general observation is that the approximation of the kernel is quite
good when it is sufficiently smeared, say $\sigma$ = 0.2.
For instance, see Fig.~\ref{fig:K(omega)q1} for the case of the
smallest non-zero momentum on this lattice, $|\bm{q}|=2\pi/L$.
As we make $\sigma$ smaller, we need higher order terms, {\it e.g.}
$N=20$ when $\sigma=0.05$, to obtain a reasonable approximation.
The same is true for larger momenta $|\bm{q}| = 2\pi/L\times 2$,
$2\pi/L\times 3$ 
(Figs.~\ref{fig:K(omega)q2} and \ref{fig:K(omega)q3}, respectively).

Even when the approximation for the kernel function is not very
accurate, the integral over $\omega$ may be obtained to a reasonable
accuracy. 
That is because the Chebyshev approximation produces a function that
oscillates around the true function.
Even in the worst case
($l=0$, $|\bm{q}|=2\pi/L\times 3$, and $\sigma=0.05$)
among those plotted in
Figs.~\ref{fig:K(omega)q1}-\ref{fig:K(omega)q3},
an integral of $\int_0^\infty d\omega K(\omega)$ deviates from its
true value by only 10\% ($N=5$), 2.7\% ($N=10$), 0.2\% ($N=20$).
Therefore, if the spectral function $W(\omega)$ is
$\omega$-independent, the error due to the Chebyshev approximation is
well under control. 
It becomes more problematic when the spectral function varies rapidly
around the upper limit of the integral.
That happens when a threshold opens near the upper limit
$M_N+|\bm{q}|$. 
For instance, the $\pi N$ threshold opens at $M_N+M_\pi$ where the
spectral function sharply increases.
The convolution integral $\int d\omega K(\omega) W(\omega)$ would then 
contain larger errors when $|\bm{q}|\sim M_\pi$, as we discuss in more
details below.
Fortunately, the single pion threshold is relatively easier to treat
theoretically using baryon chiral perturbation theory, and the
potential error may be corrected when higher precision is required.
The thresholds of more than one pions would be less problematic since
such contribution is added on the spectrum of fewer pions and
thus their impact is less significant.

Once the approximation is constructed, it is straightforward to
evaluate the $\omega$-integral.
We first extract
$\langle\psi_\mu(\bm{q})|e^{-\hat{H}t}|\psi_\nu(\bm{q})\rangle/
\langle\psi_\mu(\bm{q})|\psi_\nu(\bm{q})\rangle$
from the lattice correlator as described earlier.
By combining them we construct
$\langle\psi_\mu(\bm{q})|T_j(e^{-\hat{H}})|\psi_\nu(\bm{q})\rangle/
\langle\psi_\mu(\bm{q})|\psi_\nu(\bm{q})\rangle$,
from a ratio of correlators
$C_{\mu\nu}^{JJ}(t+2t_0;\bm{q})/C_{\mu\nu}^{JJ}(2t_0;\bm{q})$,
and then using (\ref{eq:cheb_approx}) with the coefficients $c_j^*$
determined from the form of the kernel function,
the integral is obtained.

A practical problem is that the Chebyshev polynomials $T_j^*(x)$
involve huge cancellations among different orders of $x^k$
($k=$ 0, 1, ..., $j$), since the coefficients may grow as fast as
$4^k$ with alternating signs.
As a consequence, 
$\langle\psi_\mu(\bm{q})|T_j^*(e^{-\hat{H}})|\psi_\nu(\bm{q})\rangle/
\langle\psi_\mu(\bm{q})|\psi_\nu(\bm{q})\rangle$
with large $j$ may have an error larger than 1.
Since the Chebyshev polynomials are constructed such that
$|T_j(x)|\le 1$ is satisfied, 
the terms whose magnitude is greater than 1 due to statistical
fluctuations can easily destroy the whole approximation.
In order to avoid such a problem, we should add constraints
$|\langle\psi_\mu(\bm{q})|T_j(e^{-\hat{H}})|\psi_\nu(\bm{q})\rangle/
\langle\psi_\mu(\bm{q})|\psi_\nu(\bm{q})\rangle|\le 1$,
when we determine their numerical values from the lattice data.
This can be done using a constrained fit.
Namely, we extract the Chebyshev matrix elements
$\langle\psi_\mu(\bm{q})|T_j^*(e^{-\hat{H}})|\psi_\nu(\bm{q})\rangle/
\langle\psi_\mu(\bm{q})|\psi_\nu(\bm{q})\rangle$
from the correlators
$\langle\psi_\mu(\bm{q})|e^{-\hat{H}t}|\psi_\nu(\bm{q})\rangle/
\langle\psi_\mu(\bm{q})|\psi_\nu(\bm{q})\rangle$
by a fit with the constraints.
Due to the statistical error of the correlators, the higher order
Chebyshev polynomials are not well determined but limited within
the range between $-1$ and $+1$. 
Such poorly determined higher order terms are still useful to estimate
potential errors due to the truncation of the polynomial.
(See \cite{Bailas:2020qmv} for details).

\begin{figure}[tbp]
  \centering
  \includegraphics[width=5.2cm]{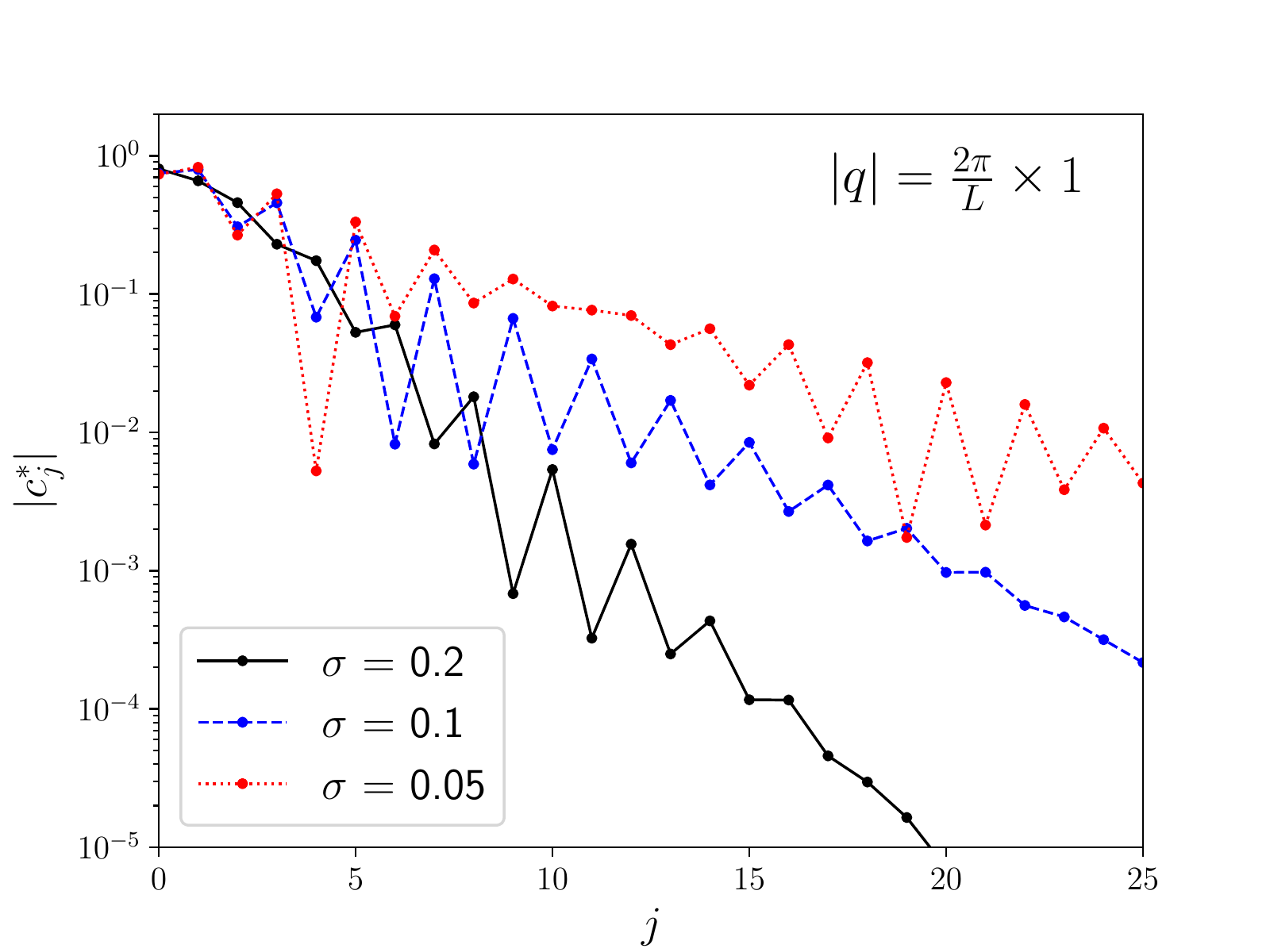}
  \includegraphics[width=5.2cm]{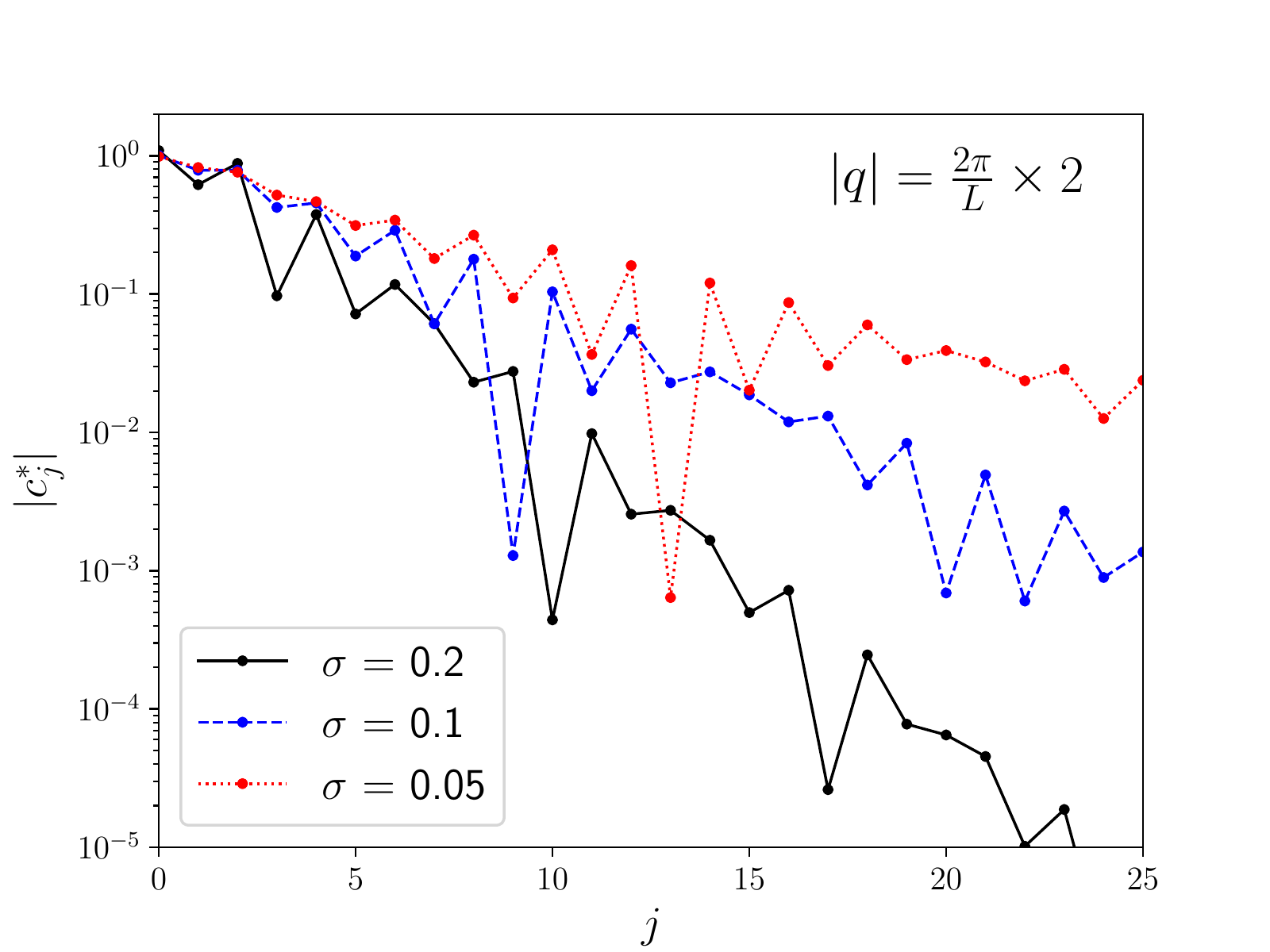}
  \includegraphics[width=5.2cm]{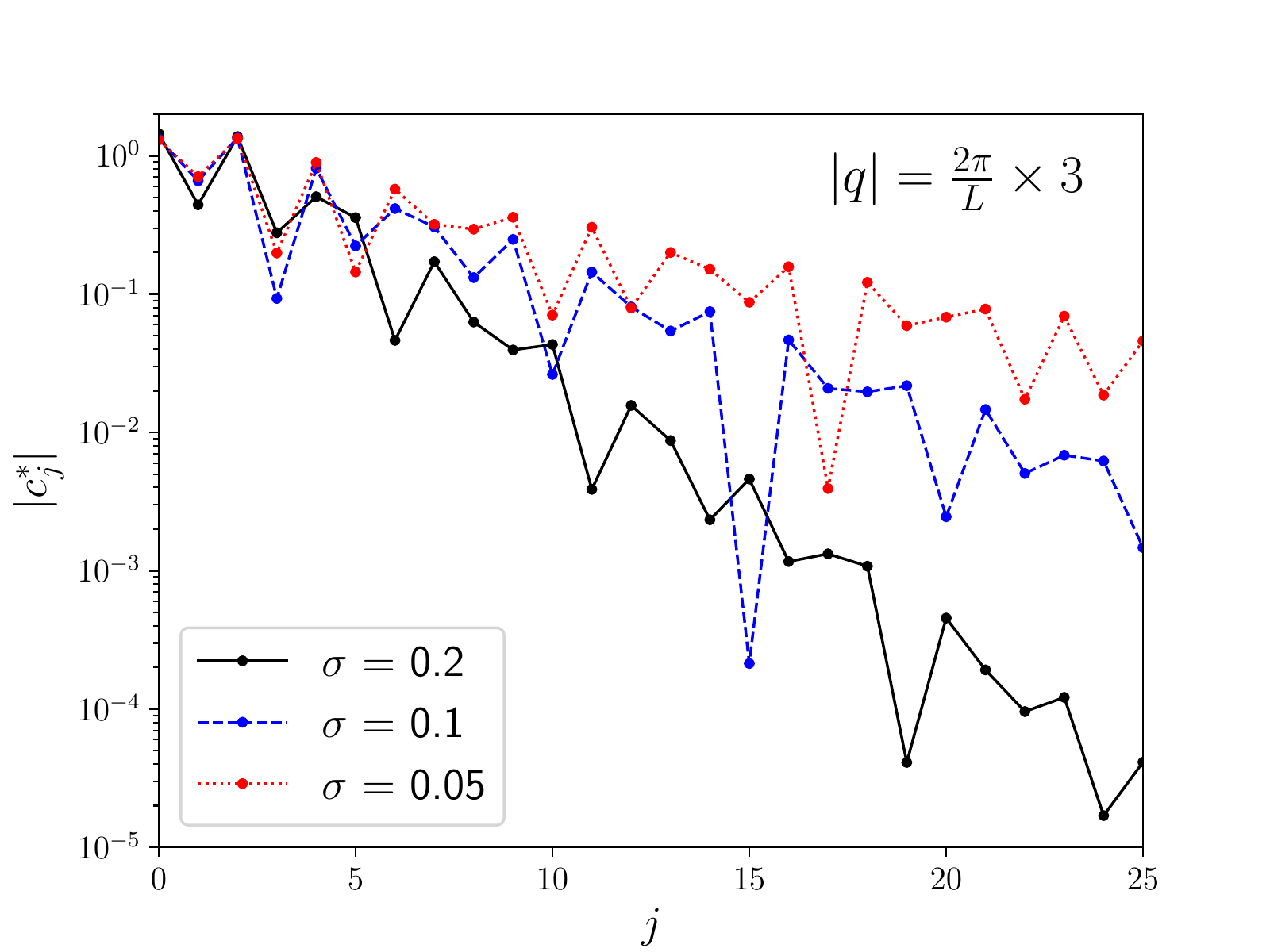}
  \includegraphics[width=5.2cm]{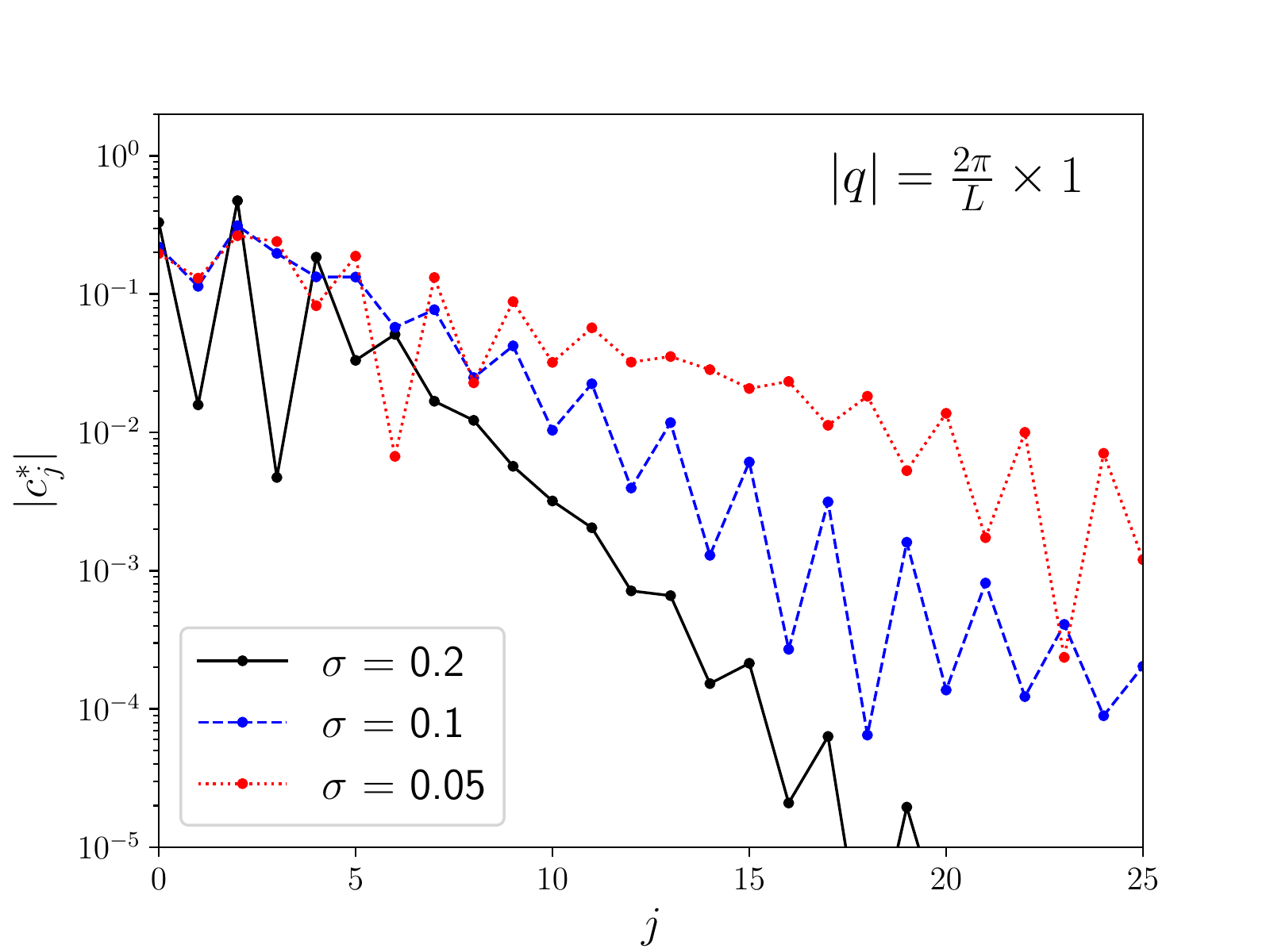}
  \includegraphics[width=5.2cm]{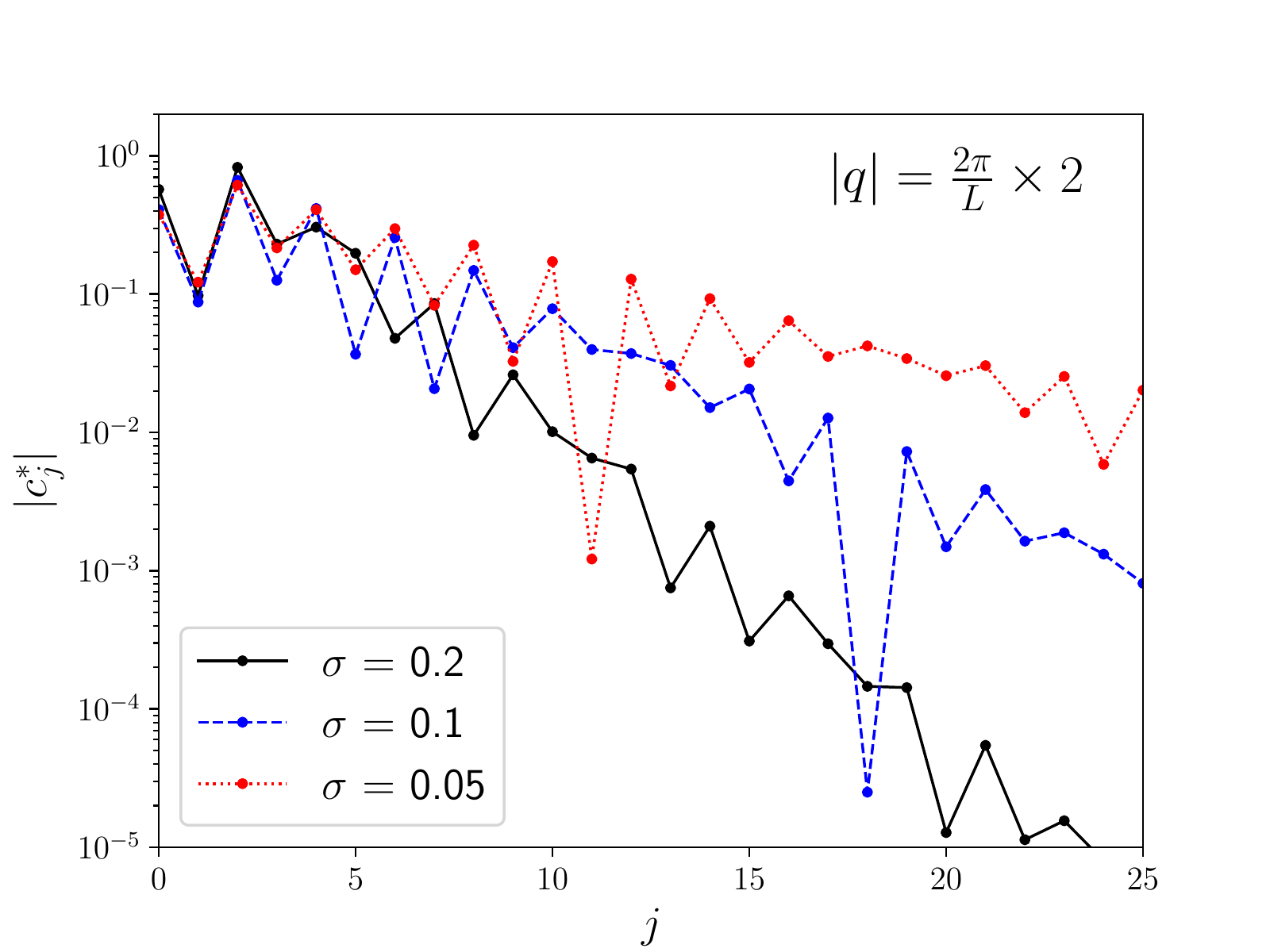}
  \includegraphics[width=5.2cm]{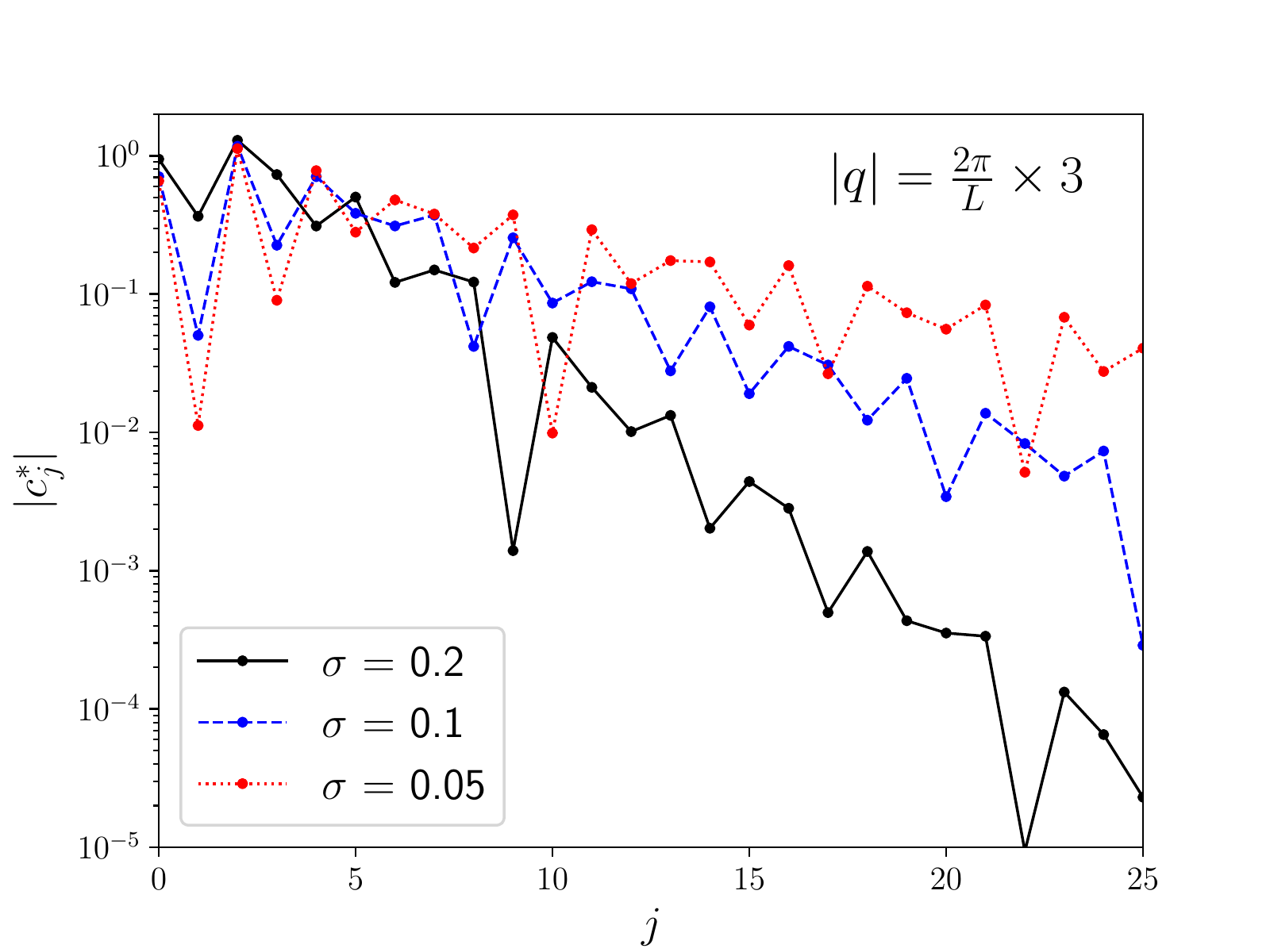}
  \includegraphics[width=5.2cm]{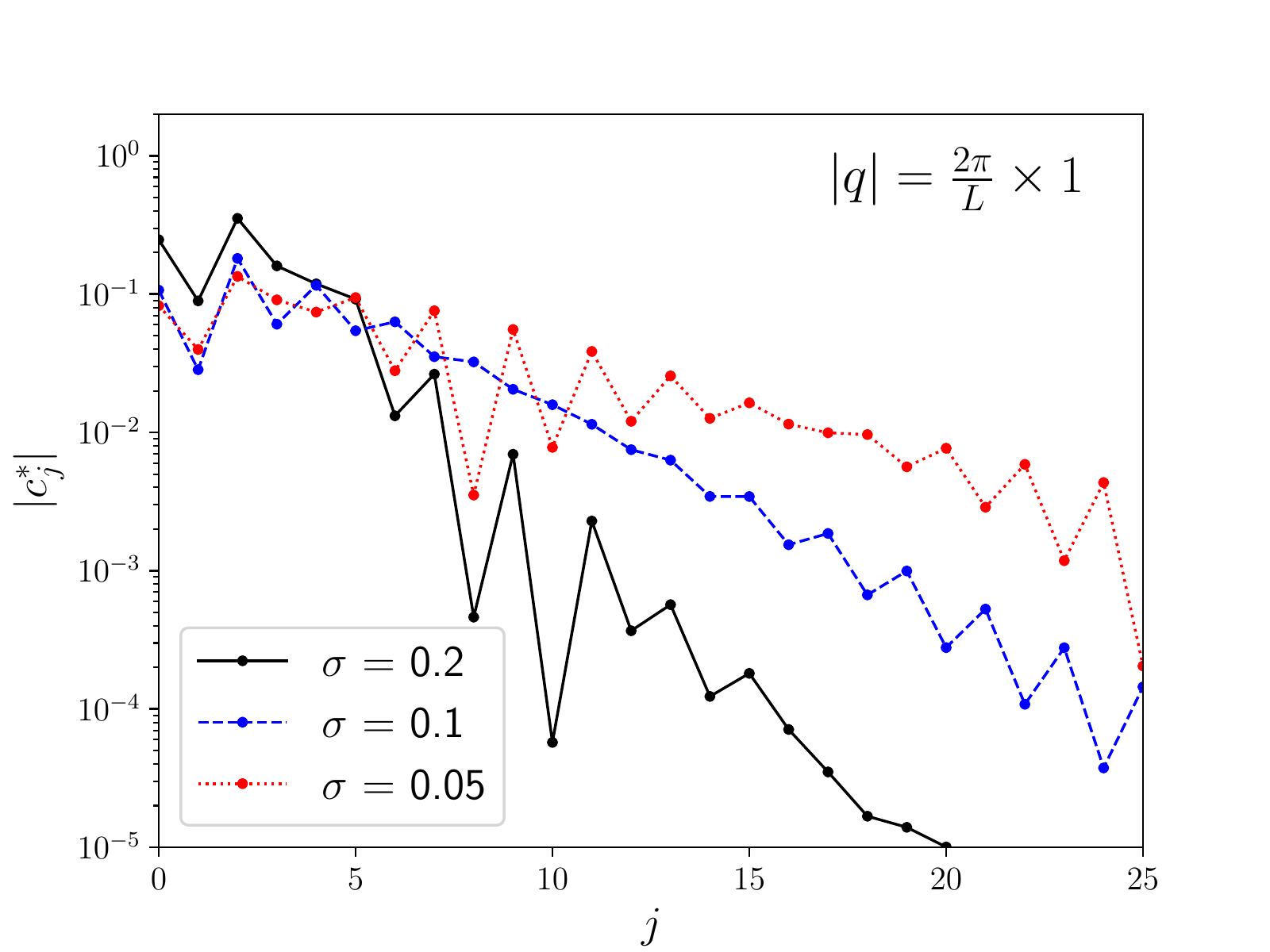}
  \includegraphics[width=5.2cm]{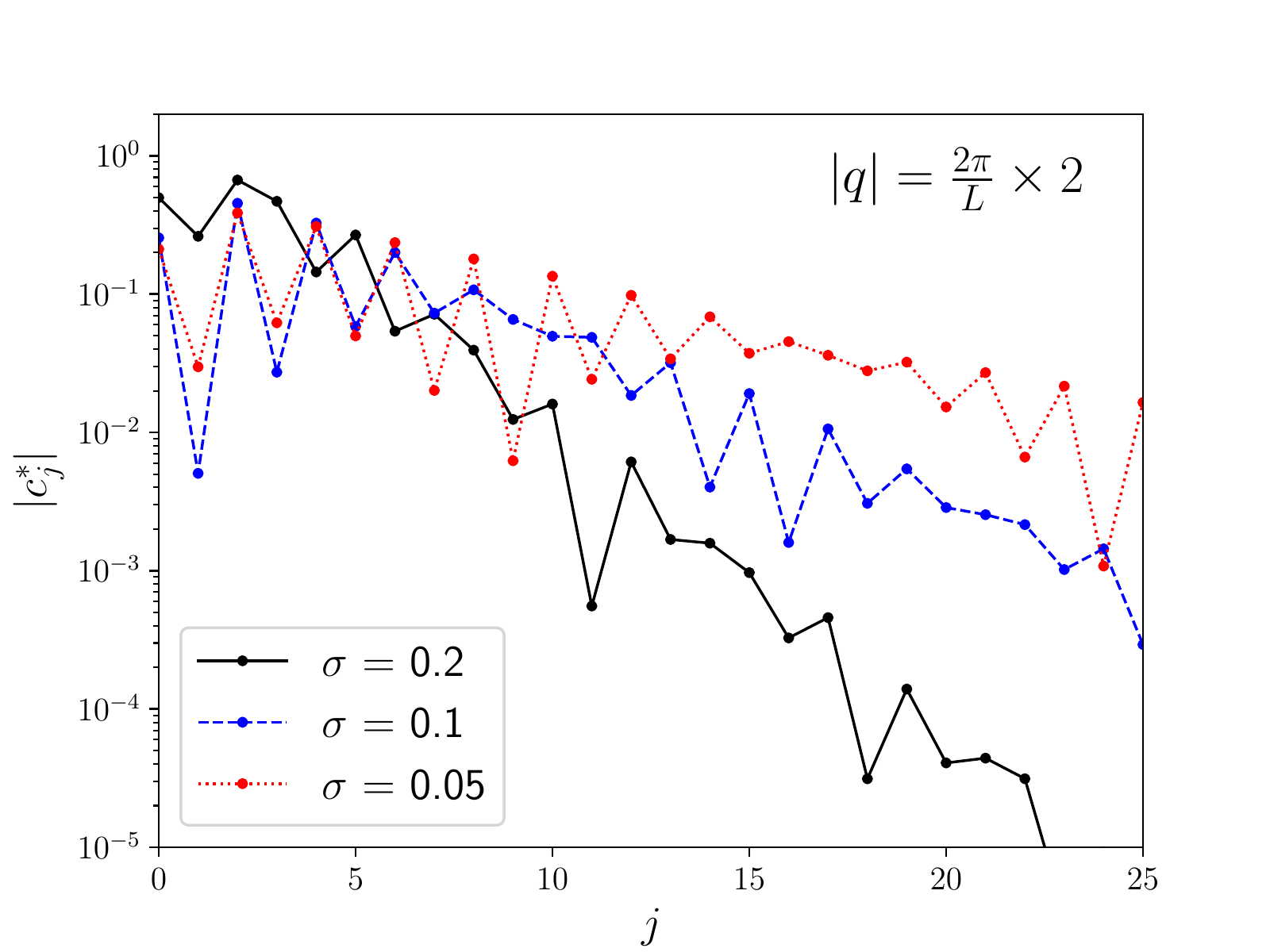}
  \includegraphics[width=5.2cm]{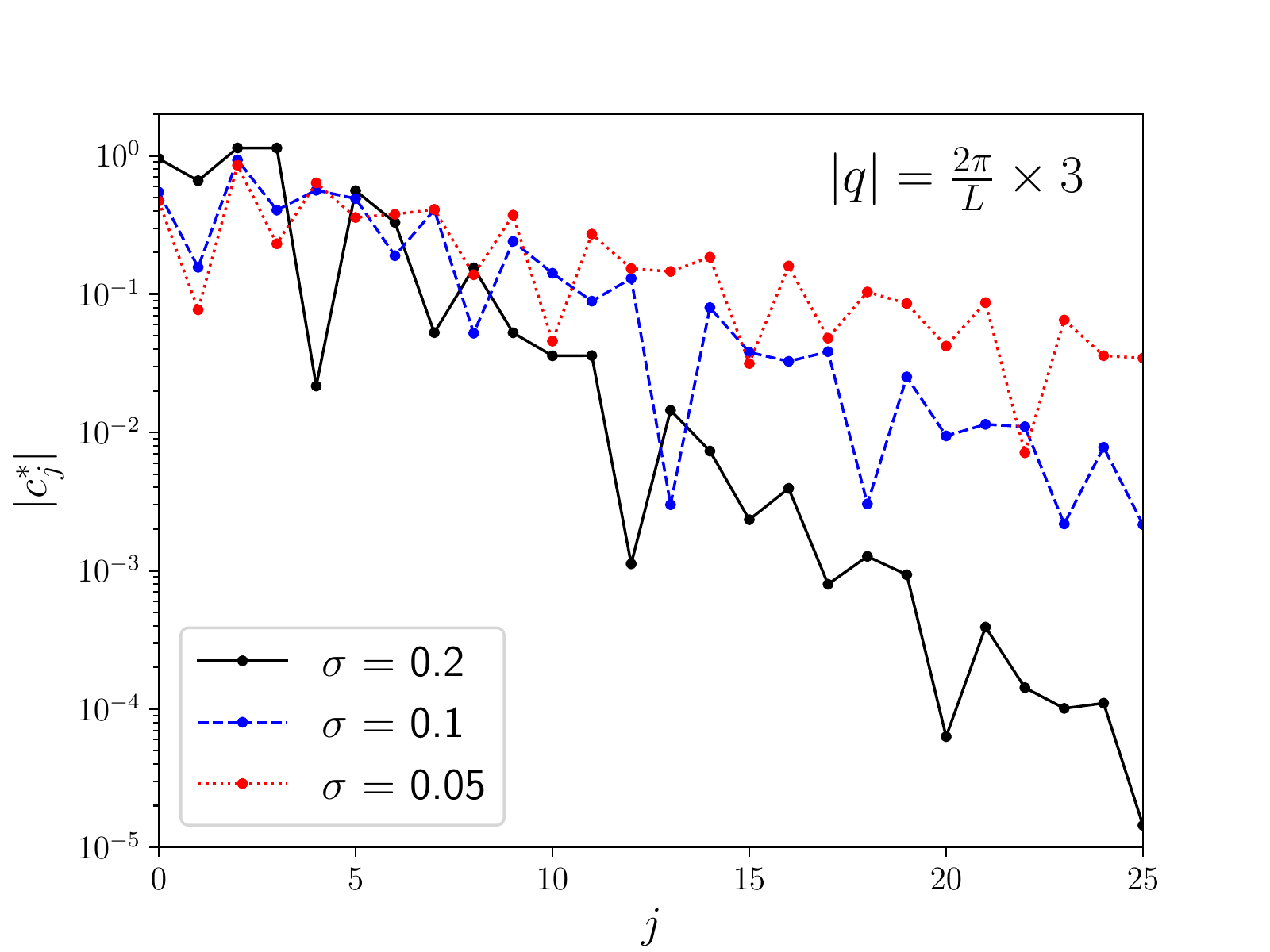}
  \caption{
    Chebyshev coefficients $|c_j^*|$.
    Top panels are for $l=0$ while the middle and bottom panels show
    them for $l=1$ and 2.
    From left to right, the inserted momentum increases:
    $|\bm{q}|=2\pi/L$, $2\pi/L\times 2$, $2\pi/L\times 3$.
  }
  \label{fig:cj}
\end{figure}

Systematic error due to ignored higher order terms in the Chebyshev
approximation can be estimated from their coefficients.
Since the shifted Chebyshev polynomials, $T_j^*(x)$, form a orthonormal
basis of functions in $0\le x\le 1$ and those of higher $j$ represents
rapid variation of the true function, we expect that the high-$j$
coefficients $c_j^*$ are suppressed when the true function varies only
slowly. 
This can be confirmed with the examples of the approximations given
above.
Figure~\ref{fig:cj} shows how $c_j^*$ decreases for higher polynomial
orders $j$.
The plots are presented for different $l$'s and momenta $|\bm{q}|$'s.
For all the cases, the coefficient $c_j^*$ basically falls
exponentially as $j$ gets higher.
The rate of decrease is faster when the smearing width $\sigma$ is
larger ($\sigma$ = 0.2), and $|c_j^*|$ becomes $O(10^{-3})$ or even
smaller already at $j=10$.
This size represents the relative error in the final result, because
the Chebyshev matrix elements
$\langle\psi_\mu(\bm{q})|T_j^*(e^{-\hat{H}})|\psi_\nu(\bm{q})\rangle/
\langle\psi_\mu(\bm{q})|\psi_\nu(\bm{q})\rangle$
are constrained between $-1$ and $+1$ by construction.
The decrease of $c_j^*$ becomes slower for smaller $\sigma$.
Thus, for more rapidly varying kernel functions, much larger polynomial
orders are necessary to achieve the same precision.
When the lattice data are available only in a limited range of the
time separation between two currents, this sets the limit of the
method, and we have to choose sufficiently large $\sigma$ such that
the truncation error is under control.
The results should then be extrapolated to the limit of $\sigma\to 0$.

An example of the extrapolation $\sigma\to 0$ is shown in Fig.~2 of
\cite{Gambino:2020crt} for the inclusive semileptonic $B$ decays.
It is for the same type of function with $l=2$, and the data show that
the dependence on the smearing parameter $\sigma$ is mild and becomes
essentially flat for some small values of $\sigma$ ($\lesssim$ 0.1 in
the lattice unit).

For the $\ell N$ scattering, for which lattice computation of the
relevant amplitude is not available in the form useful to perform 
this analysis, we consider a simple model that describes an elastic
scattering plus a single pion production processes.
The elastic channel corresponds to a delta function
$\delta(\omega-\sqrt{M_N^2+\bm{q}^2})$ in the spectral function
$W(\omega)$ in (\ref{eq:KW}).
The single pion production begins at $s=(M_N+m_\pi)^2$ for
$s=\omega^2-\bm{q}^2$.
The spectral function typically has a shape
\begin{equation}
  W(\omega) \sim
  \pi\sqrt{\left(1-\frac{M_N^2-m_\pi^2}{s}\right)^2
    - \frac{4m_\pi^2}{s}},
\end{equation}
which comes from the imaginary part of an one-loop diagram describing
a creation and annihilation of a $N\pi$ pair in the $s$ channel.
(Strictly speaking, this is valid only for two scalar particles.
To be more realistic, one should use for instance the heavy-baryon
chiral perturbation theory.
The qualitative feature near the threshold is expected to be
unchanged, though.) 
It is shown in Fig.~\ref{fig:mock} (purple curve) with an arbitrary
unit.
The relative weight between the elastic contribution (a peak at
$\omega=\sqrt{M_N^2+\bm{q}^2}$) and the $N\pi$ continuum is unknown,
so that the height of the peak is also arbitrary.
The task is then to obtain a convolution integral of $W(\omega)$ with
the (smeared) kernel $\bar{K}(\omega)$.
Apparently, the contribution of the elastic channel is underestimated
by the smearing, while the $N\pi$ contribution is likely
overestimated because the spectrum is an increasing function.

\begin{figure}[tbp]
  \centering
  \includegraphics[width=10cm]{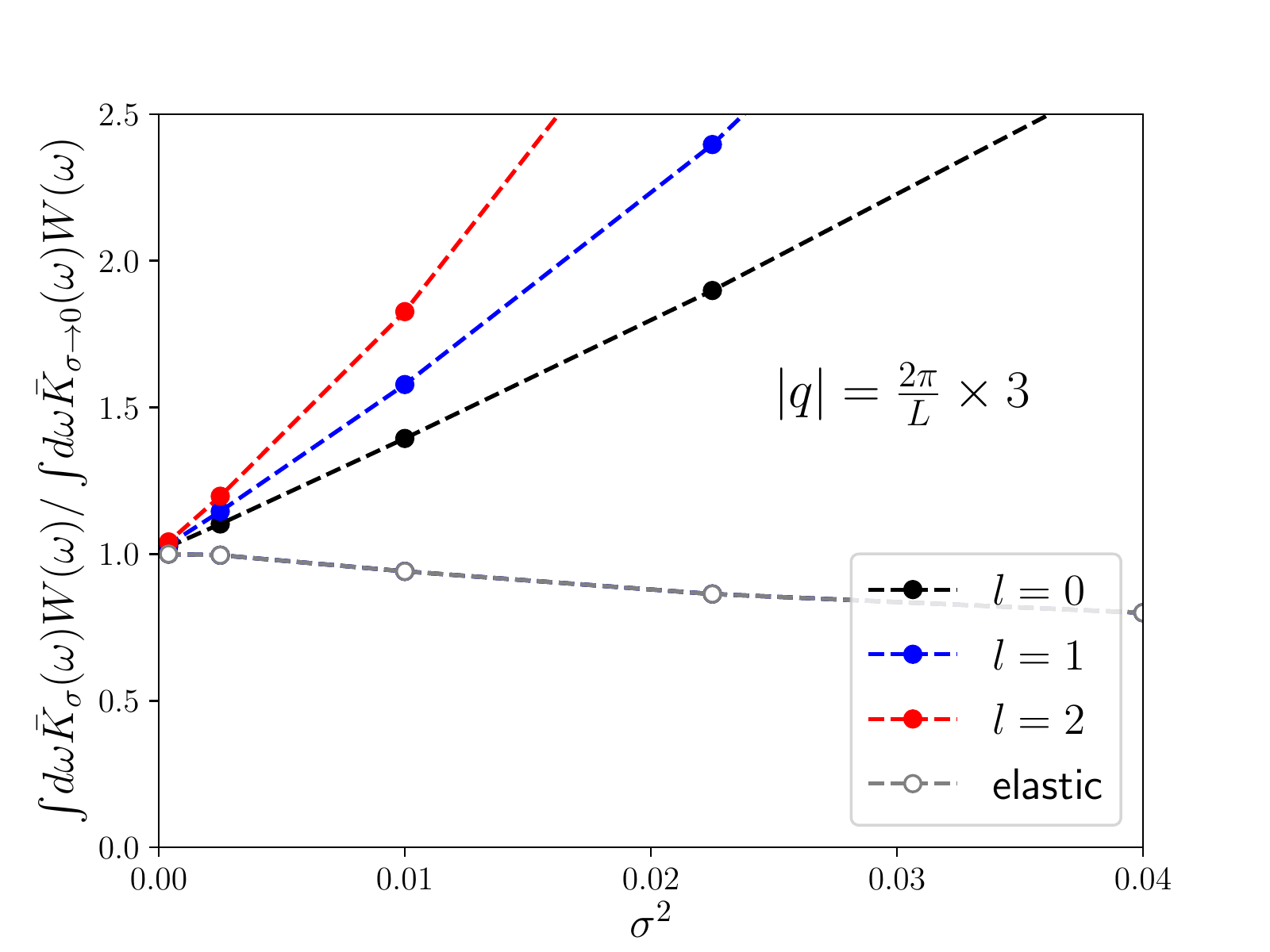}
  \caption{
    Extrapolation of the energy integral to the limit of
    $\sigma\to 0$.
    The estimate is given with a mock spectrum.
    The integral at finite $\sigma$ divided by that of $\sigma=0$ is
    shown. 
    The contribution from the elastic channel (gray), which is
    independent of $l$, extrapolates from below.
    The $N\pi$ contribution depends on $l$ ($l$ = 0 (black), 1 (blue),
    2 (red)) and is overestimated at finite $\sigma$.
    The horizontal axis is $\sigma^2$.
    The data points correspond to $\sigma$ = 0.15, 0.1, 0.05 as well as
    0.02.
    The last point is already very close to $\sigma=0$.
  }
  \label{fig:sigma_extrap}
\end{figure}

We estimate the error due to the smoothing with the model described
above.
This example, shown in Fig.~\ref{fig:mock}, at
$|\bm{q}|=2\pi/L\times 3$
is a particularly dangerous case, since the discontinuity of the
kernel is in the region where the spectrum is rapidly growing from the
$N\pi$ threshold.
The deviation from the true value at finite $\sigma$ and its
extrapolation to $\sigma\to 0$ is shown in
Figure~\ref{fig:sigma_extrap}.
The plot shows the integral of the mock data divided by the true
value for the elastic contribution as well as for the $N\pi$ states.
The elastic contribution (open circles) has the same relative error
among different $l$'s, because it is simply determined by the value of
$\bar{K}_\sigma(\omega)/\bar{K}_{\sigma\to 0}(\omega)$ at
$\omega=\sqrt{M_N^2+\bm{q}^2}$.
The $N\pi$ contribution depends on the details on the kernel function,
thus on $l$, and the error increases with $l$ (filled circles).
It appeared that the error in this particular case is quite
significant for the $N\pi$ continuum contribution, and one probably
needs $\sigma=0.1$ or smaller to control the extrapolation, which
seems to be well described by a linear dependence on $\sigma^2$.
The error would probably cancel between the elastic and $N\pi$
contributions.
When the elastic contribution is relatively large, the cancellation
becomes stronger and the total error might not be as substantial as
the estimate for the $N\pi$ contributions suggests.

The actual error depends on the details of the spectrum.
The value of $\sigma$ can be chosen at the analysis stage of the
lattice data, and thus the analysis can be repeated for various
$\sigma$ without extra computational cost.
How much one can reduce $\sigma$ depends on how large time separations
the lattice data exist at without overwhelmed by the statistical
noise.

\section{Quark-line contractions with two current insertions}
\label{sec:contraction}
The challenging part in the lattice computation of the inelastic
$\ell N$ cross section is the calculation of the forward
Compton-scattering amplitude (\ref{eq:Cmunu}), which involves two
current insertions.
It may be obtained utilizing the Feynman-Hellmann technique as applied
in \cite{Chambers:2017dov,Can:2020sxc}, but here we consider the more
conventional approach to contract the quark lines.

In order to extract the matrix element (\ref{eq:Cmunu}) 
we need to compute the following two-point and four-point functions on
the lattice.
For completeness, we also define the three-point function:
\begin{align}
  C_{2pt}(t_{sep}; \bm{p})
  =& 
     \sum_{\bm{y}}e^{-i \bm{p}\cdot(\bm{y}-\bm{x})} 
     \langle N(t_{snk},\bm{y}) \bar{N}(t_{src},\bm{x})\rangle, 
     \notag
  \\
  C_{3pt}^{\mathcal{O}_1}(t_{sep}, \tau_1; \bm{p}, \bm{q})
  =& 
     \sum_{\bm{y}, \bm{z}} e^{-i \bm{p}\cdot(\bm{y}-\bm{x})+ i\bm{q}\cdot(\bm{z}-\bm{x})} 
     \langle N(t_{snk},\bm{y}) \mathcal{O}_1(t_1,\bm{z})
     \bar{N}(t_{src},\bm{x})\rangle,
     \notag \\
  C_{4pt}^{\mathcal{O}_1, \mathcal{O}_2}(t_{sep},\tau_1,\tau_2;\bm{p},\bm{q}_1,\bm{q}_2)
  =&
     \sum_{\bm{y}, \bm{z}_1, \bm{z}_2}
     e^{-i \bm{p}\cdot(\bm{y}-\bm{x}) +
     i\bm{q}_1 \cdot(\bm{z}_1-\bm{x}) + i\bm{q}_2 \cdot(\bm{z}_2-\bm{x})} 
     \notag \\ 
   & \times
     \langle N(t_{snk},\bm{y}) \mathcal{O}_1(t_1,\bm{z}_1)
     \mathcal{O}_2(t_2,\bm{z}_2)\bar{N}(t_{src},\bm{x})\rangle,
     \notag 
\end{align}
with $t_{sep}\equiv t_{snk}-t_{src}$, 
$\tau_1\equiv t_1 -t_{src}$ and $\tau_2\equiv t_2 -t_{src}$.
The spatial momentum components are discretized on the lattice as
$p_i = 2\pi n_i/L$ with an integer $n_i$
for a lattice of the spatial extent $L$.
In the end we take the initial nucleon momentum to zero,
$\bm{p}=\bm{0}$, and the source position $\bm{x}$ can be fixed,
{\it e.g.} $\bm{x}=\bm{0}$, without loss of generality.
The operators $\mathcal{O}_1$ and $\mathcal{O}_2$ are the weak or
electromagnetic current with some Lorentz structure.

\subsection{Preparation}
The nucleon interpolating operators at source and sink
may be defined as
\begin{align}
  P_{\delta'}
  &= \epsilon^{a'b'c'} u_{\delta'}^{a'}
    [u^{b'}_{\beta'} S_{\beta'\gamma'} d^{c'}_{\gamma'}],
  &
    \bar{P}_{\delta}
  &= \epsilon^{abc}
    [\bar{d}^c_\gamma \bar{S}_{\gamma\beta} \bar{u}_\beta^b ]
    \bar{u}^a_{\delta},
    \notag \\
  N_{\delta'}
  &= \epsilon^{a'b'c'} d_{\delta'}^{a'}
    [d^{b'}_{\beta'} S_{\beta'\gamma'} u^{c'}_{\gamma'}],
  &
    \bar{N}_{\delta}
  &= \epsilon^{abc} [\bar{d}^c_\gamma\bar{S}_{\gamma\beta}\bar{d}_\beta^b]
    \bar{u}^a_{\delta}, \notag
\end{align}
for $P$ (proton) and $N$ (neutron).
Roman letters $a$, $b$, ... stand for the color index, while the Greek
indices $\alpha$, $\beta$, ... distinguish the spinor components.
$S$ is a diquark spin matrix, which we choose
$S=C\gamma_5=\gamma_1\gamma_3$ and $\bar{S}=-S$.

We define the quark propagator $\Slash{D}^{-1}$ of a flavor $q$ as
$(\Slash{D}_q^{-1})^{ab}_{\alpha\beta}(y,x) =
\langle q^a_\alpha(x) \bar{q}^b_\beta(y)\rangle$. 
The flavor $q$ may either represent up ($u$) or down ($d$) quark.
We call it the \emph{forward} propagator,
because we assume it describes a propagation in
the same direction in time as the nucleon propagator.
In the following, we use a notation
$\mathcal{F}_q(y,x) \equiv (\Slash{D}_q^{-1})^{ab}_{\alpha\beta}(y,x)$.
We will define the \emph{backward} propagator as well, shortly.

The nucleon two-point function is obtained by taking all possible
contractions as 
\begin{align}
  C_{2pt} = T_{\delta \delta'} 
  \langle
  P_{\delta'}\bar{P}_\delta 
  \rangle
  =&
     \epsilon^{a'b'c'}\epsilon^{abc}
     \left[ 
     (\mathcal{F}_u T)^{a'a}_{\alpha'\alpha} (\mathcal{F}_u)^{b'b}_{\alpha\beta} 
     (S \mathcal{F}_d \bar{S})^{c'c}_{\alpha'\beta}
     \right. \notag \\
   & 
     \phantom{\epsilon^{a'b'c'}\epsilon^{abc}}
     \left.
     + (\mathcal{F}_uT)^{a'a}_{\alpha'\alpha'} (\mathcal{F}_u)^{b'b}_{\beta'\beta} 
     (S \mathcal{F}_d \bar{S})^{c'c}_{\beta'\beta}
     \right], 
\end{align}
where $T$ is a projection matrix and the repeated indices are summed.
The projection matrix is set to extract desired nucleon state.
For instance, to sum over the nucleon spin it is set as
$T=\mathrm{diag}(1,1,0,0)$.

To compute the three-point function, our strategy is to build a
sequential propagator and then to contract at the location of the
operator. 
To do so, we define the backward (sink-sequential) propagators
$\mathcal{B}_{u,d}$, 
which depend on the flavor of the valence sequential backward quark
propagator:  
\begin{align}
  \mathcal{B}_u(z; y,x)
  =&
     \mathcal{S}_u[\mathcal{F}_u(y,x), \mathcal{F}_d(y,x)]
     \cdot \Slash{D}^{-1}_u(y,z), \notag \\
  \mathcal{B}_d(z; y,x)
  =& \mathcal{S}_d[\mathcal{F}_u(y,x), \mathcal{F}_u(y,x)]
     \cdot \Slash{D}^{-1}_d(y,z),
\end{align}
where 
$\mathcal{S}_{u,d}$'s are diquark propagators that reflect the
structure of the nucleon interpolating operator at the sink.
They are composed of $u$- and $d$- or two $u$-quark
forward propagators as
\begin{align}
  \mathcal{S}_u[\mathcal{F}_u,\mathcal{F}_d]^{aa'}_{\alpha\alpha'} 
  = &
      \epsilon^{a'b'c'}\epsilon^{abc} 
      \left[ 
      T_{\alpha\alpha'}
      (\mathcal{F}^u)^{b'b}_{\rho'\rho}
      (S\mathcal{F}^d \bar{S})^{c'c}_{\rho'\rho}
      +
      (T\mathcal{F}^u)^{b'b}_{\alpha\rho}
      (S\mathcal{F}^d \bar{S})^{c'c}_{\alpha'\rho}
      \right. \notag \\
    &
      \phantom{\epsilon^{a'b'c'}\epsilon^{abc} }
      \left. \ 
      +
      (\mathcal{F}^uT)^{b'b}_{\rho\alpha'}
      (S\mathcal{F}^d \bar{S})^{c'c}_{\rho\alpha}
      +
      (T\mathcal{F}^u)^{b'b}_{\rho\rho}
      (S\mathcal{F}^d \bar{S})^{c'c}_{\alpha'\alpha}
      \right],  \\
  \mathcal{S}_d[\mathcal{F}_{u_1},\mathcal{F}_{u_2}]^{aa'}_{\alpha\alpha'} 
  = &
      \epsilon^{a'b'c'}\epsilon^{abc} 
      \left[ 
      (\mathcal{F}_{u_1}T)^{c'c}_{\rho\rho}
      (S^T \mathcal{F}_{u_2} \bar{S}^T)^{b'b}_{\alpha'\alpha}
      + 
      (S^T \mathcal{F}_{u_1}T)^{c'c}_{\alpha'\rho}
      (\mathcal{F}_{u_2} \bar{S}^T)^{b'b}_{\rho\alpha}
      \right]. 
\end{align}
The connected three-point functions with a neutral current insertion
is then constructed as  
\begin{align}
  C_{3pt}^{\bar{u}\Gamma u}(z, y, x)
  =& 
     T_{\delta \delta'} 
     \langle 
     P_{\delta'}(y) |\bar{u}\Gamma u(z) |\bar{P}_\delta(x)
     \rangle=
     {\rm Tr}\left[ 
     \mathcal{B}_u(z; y, x) \Gamma \mathcal{F}_u(z,x)
     \right],
     \notag \\
  C_{3pt}^{\bar{d}\Gamma d}(z, y, x)
  =& 
     T_{\delta \delta'} 
     \langle 
     P_{\delta'}(y) |\bar{d}\Gamma d(z) |\bar{P}_\delta(x)
     \rangle=
     {\rm Tr}\left[ 
     \mathcal{B}_d(z; y, x) \Gamma \mathcal{F}_d(z,x)
     \right],
\end{align}
where the current operator has a $\gamma$-matrix structure $\Gamma$.

\subsection{Four-point functions}
Here we describe a scheme for the computation of the general 
four-point nucleon correlation functions for both flavor-diagonal (or
neutral) and flavor-changing (or charged) currents. 
We consider the following four-point functions, 
\begin{align}
  C_{4pt}^{J_+^{(1)} J_-^{(2)}}(z_1,z_2, y, x) 
  =&
     T_{\delta\delta'} \langle P_{\delta'}(y)| 
     J_+^{(1)}(z_1) J_-^{(2)}(z_2)
     |\bar{P}_\delta(x) \rangle,
     \notag \\
  C_{4pt}^{J_-^{(1)} J_+^{(2)}}(z_1,z_2, y, x) 
  =&
     T_{\delta\delta'} \langle P_{\delta'}(y)| 
     J_-^{(1)}(z_1) J_+^{(2)}(z_2)
     |\bar{P}_\delta(x) \rangle,
     \notag \\
  C_{4pt}^{J_+^{(1)} J_u^{(2)}}(z_1,z_2, y, x) 
  =&
     T_{\delta\delta'} \langle N_{\delta'}(y)| 
     J_+^{(1)}(z_1) J_u^{(2)}(z_2)
     |\bar{P}_\delta(x) \rangle,
     \notag \\
  C_{4pt}^{J_+^{(1)} J_d^{(2)}}(z_1,z_2, y, x) 
  =&
     T_{\delta\delta'} \langle N_{\delta'}(y)| 
     J_+^{(1)}(z_1) J_d^{(2)}(z_2)
     |\bar{P}_\delta(x) \rangle,
     \notag \\
  C_{4pt}^{J_u^{(1)} J_+^{(2)}}(z_1,z_2, y, x) 
  =&
     T_{\delta\delta'} \langle N_{\delta'}(y)| 
     J_u^{(1)}(z_1) J_+^{(2)}(z_2)
     |\bar{P}_\delta(x) \rangle,
     \notag \\
  C_{4pt}^{J_d^{(1)} J_+^{(2)}}(z_1,z_2, y, x) 
  =&
     T_{\delta\delta'} \langle N_{\delta'}(y)| 
     J_d^{(1)}(z_1) J_+^{(2)}(z_2)
     |\bar{P}_\delta(x) \rangle,
     \notag \\
  C_{4pt}^{J_u^{(1)} J_u^{(2)}}(z_1,z_2, y, x) 
  =& 
     T_{\delta\delta'} \langle P_{\delta'}(y)| 
     J_u^{(1)}(z_1) J_u^{(2)}(z_2)
     |\bar{P}_\delta(x) \rangle,
     \notag \\
  C_{4pt}^{J_d^{(1)} J_d^{(2)}}(z_1,z_2, y, x) 
  =& 
     T_{\delta\delta'} \langle P_{\delta'}(y)| 
     J_d^{(1)}(z_1) J_d^{(2)}(z_2)
     |\bar{P}_\delta(x) \rangle,
     \notag \\
  C_{4pt}^{J_u^{(1)} J_d^{(2)}}(z_1,z_2, y, x) 
  =& 
     T_{\delta\delta'} \langle P_{\delta'}(y)| 
     J_u^{(1)}(z_1) J_d^{(2)}(z_2)
     |\bar{P}_\delta(x) \rangle,
     \notag \\
  C_{4pt}^{J_d^{(1)} J_u^{(2)}}(z_1,z_2, y, x) 
  =& 
     T_{\delta\delta'} \langle P_{\delta'}(y)| 
     J_d^{(1)}(z_1) J_u^{(2)}(z_2)
     |\bar{P}_\delta(x) \rangle,
     \notag 
     \label{eq:4pt}
\end{align}
where we introduce the neutral and charged currents,
\begin{align}
  J_u^{(i)} = \bar{u}\Gamma^{(i)} u, \quad
  & J_d^{(i)} = \bar{d}\Gamma^{(i)} d, \notag \\
  J_+^{(i)} = \bar{d}\Gamma^{(i)} u, \quad
  & J_-^{(i)} = \bar{u}\Gamma^{(i)} d.
\end{align}
The superscript $(i)$ distinguishes the two operators inserted
($i=1$ or 2). 

\subsubsection{Neutral current}
\begin{figure}[tbp]
  \centering
  \includegraphics[clip,height=4cm]{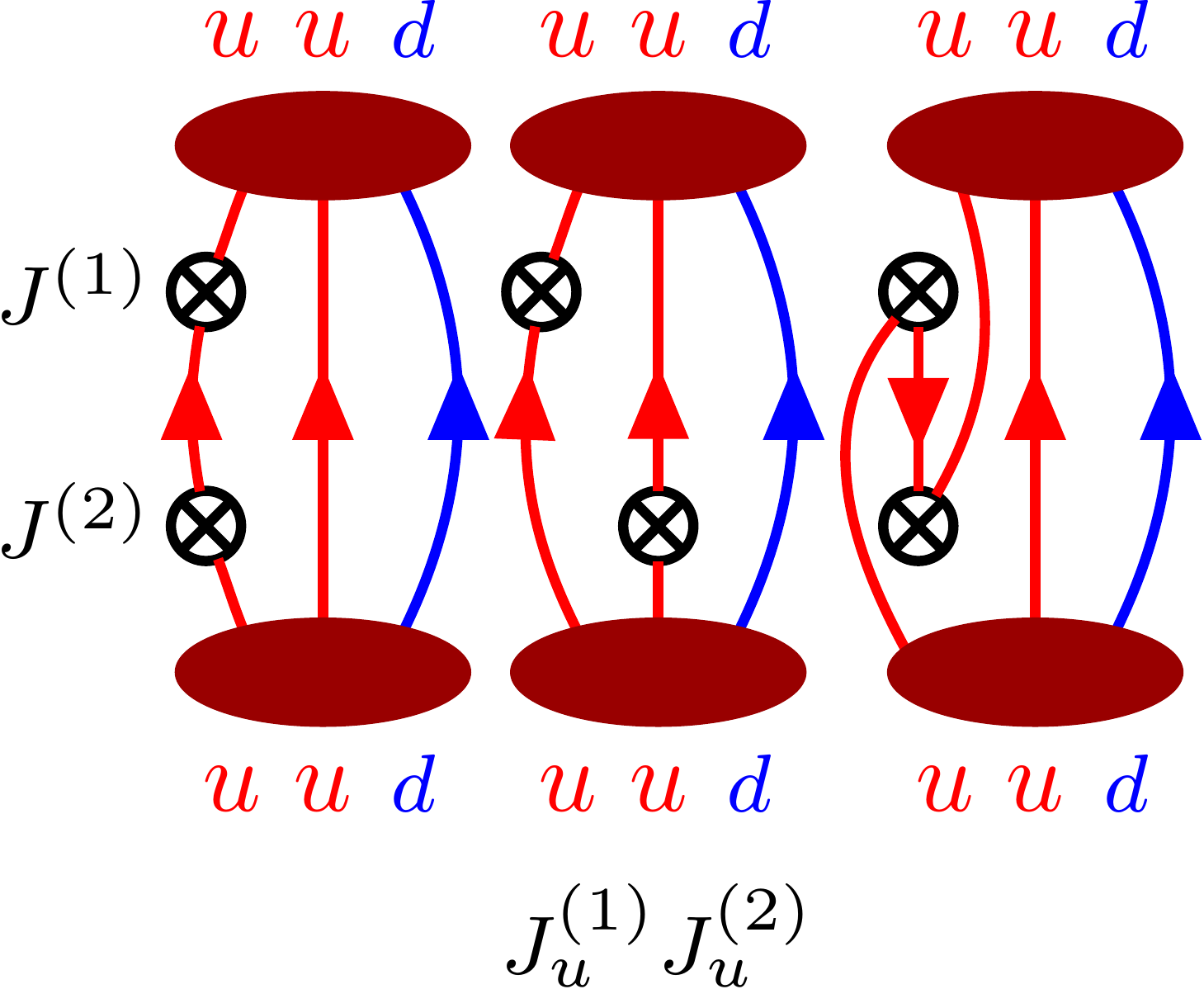}  \quad \quad \quad
  \includegraphics[clip,height=4cm]{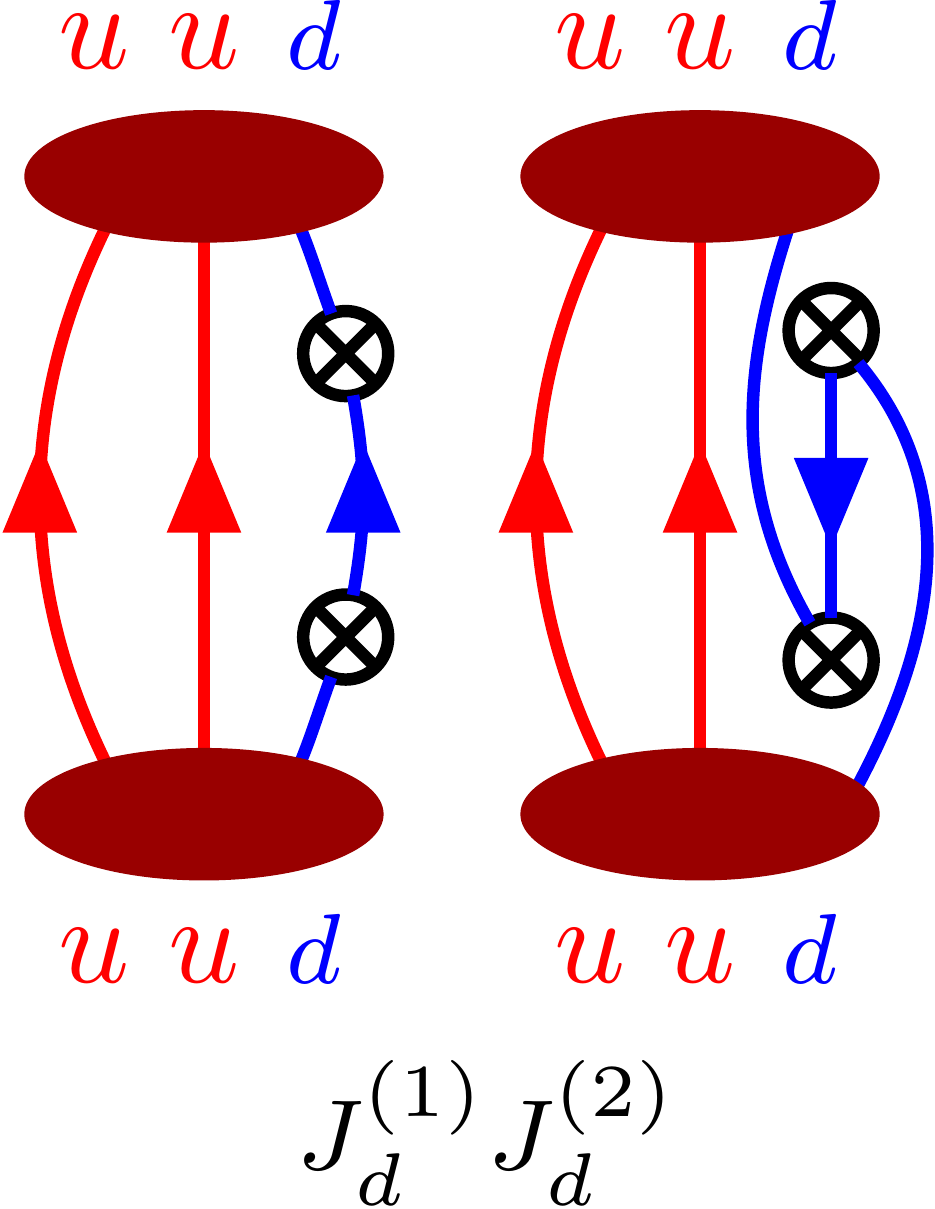}  \quad \quad \quad 
  \includegraphics[clip,height=4cm]{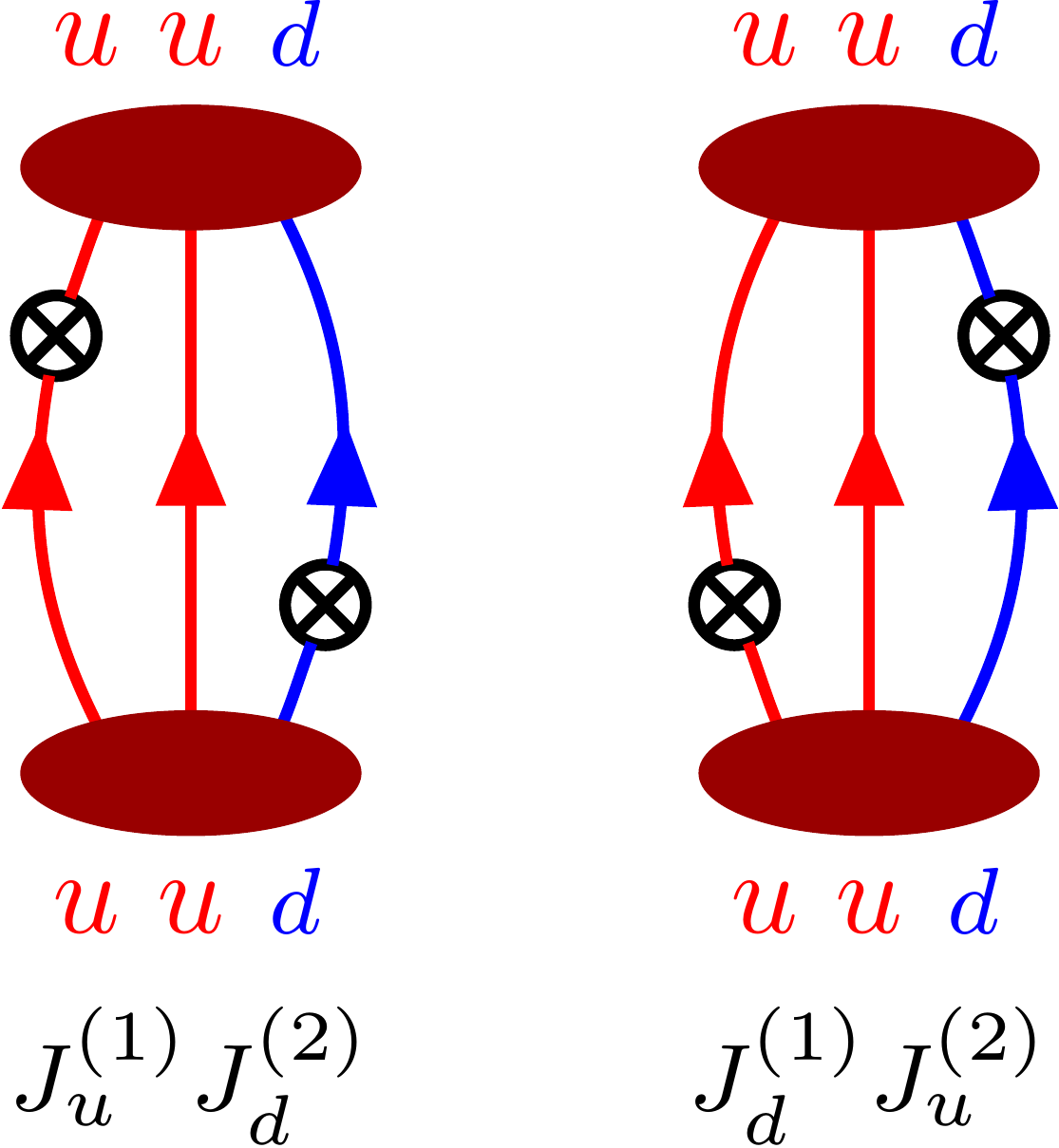} 
  \caption{\label{fig:UUDD}
    Connected diagrams for 
    the nucleon 4 point functions with 
    $J_{u}^{(1)}J_{u}^{(2)}$, $J_{d}^{(1)}J_{d}^{(2)}$, 
    $J_{u}^{(1)}J_{d}^{(2)}$, 
    and 
    $J_{d}^{(1)}J_{u}^{(2)}$ currents inserted.
  }  
\end{figure}

For the neutral currents the current insertions and the corresponding
quark lines from the source to sink nucleon operators are shown in
Fig.~\ref{fig:UUDD}. 

\begin{figure}[tbp]
  \centering
  \includegraphics[clip,height=3.5cm]{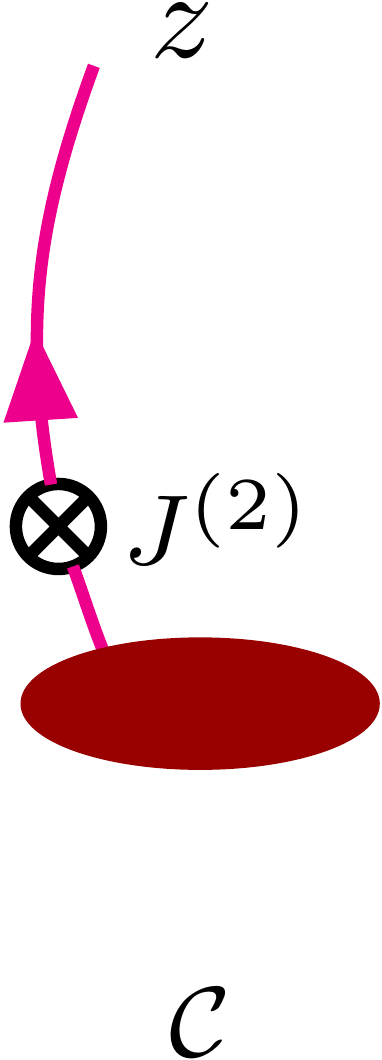}  \quad \quad \quad
  \includegraphics[clip,height=3.5cm]{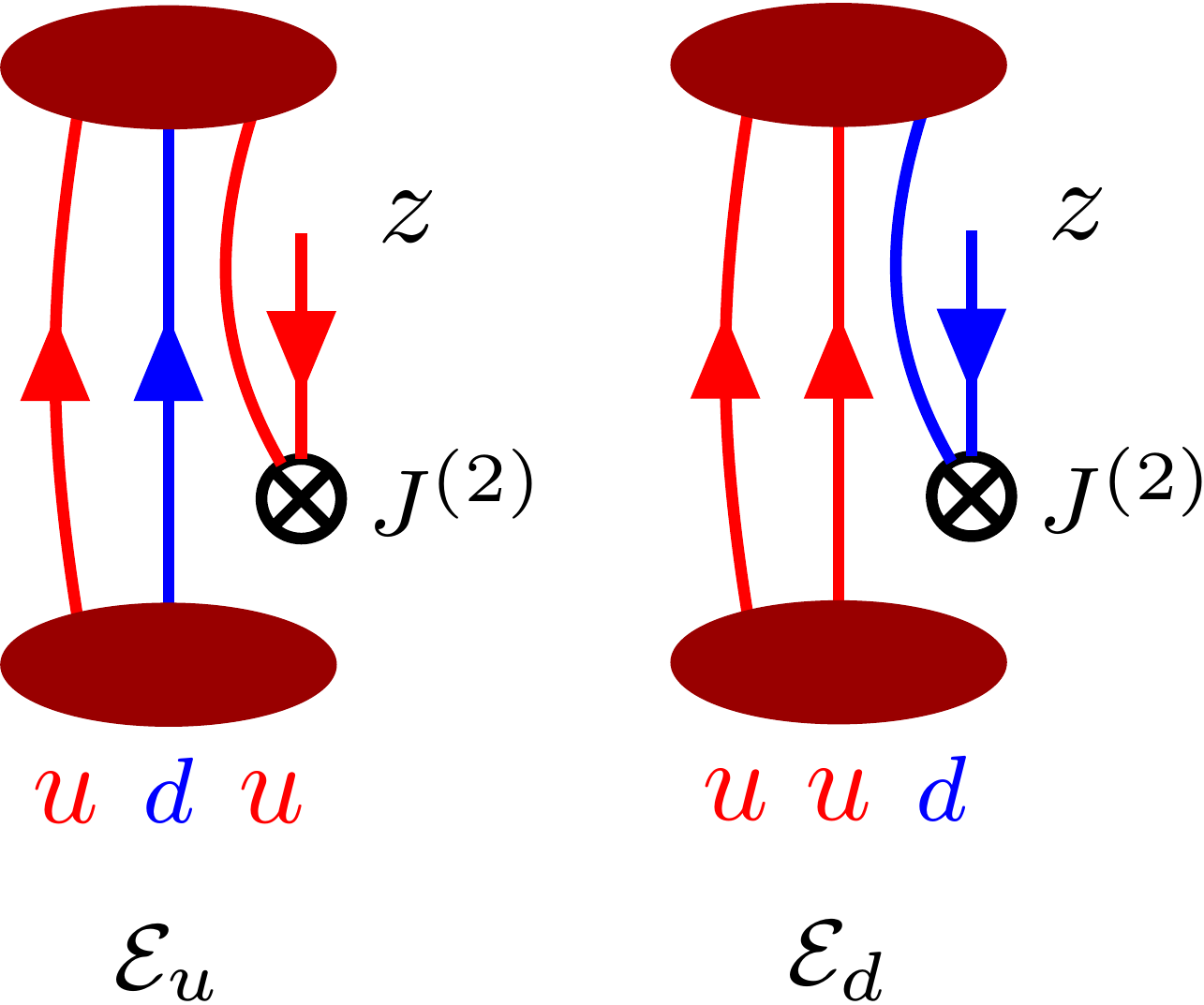}  \quad \quad
  \includegraphics[clip,height=3.5cm]{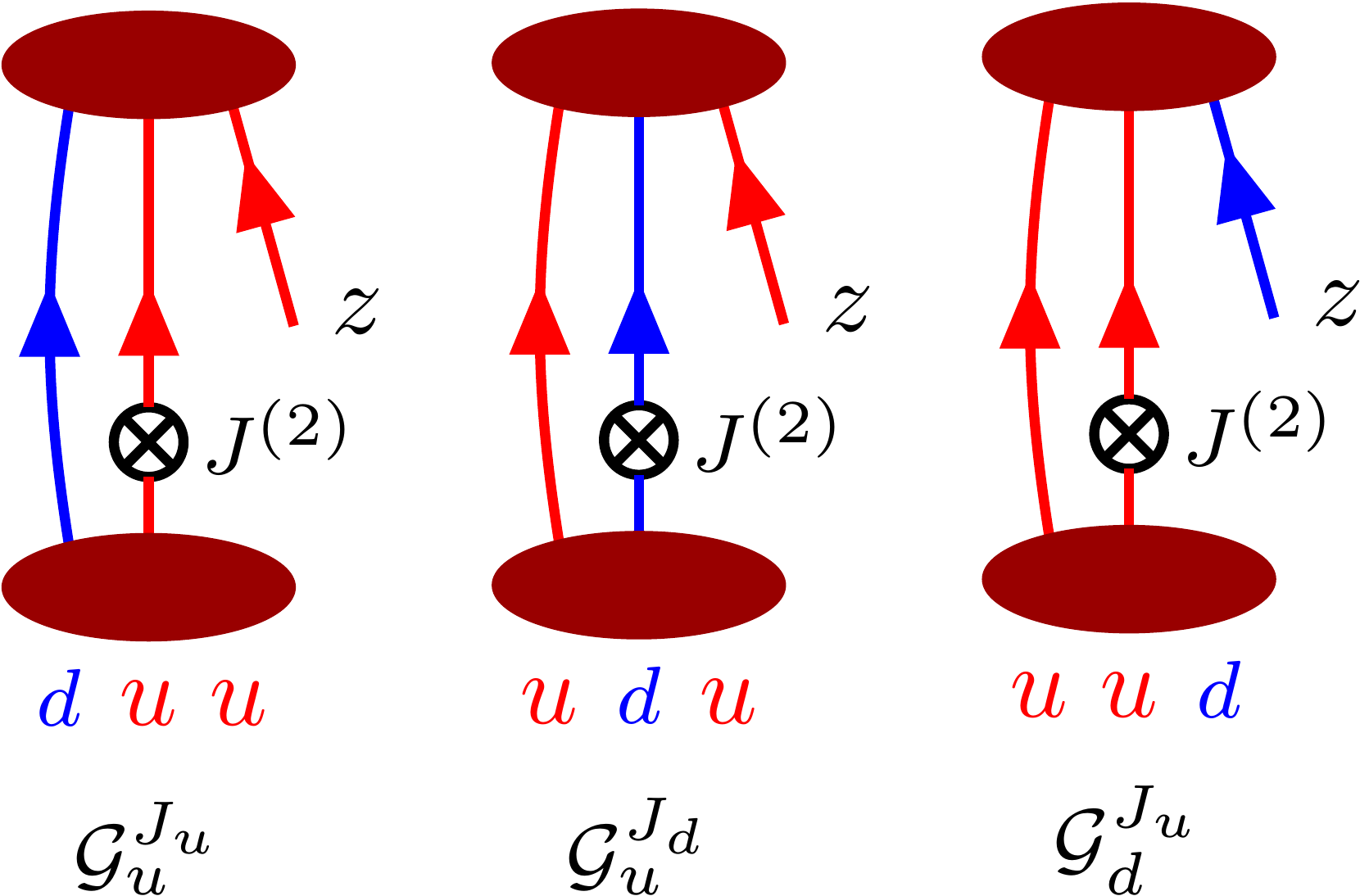}  
  \caption{\label{fig:CEG}
    Propagators for computing the connected nucleon four-point functions.}  
\end{figure}

In order to contract at the location of the current $J^{(1)}$, $z_1$,
we need to compute three different current sequential propagators. 
One is the current-sequential propagator $(\mathcal{C})$: 
\begin{align}
  \mathcal{C}_q(z, z_2, x)
  =& \Slash{D}_q^{-1}(z, z_2)\cdot \Gamma^{(2)}
     \cdot \mathcal{F}_q(z_2,x),  \notag \\
  \mathcal{C}_q(z, \bm{q}_2, x)
  =& \sum_{\bm{z}_2}\Slash{D}_q^{-1}(z, z_2)
     \cdot
     e^{i\bm{q}_2\cdot(\bm{z}_2-\bm{x})} \Gamma^{(2)} \cdot \mathcal{F}_q(z_2,x),  
\end{align}
and two others are the current-sink sequential and sink-current
sequential propagators called $\mathcal{G}$ and $\mathcal{E}$, 
which were introduced in \cite{Abramczyk:2017oxr}, 
where we should replace the CP-violating operator with the current operator $J^{(2)}$ 
for our purpose (See Fig.~\ref{fig:CEG}).  
Then, the sink-current sequential propagators $\mathcal{E}$ are defined as
\begin{align}
  \mathcal{E}_q(z; z_2, y, x)
  =& 
     \mathcal{B}_q(z_2; y, x) \cdot \Gamma^{(2)} \cdot \Slash{D}^{-1}_q(z_2, z).
\end{align}
%in the configuration space, and 
%\begin{align}
%  \mathcal{E}_q(z; \bm{p}, \bm{q}_2, x)
%  =& 
%     \sum_{\bm{z}_2} \mathcal{B}_q(\bm{z}_2; \bm{q}_2, x)
%     e^{i\bm{q}_2(\bm{z}_2-\bm{x})}\cdot \Gamma^{(2)} \cdot \Slash{D}^{-1}_q(z_2, z),
%\end{align}
%with a momentum projection.
The current-sink sequential propagators $\mathcal{G}$ can be expressed 
using the diquark propagators $\mathcal{S}$ as
%function for the sink-sequential $u$ and $d$ source
%operators  as  
\begin{align}
\mathcal{G}_u^{J_u}(z; y, x) =& 
\mathcal{S}_u[\mathcal{C}_u(y, z_2, x), \mathcal{F}_d(y,x)] \cdot \Slash{D}^{-1}_u(y, z), \notag \\
\mathcal{G}_u^{J_d}(z; y, x) =& 
\mathcal{S}_u[\mathcal{F}_u(y,x), \mathcal{C}_d(y, z_2, x)] \cdot \Slash{D}^{-1}_u(y, z), \notag \\
\mathcal{G}_d^{J_u}(z; y, x) =& 
(\mathcal{S}_d[\mathcal{C}_u(y, z_2, x), \mathcal{F}_u(y,x)] + \mathcal{S}_d[\mathcal{F}_u(y,x), \mathcal{C}_u(y, z_2, x)] )\cdot \Slash{D}^{-1}_d(y, z).  
%\notag \\
%\mathcal{G}_d^{J_d}(z; y, x) =& 0.
\end{align}
%in the configuration space, and 
%\begin{align}
%\mathcal{G}_u^{J_u}(z; \bm{p}, x) =& 
%\sum_{\bm{y}}  e^{-i\bm{p}\cdot(\bm{y}-\bm{x})} 
%\mathcal{Y}_u[\mathcal{C}_u(y, z_2, x), \mathcal{F}_d(y,x)] \cdot \Slash{D}^{-1}_u(y, z) \notag \\
%\mathcal{G}_u^{J_d}(z; \bm{p}, x) =& 
%\sum_{\bm{y}}  e^{-i\bm{p}\cdot(\bm{y}-\bm{x})} 
%\mathcal{Y}_u[\mathcal{F}_u(y,x), \mathcal{C}_d(y, z_2, x)] \cdot \Slash{D}^{-1}_u(y, z) \notag \\
%\mathcal{G}_d^{J_u}(z; \bm{p}, x) =& 
%\sum_{\bm{y}}  e^{-i\bm{p}\cdot(\bm{y}-\bm{x})} 
%(\mathcal{Y}_d[\mathcal{C}_u(y, z_2, x), \mathcal{F}_u(y,x)] + \mathcal{Y}_d[\mathcal{F}_u(y,x), \mathcal{C}_u(y, z_2, x)] )\cdot \Slash{D}^{-1}_d(y, z), \notag \\
%\mathcal{G}_d^{J_d}(z; \vec{p}, x) =& 0, 
%\end{align}
%with a momentum projection.
In what follows, we omit the arguments ($x$, $\bm{p}$, ...) in the
propagators for simplicity. 
Then, the connected four-point functions for the neutral currents are
given using $\mathcal{F}$, $\mathcal{C}$, $\mathcal{E}$, and $\mathcal{G}$ as 
\begin{align}
C_{4pt-conn}^{J_u^{(1)} J_u^{(2)}}  =&
{\rm Tr} \left[\mathcal{B}_u \cdot \Gamma^{(1)} \cdot \mathcal{C}_u\right]
+{\rm Tr} \left[(\mathcal{E}_u + \mathcal{G}_u^{J_u})\cdot \Gamma^{(1)} \cdot \mathcal{F}_u \right]
\notag \\ 
C_{4pt-conn}^{J_d^{(1)} J_d^{(2)}}  =&
{\rm Tr} \left[\mathcal{B}_d \cdot \Gamma^{(1)} \cdot \mathcal{C}_d \right]
+{\rm Tr} \left[\mathcal{E}_d \cdot \Gamma^{(1)} \cdot \mathcal{F}_d \right], 
\notag \\ 
C_{4pt-conn}^{J_u^{(1)} J_d^{(2)}}  =&
{\rm Tr} \left[\mathcal{G}_u^{J_d} \cdot \Gamma^{(1)} \cdot \mathcal{F}_u \right], 
\notag \\ 
C_{4pt-conn}^{J_d^{(1)} J_u^{(2)}}  =&
{\rm Tr} \left[\mathcal{G}_d^{J_u} \cdot \Gamma^{(1)} \cdot \mathcal{F}_d \right], 
\end{align}
where ${\rm Tr}$ denotes a trace over color, spinor indices as well as
the location of the current $z=z_1$, which is limited on the
time-slice $t_1$. 
The first term in either case corresponds to the diagram given on the
left in Fig.~\ref{fig:UUDD}.
The $\mathcal{G}$-type contribution does not appear 
for $J_d J_d$.

\subsubsection{Charged current}
\begin{figure}[tbp]
  \centering
  \includegraphics[clip,height=4cm]{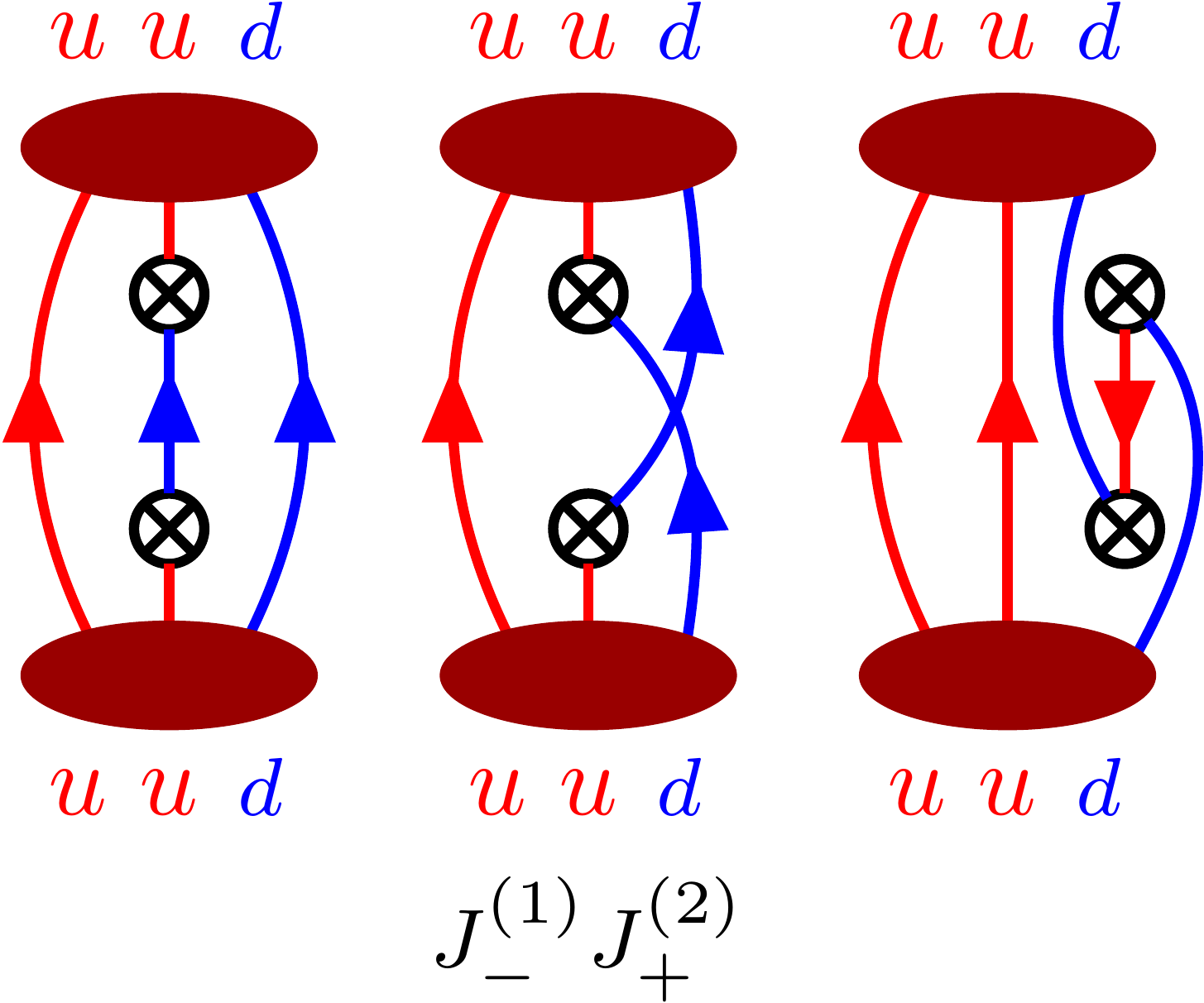}  \quad \quad
  \includegraphics[clip,height=4cm]{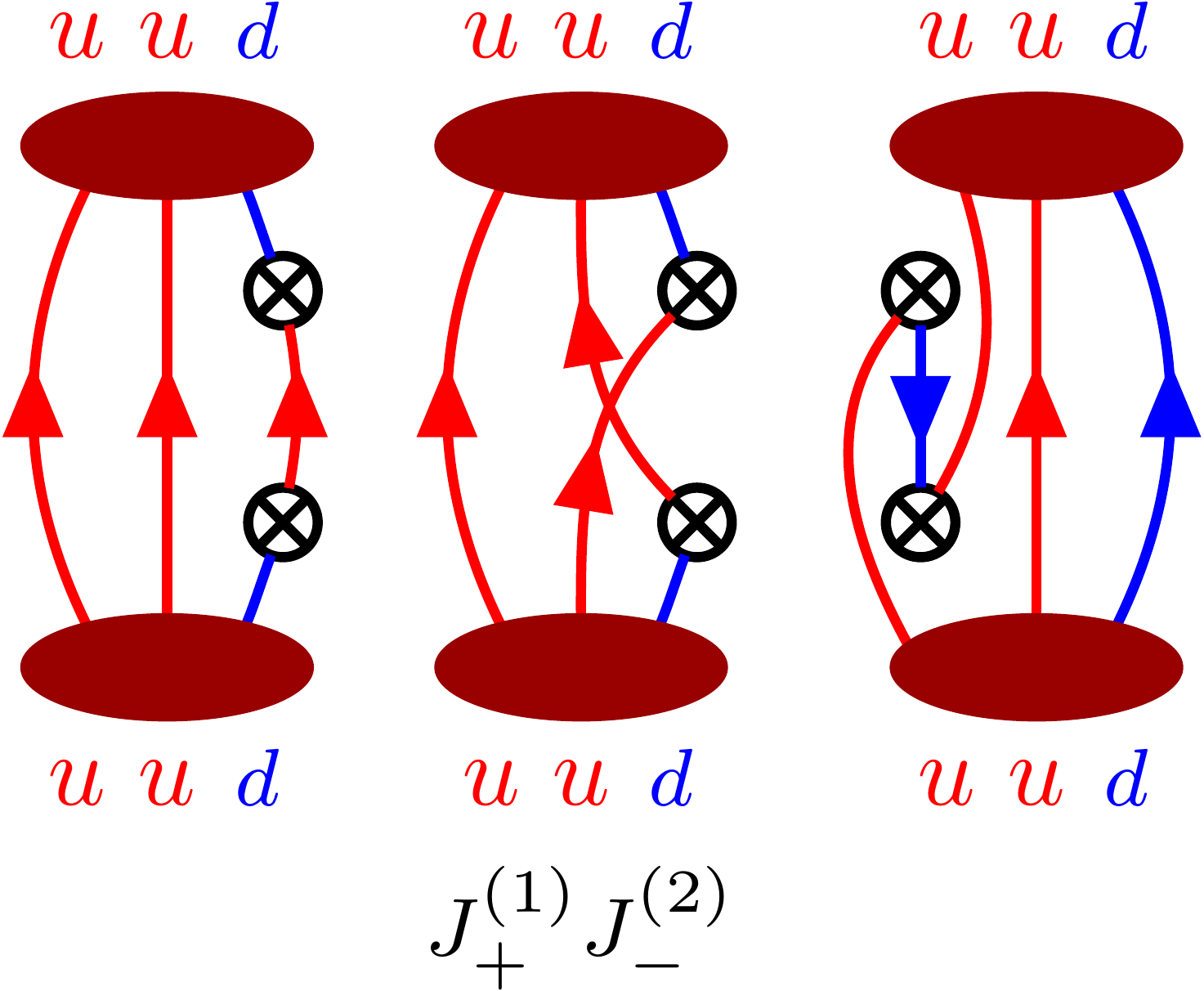}  \quad \quad
  \caption{\label{fig:MPPM}
    Connected diagrams for 
    the nucleon four-point functions with 
    $J_{-}^{(1)}J_{+}^{(2)}$ (left) and
    $J_{+}^{(1)}J_{-}^{(2)}$ (right) currents inserted.
  }  
\end{figure}

\begin{figure}[tbp]
  \centering
  \includegraphics[clip,height=3.3cm]{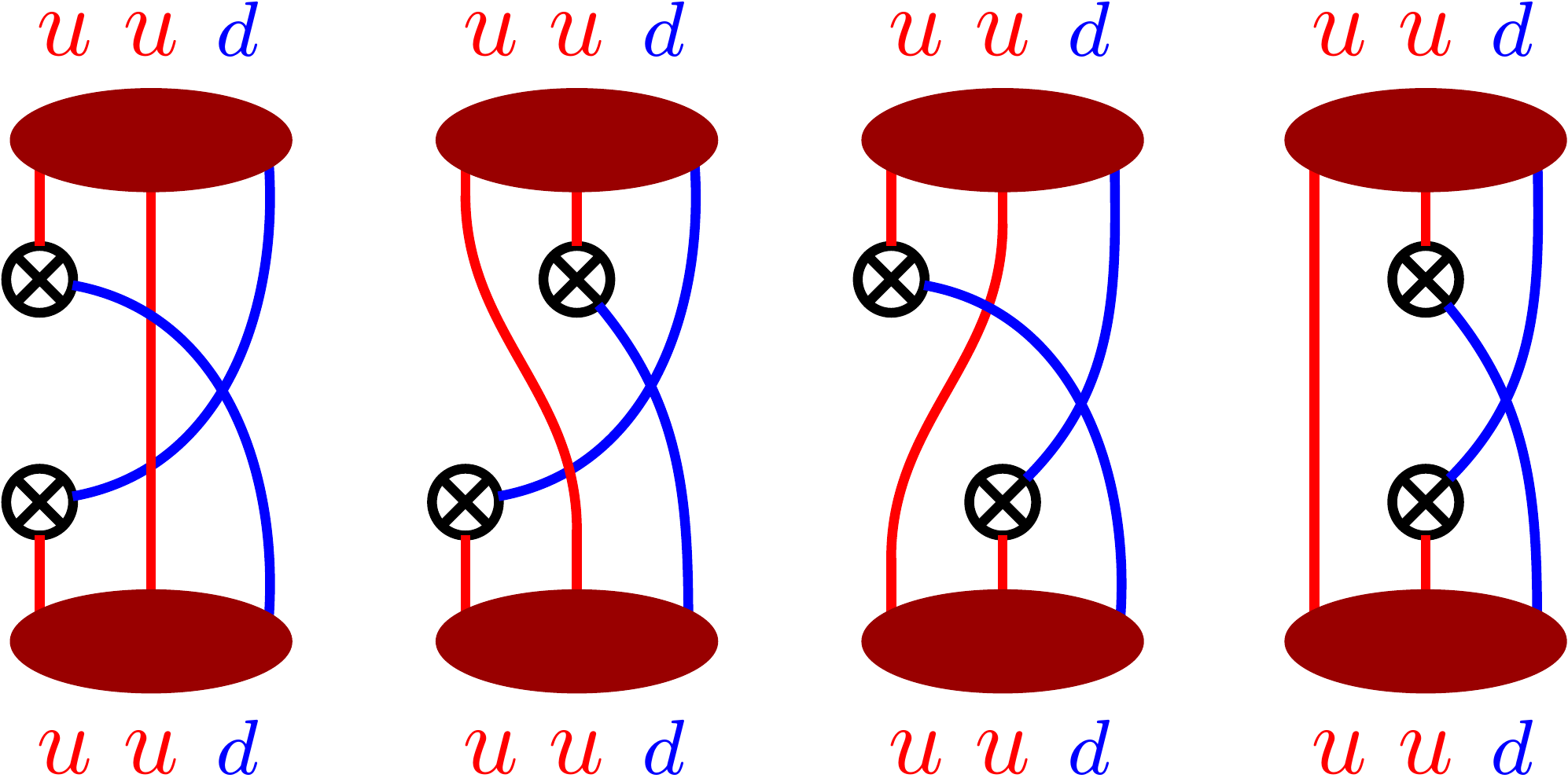} 
  \quad \quad \quad \quad 
  \includegraphics[clip,height=3.3cm]{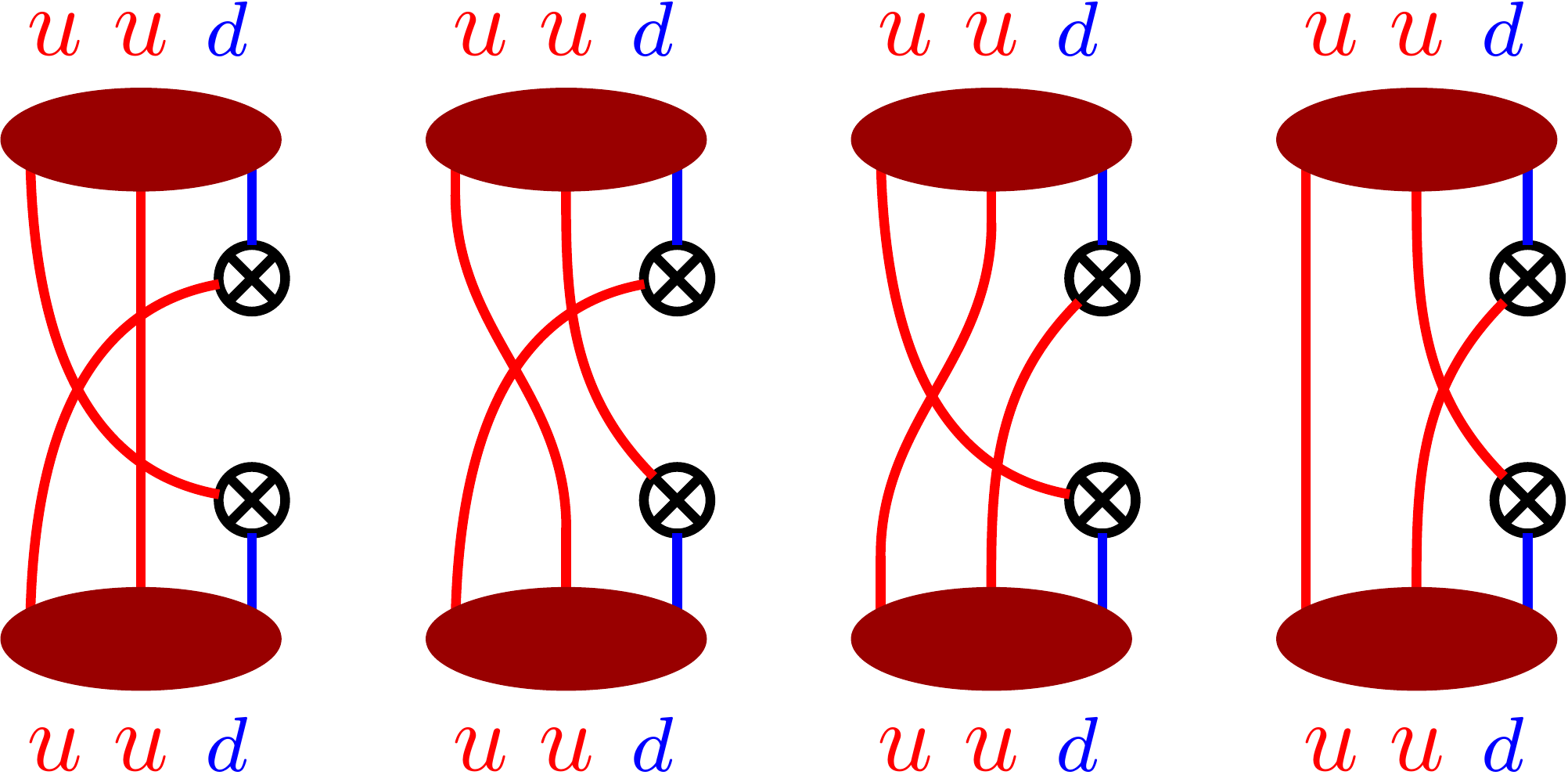} 
  \caption{\label{fig:cross}
    Crossing diagrams for the flavor changing current 
    for $J_-^{(1)}J_+^{(2)}$ (left) and
    for $J_+^{(1)} J_-^{(2)}$ (right).
  }  
\end{figure}

The quark-line contractions are more complicated 
for the charged current, for which the quark flavor changes.
All the diagrams for the charged current insertions
$J_{-}^{(1)}J_{+}^{(2)}$ and $J_{+}^{(1)}J_{-}^{(2)}$ 
are shown in Fig.~\ref{fig:MPPM}. 
The first and third diagrams in both cases resemble 
the first and third ones given in the $J_u J_u$ current
in Fig.~\ref{fig:UUDD}, 
where we can use the backward sequential propagators  
$\mathcal{B}_{q}$ and $\mathcal{E}_q$ for these diagrams.
On the other hand, for the second diagram the Wick contractions of 
the sink and source quarks are different from that of 
the $J_u J_u$ or $J_d J_d$ current (See Fig.~\ref{fig:cross}), 
so that the backward sequential propagators $\mathcal{G}$ defined
above can not be used.

To evaluate these crossing diagrams, 
it is convenient to define generalized nucleon two-point functions:
$N_{ijk}[\mathcal{F}_{q_1},\mathcal{F}_{q_2},\mathcal{F}_{q_3}] \equiv
{\rm Tr} [T\langle N_{ijk} \bar{N}_{123} \rangle]$,
$N_{ijk} \equiv
\epsilon^{a'b'c'} q_i^{a'} [q_j^{b'} S q_k^{c'}]$,
$\bar{N}_{ijk} \equiv
\epsilon^{abc} [\bar{q}_k^c \bar{S} \bar{q}_j^b ] \bar{q}^a_i$,
where we introduce fictitious valence quarks with three different
flavors $q_{i}$, ($i$, $j$, $k$ = 1 , 2, or 3).
Therefore it is understood that the Wick contraction for the function
$N_{ijk}[\mathcal{F}_{q_1},\mathcal{F}_{q_2},\mathcal{F}_{q_3}]$
is taken only for a pair of the same fictitious valence quark flavors 
$(q_i \bar{q}_i)$. 
An explicit example is like this:
\begin{align}
N_{321}[\mathcal{F}_{q_1},\mathcal{F}_{q_2},\mathcal{F}_{q_3}] =& 
T_{\delta \delta'} 
\epsilon^{a'b'c'}
\epsilon^{abc}
\langle
(q_3)_{\delta'}^{a'} [(q_2)_{\beta'}^{b'} S_{\beta'\gamma'} (q_1)_{\gamma'}^{c'}]
[(\bar{q}_3)_\gamma^{c} \bar{S}_{\gamma\beta} (\bar{q}_2)_{\beta}^{b}]  (\bar{q}_1)_{\delta}^a
\rangle \notag \\
=&
\epsilon^{a'b'c'}
\epsilon^{abc}
(S \mathcal{F}_{q_1} T)^{a'a}_{\alpha\beta} 
(\mathcal{F}_{q_2} \bar{S}^T)^{b'b}_{\alpha\gamma}
(\mathcal{F}_{q_3})^{c'c}_{\beta\gamma}.
\end{align}
Using this we can construct a generalized diquark propagator %backward sequential source operator 
for the crossing diagrams. 

Let us first focus on the case of $J_-^{(1)}J_+^{(2)}$ insertions. 
As shown in Fig.~\ref{fig:cross}, 
when the final contraction is taken at $J^{(1)}$, 
the forward propagator is always given by $\mathcal{F}_d$. 
Using this, we construct the backward sequential propagator
$\mathcal{H}^{J_-J_+}$  for the crossing diagrams,
\begin{align}
  \mathcal{H}^{J_-J_+}_u(z; y, x) =
  \mathcal{S}^{J_-J_+}[\mathcal{F}_u(y, x), \mathcal{C}_+(y, z_2, x)] \cdot \Slash{D}^{-1}_u(y,z), 
\end{align}
such that 
\begin{align}
  C_{4pt-cross}^{J_-^{(1)}J_+^{(2)}} = 
  {\rm Tr} [\mathcal{H}^{J_-J_+}_u \cdot \Gamma^{(1)} \cdot \mathcal{F}_{d}],
\end{align}
where $\mathcal{S}^{J_-J_+}[\mathcal{F}_1, \mathcal{F}_2]$ 
is a diquark propagator for crossing diagrams,   % backward current-sink sequential source operator, 
and the concrete form will be given below. 
$\mathcal{C}_+$ is a current-sequential forward propagator defined as 
\begin{align}
  \mathcal{C}_+(z, z_2, x) =
%  \sum_{z_2} 
  \Slash{D}_d^{-1}(z, z_2) \cdot \Gamma^{(2)} \cdot \mathcal{F}_u(z_2, x) . 
\end{align}
Obviously, $\mathcal{C}_+=\mathcal{C}_{u,d}$ in the isospin symmetric
limit.  
%As in the flavor diagonal case, 
We note that such diquark propagator for crossing diagrams can be
obtained by taking the functional derivative of 
the generalized nucleon two-point functions
($N_{ijk}[\mathcal{F}_{q_1},\mathcal{F}_{q_2},\mathcal{F}_{q_3}]$).  
For example, the diquark propagator corresponding %backward sequential source operator 
to the first diagram in Fig.~\ref{fig:cross} 
is given as
$\mathcal{S}_{321}^{q_3}[\mathcal{F}_{q_1}=\mathcal{C}_+, \mathcal{F}_{q_2}=\mathcal{F}_{u}]$, 
where the function $\mathcal{S}^{q_l}_{ijk}[\mathcal{F}_{q_{l'}}, \mathcal{F}_{q_{l''}}]$
is defined as 
\begin{align}
  \mathcal{S}^{q_1}_{ijk}[\mathcal{F}_{q_2}, \mathcal{F}_{q_3}]_{\alpha\alpha'}^{aa'}
  =& 
     \left\{ 
     T_{\delta \delta'} 
     (N_{ijk})_{\delta'}
     \left( \frac{\overleftarrow{\delta}}{\delta (q_1)_{\alpha'}^{a'}} 
     \frac{\overrightarrow{\delta}}{\delta (\bar{q}_1)_{\alpha}^a}\right) (\bar{N}_{123})_\delta
     \right\}, 
     \notag \\
  \mathcal{S}^{q_2}_{ijk}[\mathcal{F}_{q_1}, \mathcal{F}_{q_3}]_{\alpha\alpha'}^{aa'}
  =& 
     \left\{ 
     T_{\delta \delta'} 
     (N_{ijk})_{\delta'}
     \left( \frac{\overleftarrow{\delta}}{\delta (q_2)_{\alpha'}^{a'}} 
     \frac{\overrightarrow{\delta}}{\delta (\bar{q}_2)_{\alpha}^a}\right) (\bar{N}_{123})_\delta
     \right\},
     \notag \\
  \mathcal{S}^{q_3}_{ijk}[\mathcal{F}_{q_1}, \mathcal{F}_{q_2}]_{\alpha\alpha'}^{aa'}
  =& 
     \left\{ 
     T_{\delta \delta'} 
     (N_{ijk})_{\delta'}
     \left( \frac{\overleftarrow{\delta}}{\delta (q_3)_{\alpha'}^{a'}} 
     \frac{\overrightarrow{\delta}}{\delta (\bar{q}_3)_{\alpha}^a}\right) (\bar{N}_{123})_\delta
     \right\}.
\end{align}
For example,
$\mathcal{S}^{q_3}_{132}[\mathcal{F}_{q_1},\mathcal{F}_{q_2}]$
is 
\begin{align}
  \mathcal{S}^{q_3}_{132}[\mathcal{F}_{q_1}, \mathcal{F}_{q_2}]_{\alpha\alpha'}^{aa'}
  =& 
     \epsilon^{a'b'c'}
     \epsilon^{abc}
     \bar{S}_{\alpha\beta} (\mathcal{F}_{q_1}T)^{b'b}_{\rho\rho} (S \mathcal{F}_{q_2})^{c'c}_{\alpha'\beta}.\end{align}
The full contribution of the diquark propagator $\mathcal{S}^{J_-J_+}$ is then obtained by the sum of four diagrams,  
\begin{align}
  (\mathcal{S}^{J_-J_+}[\mathcal{F}_u, \mathcal{C}_+])^{aa'}_{\alpha\alpha'}
  =& 
     \left(\mathcal{S}^{q_3}_{132}[\mathcal{F}_{u}, \mathcal{C}_+] 
     +
     \mathcal{S}^{q_3}_{231}[\mathcal{C}_+, \mathcal{F}_{u}]
     +
     \mathcal{S}^{q_3}_{312}[\mathcal{F}_{u}, \mathcal{C}_+]
     +
     \mathcal{S}^{q_3}_{321}[\mathcal{C}_+, \mathcal{F}_{u}] \right)^{aa'}_{\alpha\alpha'} \notag \\
  =&
     \epsilon^{a'b'c'}
     \epsilon^{abc}
     \bar{S}_{\alpha\beta} \left( 
     (\mathcal{F}_u T)^{b'b}_{\rho\rho} (S \mathcal{C}_+)^{c'c}_{\alpha'\beta} 
     +
     (\mathcal{F}_u)^{c'c}_{\rho\beta} (S\mathcal{C}_+ T)_{\alpha'\rho}^{b'b} 
     \right.  \notag \\
   &
     \phantom{\epsilon^{a'b'c'}
     \epsilon^{abc}
     \bar{S}_{\alpha\beta}}
     \left. + 
     (S \mathcal{C}_+)^{c'c}_{\rho\beta} (\mathcal{F}_u T)^{b'b}_{\rho\alpha'}
     + (S \mathcal{C}_+ T)^{b'b}_{\rho\alpha'}  (\mathcal{F}_u)^{c'c}_{\rho\beta}
     \right).
\end{align}
For the other two diagrams, 
they are evaluated using the backward propagators $\mathcal{B}_u$ and
$\mathcal{E}$, since the diagrammatic structure is the same as the
flavor-diagonal case. 
However the quark flavors in the third 
($\mathcal{E}$-type) diagram in Fig.~\ref{fig:MPPM} should be
different from the original $\mathcal{E}$-type propagators,  
since the quark flavor for backward propagators will change in the
case of the charged current.
For this purpose we introduce generalized functions
$\mathcal{E}_{q_1 q_2}$ defined as 
\begin{align}
  \mathcal{E}_{q_1 q_2}(z; y, x)
  =& 
     \mathcal{B}_{q_1}(z_2; y, x) \cdot \Gamma^{(2)} \cdot \Slash{D}^{-1}_{q_2}(z_2, z).
\end{align}
Then the first diagram in Fig.~\ref{fig:MPPM} is 
${\rm Tr} [\mathcal{B}_{u} \cdot \Gamma^{(1)} \cdot \mathcal{C}_{+}]$, 
and the third diagram is 
${\rm Tr} [\mathcal{E}_{du} \cdot \Gamma^{(1)} \cdot \mathcal{F}_{d}]$. 

Next, we consider the ${J_+^{(1)}J_-^{(2)}}$ insertions. 
A similar argument to the previous analysis for ${J_-^{(1)}J_+^{(2)}}$ can apply.
As shown on the right-hand side of Fig.~\ref{fig:cross}, 
there are four crossing diagrams like those of 
${J_-^{(1)}J_+^{(2)}}$. 
Another diquark propagator for crossing diagrams %The backward current-sink sequential source function
$\mathcal{S}^{J_+J_-}$ %for crossing diagrams  
is given as 
\begin{align}
(\mathcal{S}^{J_+J_-}[\mathcal{F}_u, \mathcal{C}_-])_{\alpha \alpha'}^{aa'}=& 
\left(\mathcal{S}^{q_1}_{321}[\mathcal{F}_u, \mathcal{C}_-]
+
\mathcal{S}^{q_1}_{231}[\mathcal{F}_u, \mathcal{C}_-]
+
\mathcal{S}^{q_2}_{312}[\mathcal{F}_u, \mathcal{C}_-]
+
\mathcal{S}^{q_2}_{132}[\mathcal{F}_u, \mathcal{C}_-]\right)_{\alpha \alpha'}^{aa'}
\notag \\
=&
\epsilon^{a'b'c'}
\epsilon^{abc} 
S_{\beta\alpha'}
\left( 
(T\mathcal{C}_- \bar{S})^{b'b}_{\alpha\rho} (F_u)^{c'c}_{\beta\rho} 
+
(T \mathcal{F}_u)^{b'b}_{\alpha\rho} (\mathcal{C}_- \bar{S})_{\beta\rho}^{c'c} 
\right.  \notag \\
&
\phantom{\epsilon^{a'b'c'}
\epsilon^{abc} S_{\beta\alpha'}\  }
\left. + 
(T \mathcal{C}_-\bar{S})^{b'b}_{\rho\alpha} (\mathcal{F}_u)^{c'c}_{\beta\rho}
+ (\mathcal{F}_u T)^{b'b}_{\rho\rho}  (\mathcal{C}_-\bar{S})^{c'c}_{\beta\alpha}
\right),
\end{align}
where $C_-$ is the current-sequential propagator
\begin{align}
\mathcal{C}_-(z, z_2, x) =& \Slash{D}_u^{-1}(z, z_2) \cdot \Gamma^{(2)} \cdot \mathcal{F}_d(z_2, x).
\end{align}
Then the backward current-sink sequential propagator 
for the crossing diagrams $\mathcal{H}^{J_+J_-}_d$ is given in terms of $\mathcal{S}^{J_+J_-}$, 
\begin{align}
\mathcal{H}^{J_+J_-}_{d}(z; y, x) =& \mathcal{S}^{J_+J_-}[\mathcal{F}_u(y, x), \mathcal{C}_-(y, z_2, x)] \cdot \Slash{D}^{-1}_d(y, z).
\end{align}
In summary, the full connected contribution of 
the nucleon ${J_-^{(1)}J_+^{(2)}}$ and ${J_+^{(1)}J_-^{(2)}}$
correlation functions are written as 
\begin{align}
C_{4pt-conn}^{J_-^{(1)}J_+^{(2)}} =&  
{\rm Tr} [\mathcal{B}_{u} \cdot \Gamma^{(1)} \cdot \mathcal{C}_{+}]
+ 
{\rm Tr} [(\mathcal{E}_{du} + \mathcal{H}^{J_-J_+}_u) \cdot \Gamma^{(1)} \cdot \mathcal{F}_d],
\notag  \\
C_{4pt-conn}^{J_+^{(1)}J_-^{(2)}} =& 
{\rm Tr} [\mathcal{B}_{d} \cdot \Gamma^{(1)} \cdot \mathcal{C}_{-}]
+ 
{\rm Tr} [(\mathcal{E}_{ud} + \mathcal{H}^{J_+J_-}_d) \cdot \Gamma^{(1)} \cdot \mathcal{F}_u],
\end{align}
where the first term in either case corresponds to 
the diagram on the left in Fig.~\ref{fig:MPPM}.

\begin{figure}[tbp]
\centering
\includegraphics[clip,height=4cm]{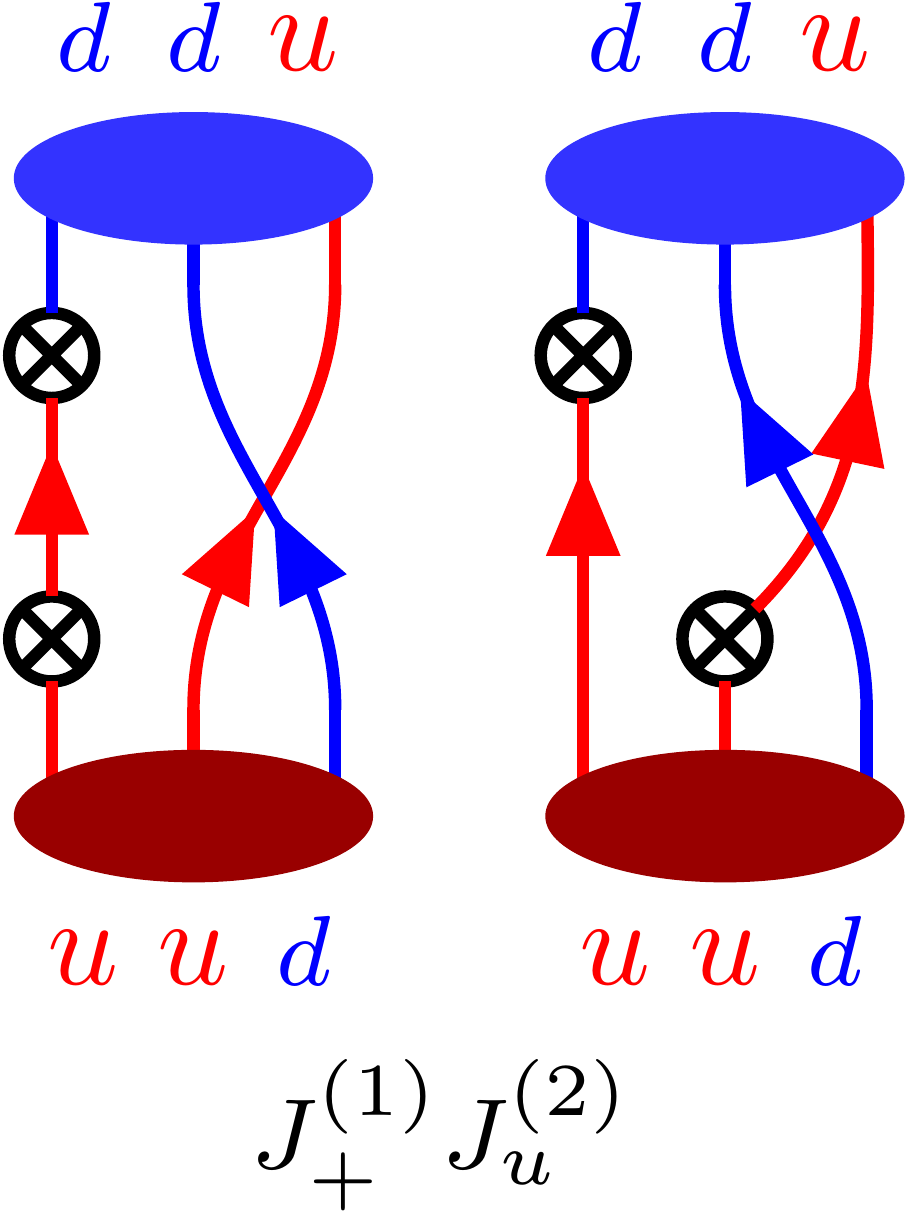} \hspace{5mm}
\includegraphics[clip,height=4cm]{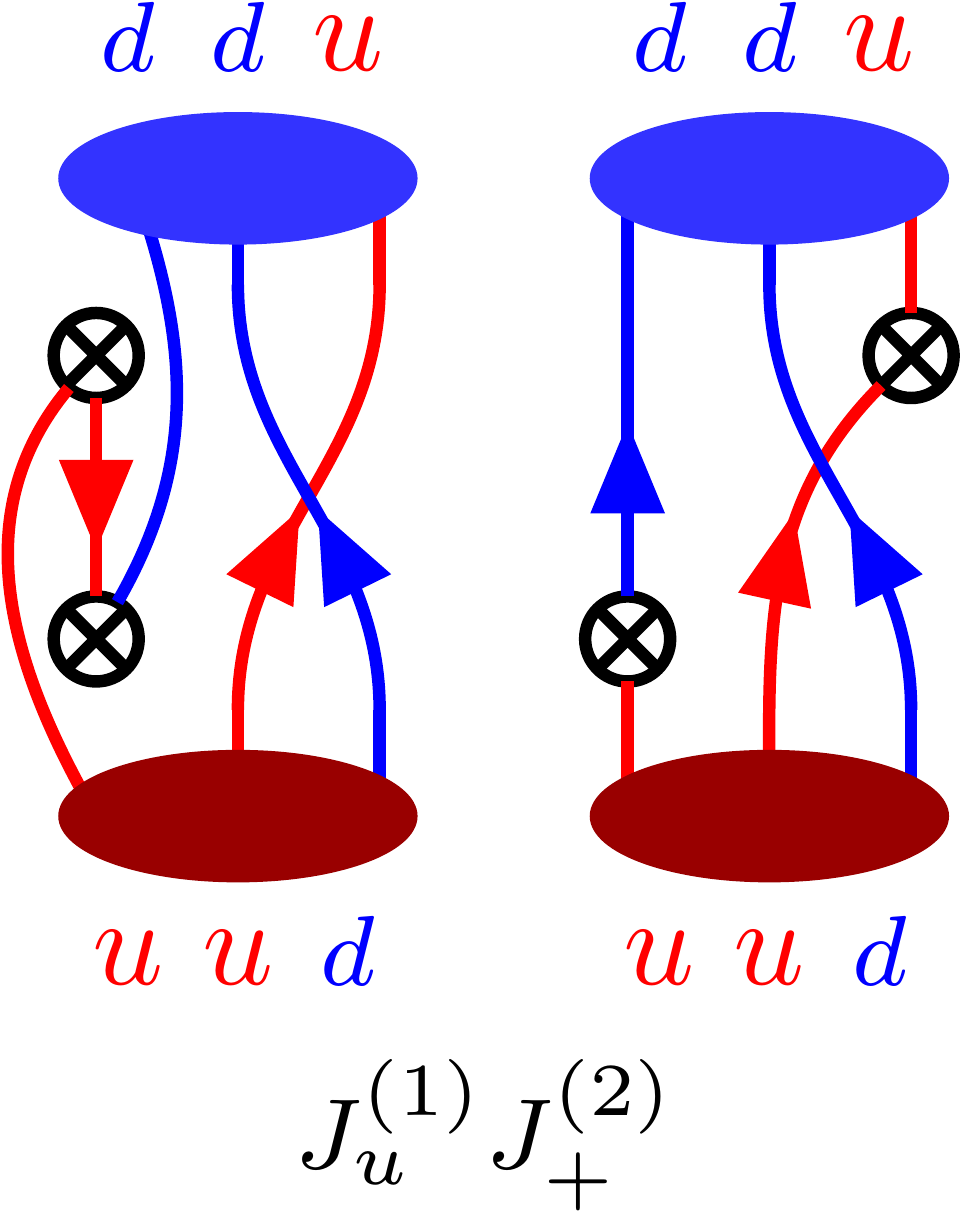} \hspace{5mm}
\includegraphics[clip,height=4cm]{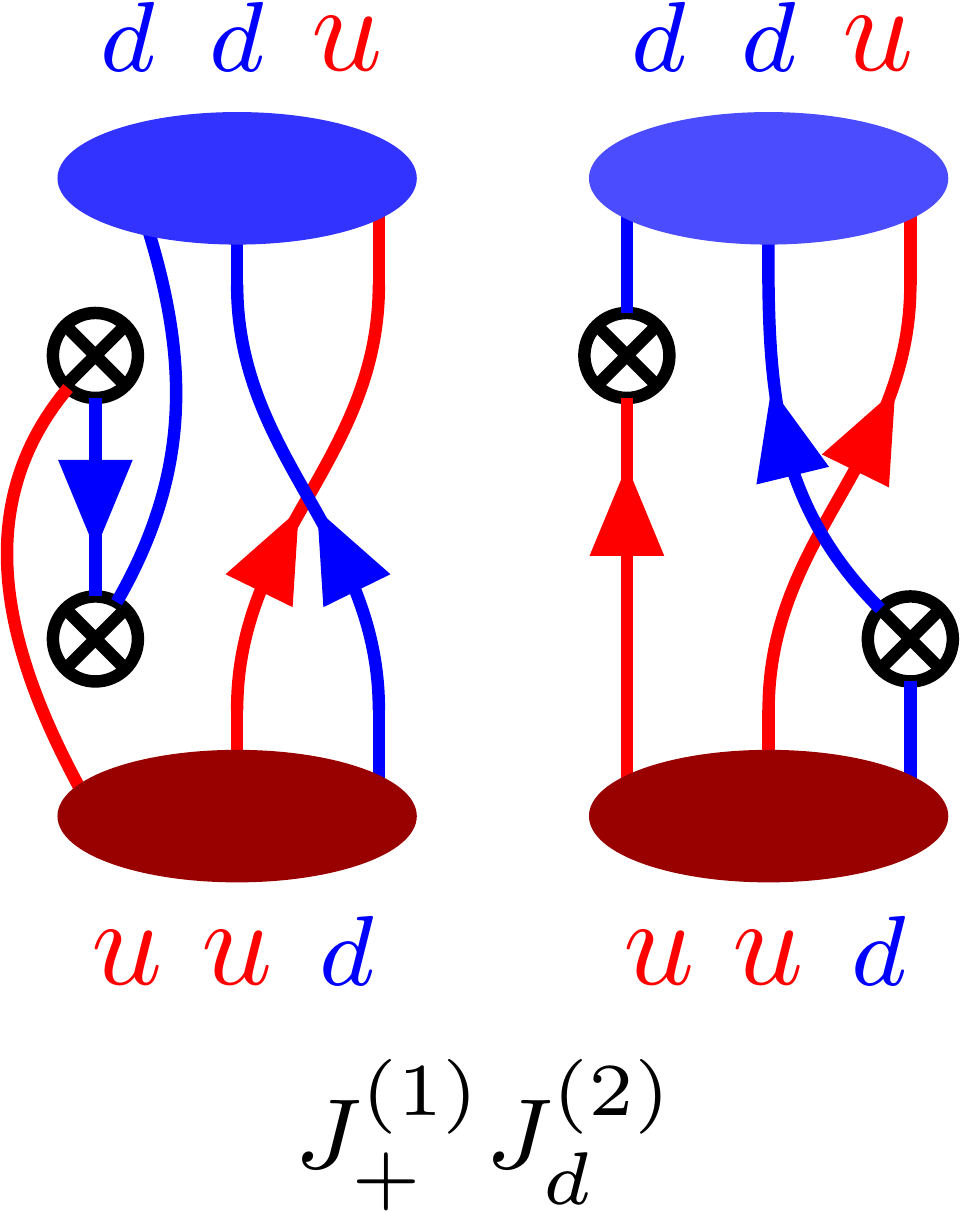} \hspace{5mm}
\includegraphics[clip,height=4cm]{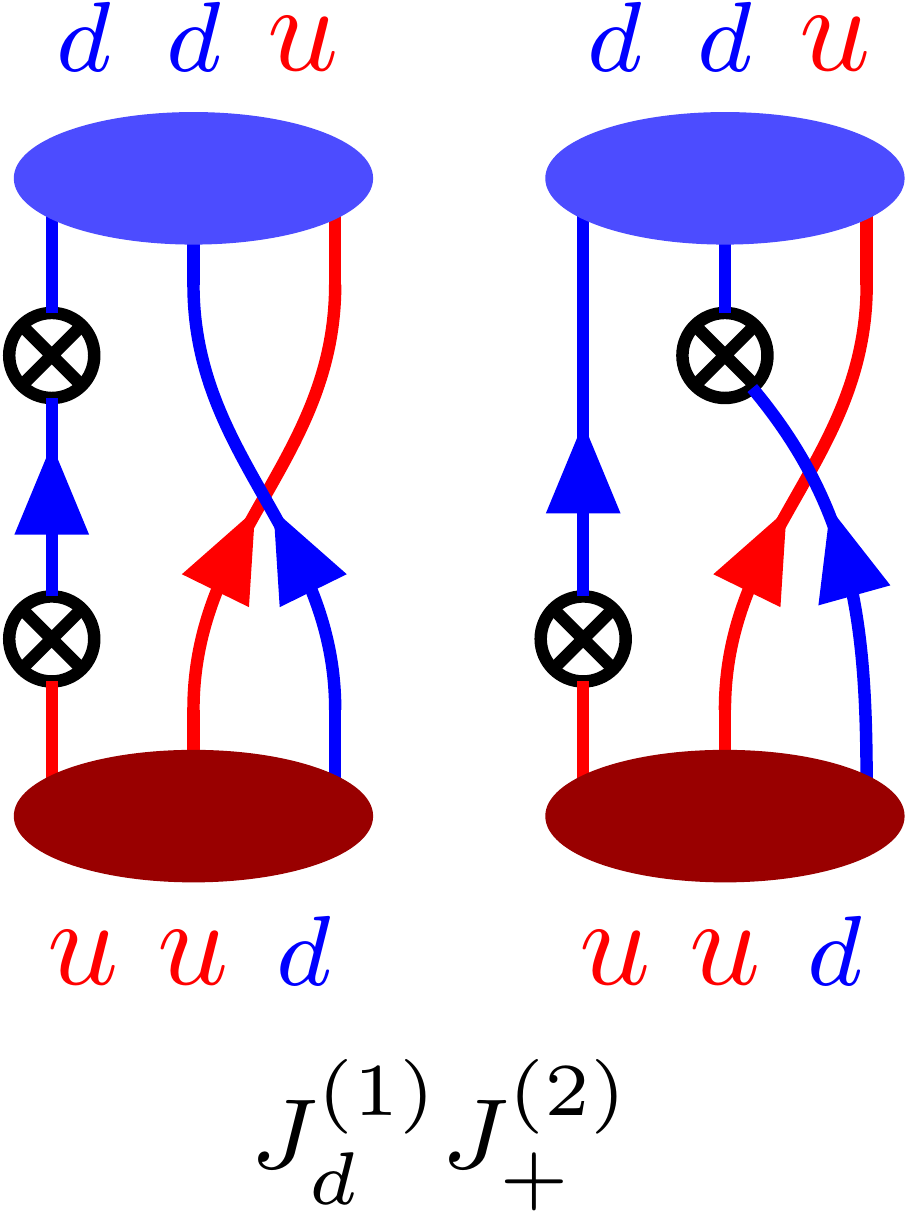} 
\caption{\label{fig:Pu-uP-Pd-dP}
Diagrams for proton to neutron transitions
for $J_+^{(1)} J_u^{(2)}$ 
$J_u^{(1)} J_+^{(2)}$, $J_+^{(1)} J_d^{(2)}$, 
and $J_d^{(1)} J_+^{(2)}$. 
}  
\end{figure}

\subsubsection{$\beta$-decay amplitude}
Finally, we also consider the proton to neutron transition amplitudes 
that are given by four-point functions of 
$J_+^{(1)} J_u^{(2)}$, $J_u^{(1)} J_+^{(2)}$, 
$J_+^{(1)} J_d^{(2)}$, and $J_d^{(1)} J_+^{(2)}$.
Again, the crossing diagrams exist for these correlation
functions. 
As shown in Fig.~\ref{fig:Pu-uP-Pd-dP}, 
we have two types of diagrams for each operator combination.
We also note that there are four different types of the contractions for 
each diagram, which are common 
to all diagrams, 
since the flavor changing current is commonly given by $J_{+}$.

\begin{figure}[tbp]
\centering
\includegraphics[clip,height=3.3cm]{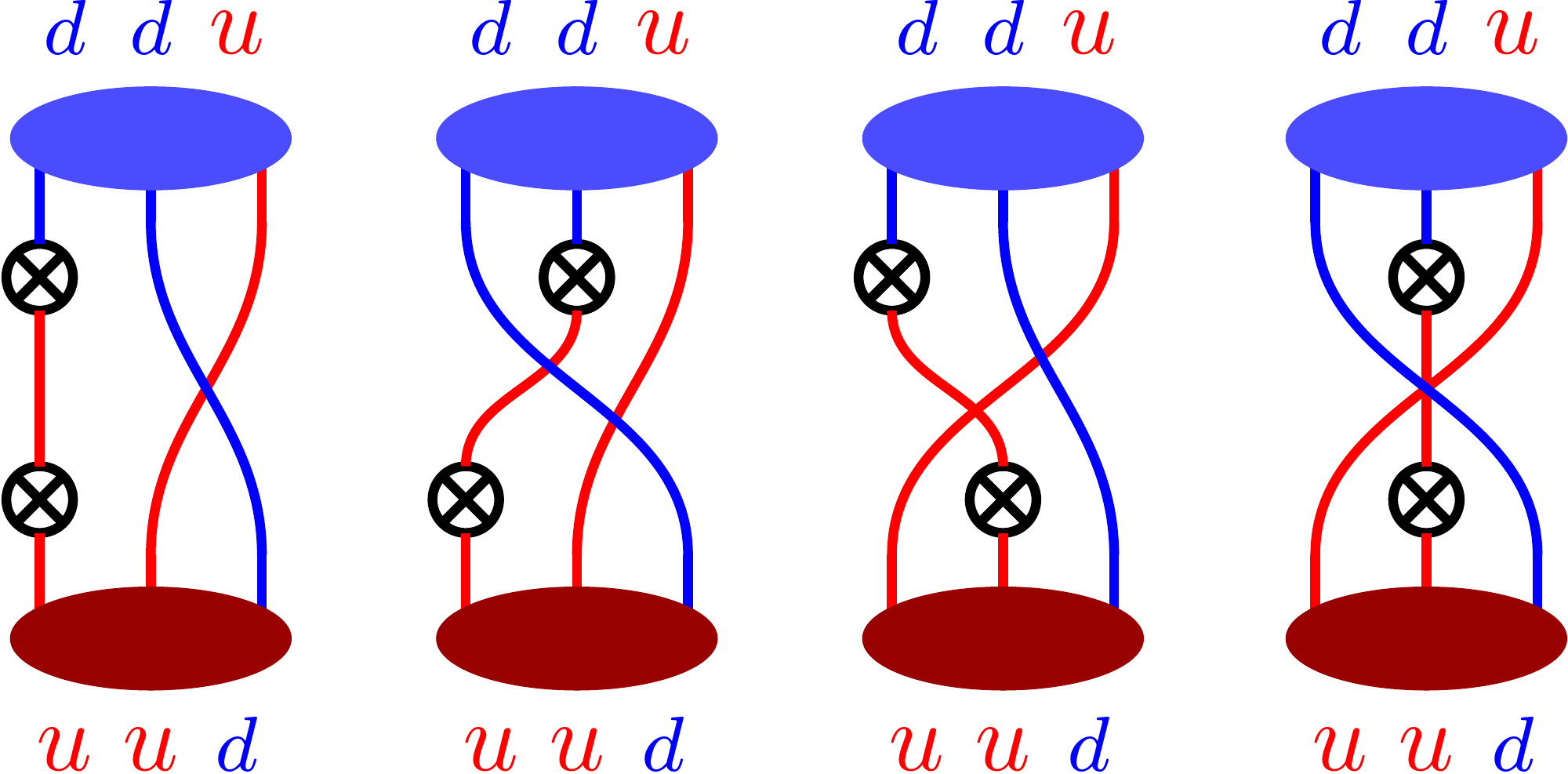} 
\caption{\label{fig:NP_cross}
All contractions that contribute to the first diagram in Fig.~\ref{fig:Pu-uP-Pd-dP}.}  
\end{figure}

As for the diagram on the left of each operator, 
it is convenient to introduce the following diquark propagator for the proton to neutron transition 
%backward sequential source operators 
\begin{align}
\left(\mathcal{S}^{NP}[\mathcal{F}_u, \mathcal{F}_d]\right)_{\alpha \alpha'}^{aa'}=& 
\left(\mathcal{S}^{q_1}_{132}[\mathcal{F}_u, \mathcal{F}_d]
+
\mathcal{S}^{q_1}_{312}[\mathcal{F}_u, \mathcal{F}_d]
+
\mathcal{S}^{q_2}_{231}[\mathcal{F}_u, \mathcal{F}_d]
+
\mathcal{S}^{q_2}_{321}[\mathcal{F}_u, \mathcal{F}_d]\right)_{\alpha \alpha'}^{aa'}.
\end{align}
For example, the diagrams given in Fig.~\ref{fig:NP_cross} are obtained by 
\begin{align}
{\rm Tr}\left[\mathcal{B}^{NP}_d\cdot \Gamma^{(1)}\cdot \mathcal{C}_u \right],
\end{align}
where $\mathcal{B}^{NP}_d$ is the sequential backward propagator of $d$ quark, 
\begin{align}
\mathcal{B}^{NP}_d(z; y, x) = \mathcal{S}^{NP}[\mathcal{F}_u(y, x), \mathcal{F}_d(y, x)]\cdot \Slash{D}^{-1}_d(y, z).
\end{align}
Using $\mathcal{B}^{NP}_d$ we also obtain the backward sink-current sequential propagators
for the left diagrams of  $J_u^{(1)} J_+^{(2)}$, and $J_+^{(1)} J_d^{(2)}$ in Fig.~\ref{fig:Pu-uP-Pd-dP}, 
\begin{align}
\mathcal{E}^{NP}_{q}(z; y, x)=&\mathcal{B}^{NP}_d(z_2; y, x) \cdot \Gamma^{(2)} \cdot \Slash{D}^{-1}_q(z_2, z). 
\end{align}

For the diagrams on the right for each operator combination in
Fig.~\ref{fig:Pu-uP-Pd-dP},  
we use the following four different types of the backward current-sink sequential propagators, 
\begin{align}
\mathcal{H}^{J_+J_u}_d=&\mathcal{S}^{J_+J_u} \cdot \Slash{D}^{-1}_d, 
&
\mathcal{H}^{J_uJ_+}_u=&\mathcal{S}^{J_uJ_+} \cdot \Slash{D}^{-1}_u, 
\notag \\
\mathcal{H}^{J_+J_d}_d=&\mathcal{S}^{J_+J_d} \cdot \Slash{D}^{-1}_d, 
&
\mathcal{H}^{J_dJ_+}_d=&\mathcal{S}^{J_dJ_+} \cdot \Slash{D}^{-1}_d, 
\end{align}
where each of diquark propagators is defined as 
\begin{align}
\left(\mathcal{S}^{J_+J_u}\right)_{\alpha \alpha'}^{aa'}=& 
\left(\mathcal{S}^{q_1}_{132}[\mathcal{C}_u, \mathcal{F}_d]
+
\mathcal{S}^{q_1}_{312}[\mathcal{C}_u, \mathcal{F}_d]
+
\mathcal{S}^{q_2}_{231}[\mathcal{C}_u, \mathcal{F}_d]
+
\mathcal{S}^{q_2}_{321}[\mathcal{C}_u, \mathcal{F}_d]\right)_{\alpha \alpha'}^{aa'},
\notag \\
\left(\mathcal{S}^{J_uJ_+}\right)_{\alpha \alpha'}^{aa'}=& 
\left(\mathcal{S}^{q_2}_{132}[\mathcal{C}_+, \mathcal{F}_d]
+
\mathcal{S}^{q_2}_{312}[\mathcal{C}_+, \mathcal{F}_d]
+
\mathcal{S}^{q_1}_{231}[\mathcal{C}_+, \mathcal{F}_d]
+
\mathcal{S}^{q_1}_{321}[\mathcal{C}_+, \mathcal{F}_d]\right)_{\alpha \alpha'}^{aa'},
\notag \\
\left(\mathcal{S}^{J_+J_d}\right)_{\alpha \alpha'}^{aa'}=& 
\left(\mathcal{S}^{q_1}_{132}[\mathcal{F}_u, \mathcal{C}_d]
+
\mathcal{S}^{q_1}_{312}[\mathcal{F}_u, \mathcal{C}_d]
+
\mathcal{S}^{q_2}_{231}[\mathcal{F}_u, \mathcal{C}_d]
+
\mathcal{S}^{q_2}_{321}[\mathcal{F}_u, \mathcal{C}_d]\right)_{\alpha \alpha'}^{aa'},
\notag \\
\left(\mathcal{S}^{J_dJ_+}\right)_{\alpha \alpha'}^{aa'}=& 
\left(\mathcal{S}^{q_3}_{132}[\mathcal{C}_+, \mathcal{F}_u]
+
\mathcal{S}^{q_3}_{312}[\mathcal{C}_+, \mathcal{F}_u]
+
\mathcal{S}^{q_3}_{231}[\mathcal{F}_u, \mathcal{C}_+]
+
\mathcal{S}^{q_3}_{321}[\mathcal{F}_u, \mathcal{C}_+]\right)_{\alpha \alpha'}^{aa'}.
\end{align}
Using these propagators, 
the full connected contribution of 
the nucleon $J_+^{(1)} J_u^{(2)}$, $J_u^{(1)} J_+^{(2)}$, 
$J_+^{(1)} J_d^{(2)}$, and $J_d^{(1)} J_+^{(2)}$ correlation functions are
written as 
\begin{align}
C_{4pt-conn}^{J_+^{(1)} J_u^{(2)}} =&  
{\rm Tr} [\mathcal{B}_{d}^{NP} \cdot \Gamma^{(1)} \cdot \mathcal{C}_{u}] + 
{\rm Tr} [\mathcal{H}^{J_+J_u}_d \cdot \Gamma^{(1)} \cdot \mathcal{F}_u],
\notag \\
C_{4pt-conn}^{J_u^{(1)} J_+^{(2)}} =&  
{\rm Tr} [(\mathcal{E}_{u}^{NP} + \mathcal{H}^{J_uJ_+}_{u})\cdot \Gamma^{(1)} \cdot \mathcal{F}_{u}],
\notag \\
C_{4pt-conn}^{J_+^{(1)} J_d^{(2)}} =&  
{\rm Tr} [(\mathcal{E}_{d}^{NP} + \mathcal{H}^{J_+J_d}_{d})\cdot \Gamma^{(1)} \cdot \mathcal{F}_{u}],
\notag \\
C_{4pt-conn}^{J_d^{(1)} J_+^{(2)}} =&  
{\rm Tr} [\mathcal{B}_{d}^{NP} \cdot \Gamma^{(1)} \cdot \mathcal{C}_{+}] + 
{\rm Tr} [\mathcal{H}^{J_dJ_+}_d \cdot \Gamma^{(1)} \cdot \mathcal{F}_d].
\end{align}

We note that the neutron amplitudes of the Compton scattering as well
as the $\beta^-$ decay are also obtained from these functions by
interchanging $u$ and $d$ flavor indices. 
Therefore, all the necessary quark-line contractions for the
$\beta$-decay are given by these formulae. 

As implemented in \cite{Chambers:2017dov,Can:2020sxc}, 
some of these matrix elements may be obtained utilizing the 
Feynman-Hellmann technique.
However, such numerical implementation would be more demanding.
For example, the four-point functions with an insertion of
$J_-^{(1)}J_+^{(2)}$ 
are obtained from a second derivative
of the two-point function in the presence of two external source terms
$\epsilon_1 J_-^{(1)}+\epsilon_2 J_+^{(2)}$.
When the currents are flavor-changing, it corresponds to a
modification of the Dirac operator to a flavor-dependent one
$\Slash{D}(m_q)\to\left(
  \begin{smallmatrix} \Slash{D}(m_u) & \epsilon_1 \Gamma^{(1)} \\
    \epsilon_2 \Gamma^{(2)} & \Slash{D}(m_d) \end{smallmatrix}
\right)$,
and one has to compute its inverse at various values of $\epsilon_{1,2}$
to take a numerical derivative and then an extrapolation 
$\epsilon_{1,2}\to 0$.
Furthermore, in order to control the time separation between the two
currents one has to repeat the method with the source terms only at a
given time slice, so that the total cost may be as high as the direct
computation of the four-point function.

We use this technique to verify the quark-line contraction codes
with the flavor changing currents.
We place the external field corresponding to $J^{(1,2)}$ multiplied by
a small parameter $\epsilon_{1,2}$ at a given time slice $\tau_{1,2}$
and compute the two-point function. 
By taking a numerical derivative we confirm that the results agree
with what we obtain with the four-point functions including two
current insertions.  
The details of the analyses and numerical results are shown in
Appendix~\ref{app:numerical_check}.

\section{Discussions}
\label{sec:discussions}

This work proposes a method to compute the inelastic $\ell N$
scattering cross section using lattice QCD.
The key steps for computing the total cross section are to decompose
the forward Compton-scattering amplitude into the contributions of
different energies and then to integrate with an appropriate weight
factor that represents the phase space.
Instead of literally performing this program, we consider the spectral 
decomposition only virtually and realize the energy integral by
identifying the weight factor as an operator constructed from the
Hamiltonian.
This weight \emph{operator} between the two currents in the Compton
amplitude can be reconstructed from the corresponding lattice
computation.
The essential point is that we can avoid the explicit spectral
decomposition, which is a well-known example of the ill-posed
inverse problem. 

In the standard analyses of DIS, the cross section is measured
depending on $Q^2$ and $x$, from which one can determine the structure
functions $F_i(x,Q^2)$. 
(The subscript distinguishes distinct kinematical structures, but the
details are not important in this discussion.)
In the parton model, the structure functions are further decomposed
into the contributions of partons (quarks and gluons) and written in
terms of the parton distribution functions (PDFs).
The basic assumption here is that the power corrections of the form
$1/Q^2$ and higher can be neglected, which is justified only above
$Q^2\sim$ several GeV$^2$.
In fact, in the low $Q^2$ region, there are resonance structures in
$F_i(x,Q^2)$ near $x\sim 1$ due to low-lying energy states.
Such resonances may not be treated by perturbation theory.
Even using the lattice QCD calculation, treating the individual
excited states is a very challenging task, since they are actually 
multi-particle states like $N\pi$, $N\pi\pi$, $\cdots$, whose spectrum
becomes dense on large volumes.
Our proposal is to consider only a sum over such states, which
is an well-defined quantity and the correspondence with the lattice
observable can be established.
In other words, instead of the structure functions $F_i(x,Q^2)$,
we only analyze their weighted integrals.

The problem of the standard analysis at low $Q^2$ is related to the
assumption of quark-hadron duality.
In the perturbative QCD calculation one takes the quark and gluon
external states, which are unphysical, instead of hadronic states.
Such an assumption can be justified when one sums over a certain range
of kinematical variables because the non-perturbative effects which
are enhanced near the resonances can be avoided \cite{Poggio:1975af}.
In the analysis of DIS, the duality has been observed to be satisfied
after averaging over some appropriate range of $x$ 
\cite{Bloom:1970xb,Bloom:1971ye,Niculescu:2000tk}
(see also \cite{Melnitchouk:2005zr} for a review).
Conversely, some sort of smearing (or averaging) of the experimental
data is necessary to compare with perturbative QCD,
and it becomes more prominent in the low $Q^2$ region.
So far, no quantitative measure on how much smearing has to be
introduced for a desired precision has been known.
Our proposal is one way to define such smearing on a theoretically
solid foundation.
In this paper, we have focused on the calculation of the total cross
section, but it can be easily extended to the cases of partially
integrated cross section, which plays the role of the smearing.
Fully non-perturbative computation is possible in our framework, and
no assumption of duality is necessary.

One of the main themes in the study of nucleon structure is the
determination of PDFs.
With our method, the $x$-dependence of PDFs is not accessible as we
need an integral over the energy of final hadronic system.
Still, some moments of the cross section can be computed by choosing
an appropriate weight function including some power of $x$, for
instance.
Once various moments are obtained, they can be used to constrain the
overall shape of PDFs.
We have to be careful, though, because the integral over $\omega$
while fixing $\bm{q}^2$ corresponds to an integral along a curve on a
$(x,Q^2)$ plane as shown in Figure~\ref{fig:kin}.
They cannot be simply written using the moments conventionally defined
as $\int_0^1 dx x^n F_i(x,Q^2)$ at a fixed $Q^2$.
One may consider instead integrated moments of the form
$\int_{Q^2_{\mathrm{min}}}^\infty 1/(Q^2)^m \int_0^1dx x^n F_i(x,Q^2)$,
which can be calculated using the method introduced in this work.
Given the fact that the $Q^2$ dependence of $F_i(x,Q^2)$ is a
subdominant effect and can be understood using evolution equations,
one could still gain some information from such analyses.

Towards realistic computation of the inelastic $\ell N$ scattering
cross section, there are a couple of challenges we would be confronted.
One of the major limitations, which is actually common in lattice QCD
calculations, is that the spatial momentum $|\bm{q}|$ one can reach
would be less than a few GeV$/c$.
The momentum is of course limited due to the discretization effects,
but in practice the limitation rather comes from large statistical
noise of higher momentum correlators, for which the signal is rapidly
overwhelmed by the noise. 
As a consequence, it will be very challenging for the lattice
computations to reach the DIS kinematical region, $\gg$ a few
GeV$/c$.
The energy region of interest for the neutrino experiments, T2K and
DUNE, on the other hand, is below a few GeV, which would be within
reach. 

Even at small momenta, the computation of the forward Compton
amplitude is a challenging task, because it requires a saturation of
the ground state nucleon on the both ends to prepare $\langle N|$ and
$|N\rangle$.
The signal-to-noise ratio for nucleon decreases as
$\exp[-(m_N-3m_\pi/2)t]$ for large time separations where the ground
state would dominate \cite{Lepage:1989hd}.
Even for a nucleon two-point function, it requires a lot of effort to
make sure that the ground state has been reached, and there have been
many studies to investigate the ground-state saturation for
three-point functions, which are relevant for various nucleon charges.
(See, for instance, \cite{Capitani:2012gj,Yoon:2016dij,Green:2019zhh}.
More details and a full list of references may be found in
\cite{Lin:2017snn}.)
The up-to-date simulations have a time separation between the nucleon
source and sink operators about 1~fm, which is not ideal to
sufficiently suppress the excited states of mass gap about the pion
mass but is a necessary compromise.
The insertion of two currents separated from each other in the time
direction requires even larger time separation than the computation of
the three-point functions.
The computations carried out so far
\cite{Chambers:2017dov,Liang:2019frk,Can:2020sxc}
do not include extensive tests of the ground-state saturation.
We emphasize that the signal-to-noise problem is common for all
lattice calculations, especially for those of nucleon properties, and
various methods are being studied to improve the situation.

\section{Conclusions}
\label{sec:conclusions}
This paper describes a new method that enables us to explore
a class of new applications of lattice QCD.
It is about inclusive processes such as the lepton-nucleon scattering
without specifying the final hadronic states.
On the lattice, the corresponding quantity is the forward
Compton-scattering amplitude calculated at various Euclidean time
separations between the two inserted currents.
The necessary quark-line contractions are summarized.
The lattice observable can be related to the physical cross section.
The method has been already successfully tested for inclusive $B$
meson decays \cite{Gambino:2020crt}.

The method opens a new possibility to study the $\nu N$ scattering in
the low-energy region, which is relevant to the neutrino oscillation
experiments, such as T2K and DUNE.
So far, theoretically solid analysis has been possible only for
elastic scattering, for which the form factors computed on the lattice
can be used, and for deep inelastic scattering, which is described by
perturbation theory.
The region between these lowest and high energy scales can be treated
fully non-perturbatively using the technique proposed in this work.

Although this paper describes only the total cross section in order to
be explicit, the proposed method can be utilized to compute other
related quantities, such as moments of $x$ or other variables.
The change of the analysis is only to modify the smearing kernel, and
the extension is straightforward as long as it does not introduce more
discontinuities.
Through such moments, one can extract more information about the
process and the nuclear structure.

The actual computation is yet to be carried out.
The computational cost is significantly more demanding for the forward
Compton-scattering compared to that of the three-point function used
to extract the form factors.
Furthermore, the signal-to-noise problem is severer, so that the
realistic calculation would be a substantial challenge in lattice QCD
in the next decade.

\begin{acknowledgments}
  We thank the members of the JLQCD collaboration for discussions.
  Numerical tests have been performed 
  on SX-Aurora TSUBASA at High Energy Accelerator Research
  Organization (KEK) under its Particle, Nuclear and Astro Physics
  Simulation Program, as well as
  on HOKUSAI supercomputer of the RIKEN ACCC facility.
  This work is supported in part by JSPS KAKENHI Grant Number
  17K14309, 18H03710, 18H01216, 18H04484,
  and by the Post-K and Fugaku supercomputer project through the Joint
  Institute for Computational Fundamental Science (JICFuS).
  A part of the numerical tests was done using the Qlua
  software suite \cite{Qlua}.
  The Feynman diagrams are written using TikZ-Feynman
  \cite{Ellis:2016jkw} and TikZ-FeynHand \cite{Dohse:2018vqo}.
\end{acknowledgments}

\appendix
\section{Numerical check of the contraction}
\label{app:numerical_check}
Here we present a numerical check of the current contractions
described in Sec.~\ref{sec:contraction}.

Let us consider the current with an external field, $\mathcal{A}_\mu$, 
so that the effective action can be expressed as 
$\mathcal{L}_{eff} = \mathcal{L}_{QCD} + \epsilon_1 J_\mu^{(1)} \mathcal{A}^\mu + 
\epsilon_2 J_\mu^{(2)} \mathcal{A}^\mu$
with some small parameters $\epsilon_{1,2}$.
On this background field, 
we replace the Dirac operator $\Slash{D}$ by
\begin{align}
\Slash{D} \to (\Slash{D} + \epsilon_1 \Gamma^{(1)} + \epsilon_2 \Gamma^{(2)} ).
\end{align}
Then, the quark propagation with operator insertions is included 
in the forward-sequential propagator as a small perturbation. 
For simplicity, in the following analysis we consider the
iso-symmetric limit, 
{\it i.e.} $\Slash{D}_u^{-1} = \Slash{D}_d^{-1}$.

As an example, we consider the case of $J_u^{(1)} J_u^{(2)}$. 
The two currents couple only to the up-quark propagator, 
so that we should modify the up-type Dirac operator as, 
\begin{align}
\Slash{D}_u^{-1} \to &  (\Slash{D}_u + \epsilon_1 \Gamma^{(1)}\delta(t-t_{2}) 
+ \epsilon_2 \Gamma^{(2)} \delta(t-t_{1}) )^{-1} \notag \\
= & 
\Slash{D}_u^{-1} - \epsilon_1 \Slash{D}_u^{-1} \Gamma^{(1)} \Slash{D}_u^{-1}
- \epsilon_2 \Slash{D}_u^{-1} \Gamma^{(2)} \Slash{D}_u^{-1} 
\notag \\
& + \epsilon_1 \epsilon_2 (
\Slash{D}_u^{-1} \Gamma^{(1)} \Slash{D}_u^{-1} \Gamma^{(2)} \Slash{D}_u^{-1}
+
\Slash{D}_u^{-1} \Gamma^{(2)} \Slash{D}_u^{-1} \Gamma^{(1)} \Slash{D}_u^{-1} )
+ \mathcal{O}(\epsilon_1^2, \epsilon_2^2).
\end{align}
Thus we define the following ``perturbed'' propagator,  
\begin{align}
\mathcal{F}^{\epsilon_1+\epsilon_2}(y,x) =& 
\mathcal{F}(y, x) 
- \epsilon_1 
\Slash{D}^{-1}(y,z_1)   \cdot
\Gamma^{(1)} \cdot  \mathcal{F}(z_1, x) 
- \epsilon_2 
\Slash{D}^{-1}(y,z_2)   \cdot
\Gamma^{(2)} \cdot  \mathcal{F}(z_2, x)  \notag \\
&+ \epsilon_1 \epsilon_2  (
\Slash{D}^{-1}(y,z_1)   \cdot \Gamma^{(1)} \cdot 
\Slash{D}^{-1}(z_1,z_2)   \cdot \Gamma^{(2)} \cdot  \mathcal{F}(z_2, x)  \notag \\
& \quad \quad \quad \quad
 + 
 \Slash{D}^{-1}(y,z_2)   \cdot \Gamma^{(2)} \cdot 
\Slash{D}^{-1}(z_2,z_1)   \cdot \Gamma^{(1)} \cdot  \mathcal{F}(z_1, x)  ). \notag 
\end{align}
Using this ``perturbed'' propagator, 
the connected diagrams for the current-current nucleon four-point 
function can be evaluated from the ordinary nucleon two-point function
$N[\mathcal{F}^{\epsilon_1+\epsilon_2}, \mathcal{F}_d]$. 
The four-point correlator $C_{4pt-conn}^{J_u^{(1)} J_u^{(2)}}(z_1,z_2, y, x)$ 
can be extracted from a contribution proportional to $\epsilon_1 \epsilon_2$ in $N[\mathcal{F}^{\epsilon_1+\epsilon_2}, \mathcal{F}_d]$, 
or equivalently one should take
$\frac{\partial^2}{\partial  \epsilon_1 \partial \epsilon_2}N[\mathcal{F}^{\epsilon_1+\epsilon_2}, \mathcal{F}_d]|_{\epsilon_{1,2}\to 0}$.
Here, $N[\mathcal{F}_{(1)}, \mathcal{F}_{(2)}]$ is 
a general nucleon two-point correlation function computed as
\begin{align}
N[\mathcal{F}_{(1)}, \mathcal{F}_{(2)}] = &
\epsilon^{a'b'c'}\epsilon^{abc}
\left[ 
(\mathcal{F}_{(1)} T)^{a'a}_{\alpha'\alpha} (\mathcal{F}_{(1)})^{b'b}_{\alpha\beta} 
(S \mathcal{F}_{(2)} \bar{S})^{c'c}_{\alpha'\beta}
\right. \notag \\
& 
\phantom{\epsilon^{a'b'c'}\epsilon^{abc}}
\left.
+ (\mathcal{F}_{(1)}T)^{a'a}_{\alpha'\alpha'} (\mathcal{F}_{(1)})^{b'b}_{\beta'\beta} 
(S \mathcal{F}_{(2)} \bar{S})^{c'c}_{\beta'\beta}
\right], 
\end{align}
so that $C_{2pt} = N[\mathcal{F}_u,\mathcal{F}_d]$.

In order to compute the correlation function with the flavor-changing
currents, we define additional perturbed propagators  
\begin{align}
\mathcal{F}^{\epsilon_1}(y,x) =&
\mathcal{F}(y, x) 
- \epsilon_1 
\Slash{D}^{-1}(y,z_1)   \cdot
\Gamma^{(1)} \cdot  \mathcal{F}(z_1, x), \notag \\
\mathcal{F}^{\epsilon_2}(y,x) =&
\mathcal{F}(y, x) 
- \epsilon_2 
\Slash{D}^{-1}(y,z_2)   \cdot
\Gamma^{(2)} \cdot  \mathcal{F}(z_2, x), \notag \\
\mathcal{F}^{\epsilon_1\epsilon_2}(y,x) =&
\mathcal{F}(y, x) 
+ \epsilon_1 \epsilon_2 
\Slash{D}^{-1}(y,z_1) \cdot \Gamma^{(1)} \cdot  
\Slash{D}^{-1}(z_1, z_2) \cdot \Gamma^{(2)} \cdot  
\mathcal{F}(z_2, x), \notag \\
\mathcal{F}^{\epsilon_2\epsilon_1}(y,x) =&
\mathcal{F}(y, x) 
+ \epsilon_1 \epsilon_2 
\Slash{D}^{-1}(y,z_2) \cdot \Gamma^{(2)} \cdot  
\Slash{D}^{-1}(z_2,z_1) \cdot \Gamma^{(1)} \cdot  
\mathcal{F}(z_1, x). 
\end{align}
From the diagrams in Fig.~\ref{fig:MPPM}, 
the correlation function with charged currents $J_-^{(1)}J_+^{(2)}$ can be
obtained from
$N[\mathcal{F}^{\epsilon_1\epsilon_2}, \mathcal{F}^{\epsilon_2\epsilon_1}] 
+ N_{132}[\mathcal{F}, \mathcal{F}^{\epsilon_2},\mathcal{F}^{\epsilon_1}]
+ N_{231}[\mathcal{F}^{\epsilon_2}, \mathcal{F},\mathcal{F}^{\epsilon_1}]
+ N_{312}[\mathcal{F}, \mathcal{F}^{\epsilon_2},\mathcal{F}^{\epsilon_1}]
+ N_{321}[\mathcal{F}^{\epsilon_2}, \mathcal{F},\mathcal{F}^{\epsilon_1}]$,
where the first term corresponds to the sum of the first and third
diagrams in Fig.~\ref{fig:MPPM},  and each of the other four terms
corresponds each of the crossing diagrams in Fig.~\ref{fig:cross}. 

In summary, 
the four types of the nucleon ``perturbed" two-point functions are
given as 
\begin{align}
C_{2pt-conn}^{J_u^{(1)} J_u^{(2)}} =& N[\mathcal{F}^{\epsilon_1+\epsilon_2}, \mathcal{F}], \notag \\
C_{2pt-conn}^{J_d^{(1)} J_d^{(2)}} =& N[\mathcal{F}, \mathcal{F}^{\epsilon_1+\epsilon_2}], \notag \\
C_{2pt-conn}^{J_-^{(1)} J_+^{(2)}}
=& 
N[\mathcal{F}^{\epsilon_1\epsilon_2}, \mathcal{F}^{\epsilon_2\epsilon_1}] 
+ N_{132}[\mathcal{F}, \mathcal{F}^{\epsilon_2},\mathcal{F}^{\epsilon_1}]
+ N_{231}[\mathcal{F}^{\epsilon_2}, \mathcal{F},\mathcal{F}^{\epsilon_1}] 
\notag \\ 
&
+ N_{312}[\mathcal{F}, \mathcal{F}^{\epsilon_2},\mathcal{F}^{\epsilon_1}]
+ N_{321}[\mathcal{F}^{\epsilon_2}, \mathcal{F},\mathcal{F}^{\epsilon_1}],
\notag \\ 
C_{2pt-conn}^{J_+^{(1)} J_-^{(2)}} 
=& 
N[\mathcal{F}^{\epsilon_2\epsilon_1}, \mathcal{F}^{\epsilon_1\epsilon_2}] 
+ N_{132}[\mathcal{F}, \mathcal{F}^{\epsilon_1},\mathcal{F}^{\epsilon_2}]
+ N_{231}[\mathcal{F}^{\epsilon_1}, \mathcal{F},\mathcal{F}^{\epsilon_2}] 
\notag \\ 
&
+ N_{312}[\mathcal{F}, \mathcal{F}^{\epsilon_1},\mathcal{F}^{\epsilon_2}]
+ N_{321}[\mathcal{F}^{\epsilon_1}, \mathcal{F},\mathcal{F}^{\epsilon_2}],
\end{align}
and the four-point functions may be obtained as
\begin{equation}
  C_{4pt-conn}^{J^{(1)}J^{(2)}}(z_1,z_2, y, x) \sim
  \left.
  \frac{\partial^2}{\partial\epsilon_1 \partial\epsilon_2} 
  C_{2pt-conn}^{J^{(1)}J^{(2)}}
  \right|_{\epsilon_{1,2}\to 0}, 
\end{equation}
where $J^{(1)}J^{(2)}$ represents one of the four combinations:
$J_u^{(1)}J_u^{(2)}$, $J_d^{(1)}J_d^{(2)}$,
$J_-^{(1)}J_+^{(2)}$, $J_+^{(1)}J_-^{(2)}$.

\begin{figure}[tbp]
\centering
 \includegraphics[clip,width=8cm]{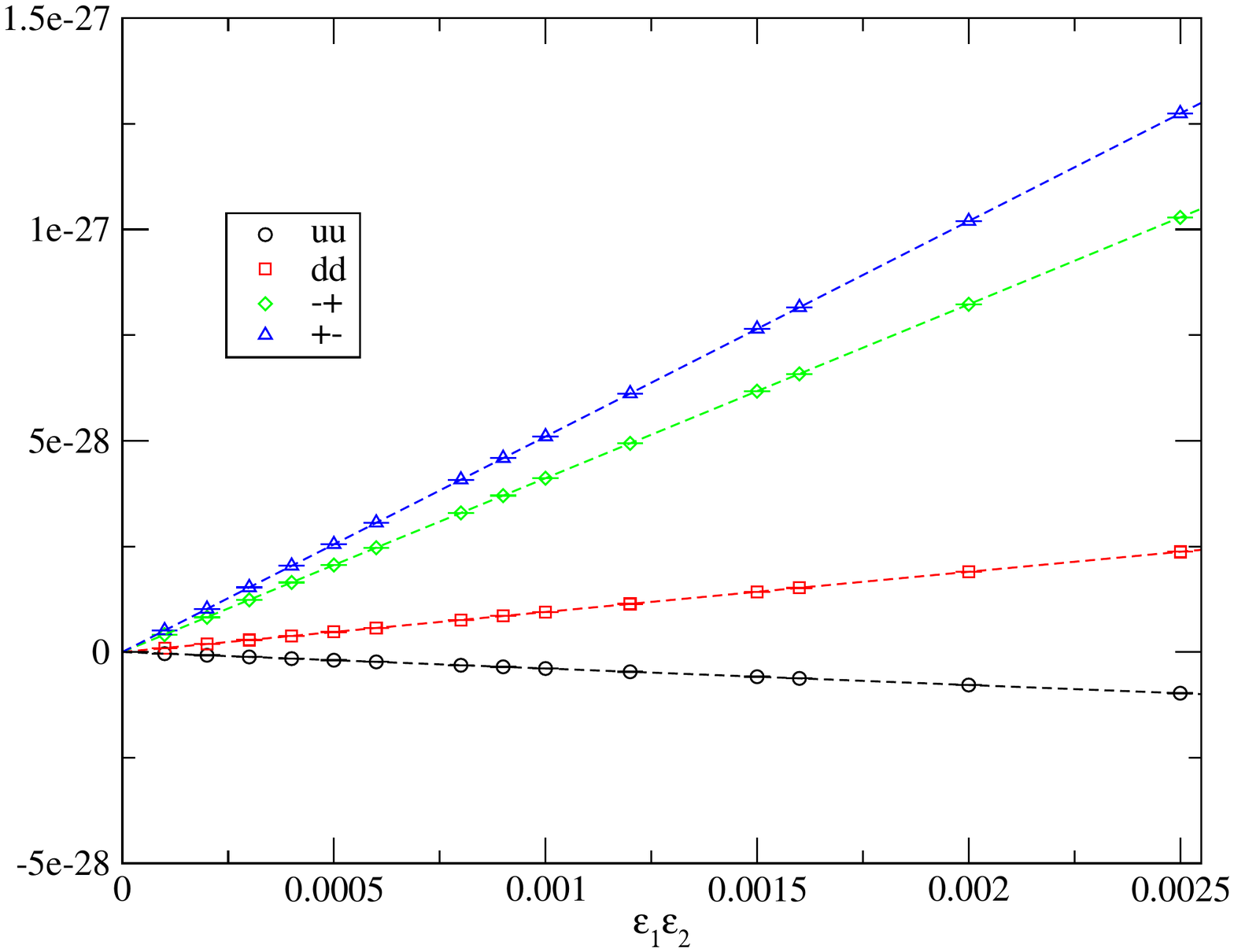}
 \includegraphics[clip,width=8cm]{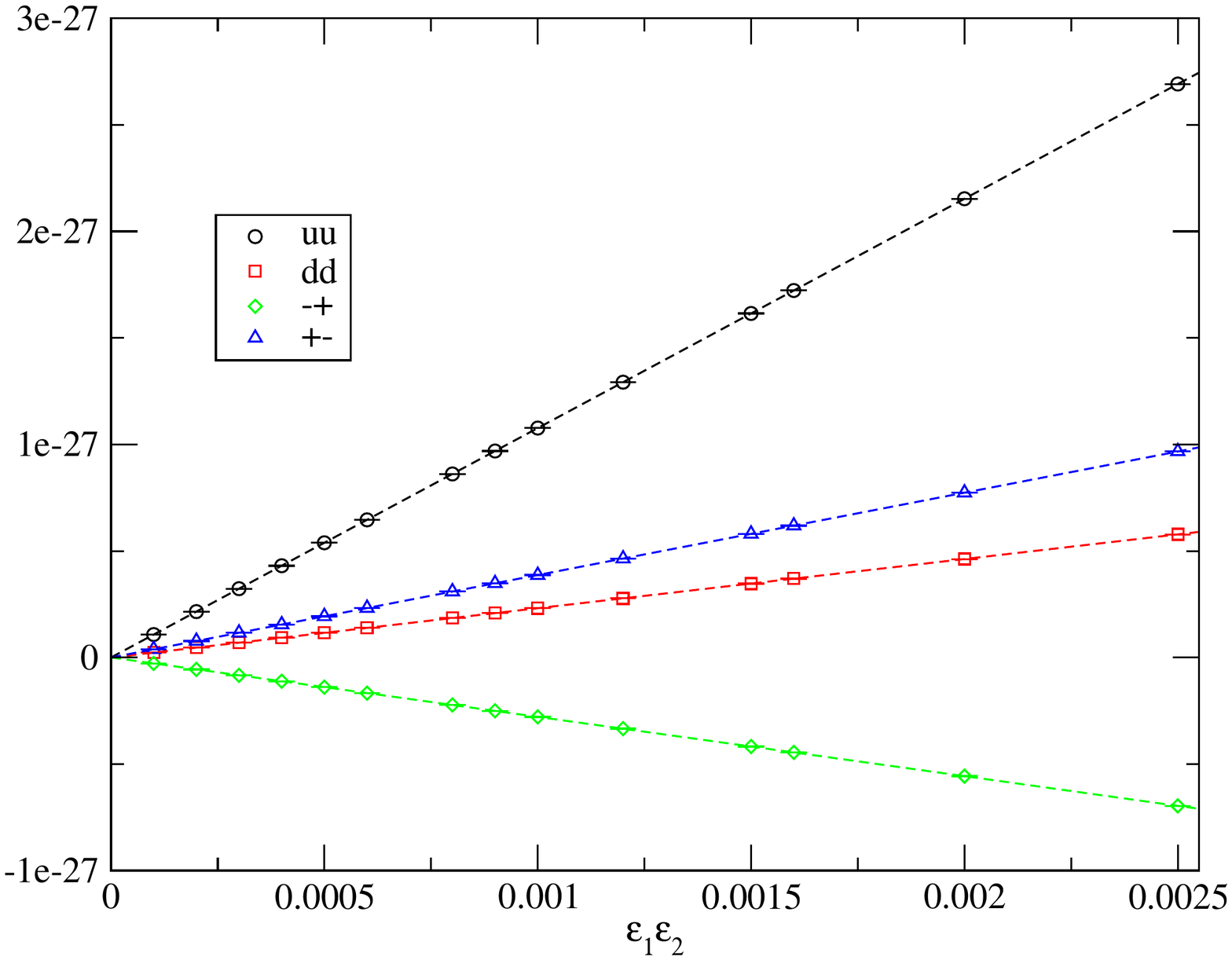} 
\caption{\label{fig:4pt} 
  Results for $\tilde{C}_{2pt-conn}$. 
  The linear fits to the lattice data are also shown. }
\end{figure}

\begin{table}[tbp]
  \centering
  \begin{tabular}{l | c c c c}
    & $J_u^{(1)} J_u^{(2)}$ & $J_d^{(1)} J_d^{(2)}$ & $J_-^{(1)} J_+^{(2)
}$ & $J_+^{(1)} J_-^{(2)}$  \\ ¥hline
    ${\rm Re} (C_{4pt-conn})$  & $-3.908$ & $0.9488$ & $4.114$ & $5.098$
\\
    ${\rm Re} (f_{\epsilon_1 \epsilon_2})$ & $-3.907$ & $0.9487$ & $4.
114$ & $5.098$ \\
    \hline
    ${\rm Im} (C_{4pt-conn})$  & $10.77$ & $2.312$ & $-2.786$ & $3.868$
\\
    ${\rm Im} (f_{\epsilon_1 \epsilon_2})$ & $10.77$ & $2.312$ & $-2.786
$ & $3.868$\\
  \end{tabular}
  \caption{\label{tab:4pt}
    Comparison of the results from the direct and the background field
    propagator methods.
    The correlators at a certain lattice point are listed in the unit of
    $10^{-25}$.
  }
\end{table}

We carry out a numerical test of these correspondences using a lattice
calculation on a $4^3 \times 8$ lattice gauge configuration.
In the background field method, we specify the parameters as follows.
The source position $x_{src}$ is $x_{src}=(\bm{x},t) = (1,2,3,6)$,
$t_{sep}=4$, $\bm{p}=(0,0,0)$, $\bm{q}_{1}=(2,-2,-1)$,
$\bm{q}_{2}=(1,2,3)$,
$\tau_1=2$, $\tau_2=3$,  
$\Gamma^{(1)}=\gamma_t$, $\Gamma^{(2)}=\gamma_x$.
Since the ``perturbed'' two-point functions
$C_{2pt-conn}^{J^{(1)} J^{(2)}}$ can be expressed as
$C_{2pt-conn}^{J^{(1)} J^{(2)}}= f(\epsilon_1, \epsilon_2)$,
with
$f(\epsilon_1, \epsilon_2)=f_0 + \epsilon_1 f_{\epsilon_1} +\epsilon_2 f_{\epsilon_2} + \epsilon_1^2 f_{\epsilon_1^2} + \epsilon_2^2 f_{\epsilon_2^2} +\epsilon_1  \epsilon_2 f_{\epsilon_1\epsilon_2}
+\mathcal{O}(\epsilon^3)$,
it is useful to consider the following combination
\begin{align}
  \tilde{f}(\epsilon_1,\epsilon_2) =
  &
    f(0,0)
    +\frac{f(\epsilon_1,\epsilon_2)+f(-\epsilon_1,-\epsilon_2) } {2}  
    -\frac{f(\epsilon_1,0)+f(-\epsilon_1,0) } {2}  
    -\frac{f(0,\epsilon_2)+f(0,-\epsilon_2) } {2}  
    \notag \\
  =&
     \epsilon_1  \epsilon_2 f_{\epsilon_1\epsilon_2} + \mathcal{O}(\epsilon^4).
\end{align}
Since in the numerical calculations we only have the data at some
discrete values of $\epsilon_1 \epsilon_2$,  
we fit the data to obtain the linear coefficient of the
$\epsilon_1\epsilon_2$ term in
$\tilde{C}_{2pt-conn}^{J^{(1)} J^{(2)}} = \tilde{f}(\epsilon_1,\epsilon_2)$.
The fits are shown in Fig.~\ref{fig:4pt}, where an excellent linear
dependence on $\epsilon_1\epsilon_2$ may be observed.
The comparison between the direct and the background field method is
shown in Table~\ref{tab:4pt}.
We can confirm a precise correspondence.

\bibliography{lNscat}

\end{document}